\documentclass{aa} 
\usepackage{txfonts} 
\usepackage{graphicx}
\usepackage{natbib}

\bibpunct{(}{)}{;}{a}{}{,}

\begin{document}

\title{Designing future dark energy space missions: II. Photometric redshift of space weak lensing optimized surveys} 

\author{S. Jouvel \inst{1,2} \and J-P. Kneib \inst{1} \and G. Bernstein
  \inst{3} \and O. Ilbert \inst{4,1} \and P. Jelinsky \inst{5} \and
  B. Milliard \inst{1} \and A. Ealet \inst{6,1} \and C. Schimd \inst{1} \and T. Dahlen
  \inst{7} \and S. Arnouts \inst{1,8}} 
  \institute{
  Laboratoire d'Astrophysique de Marseille, CNRS-Universit\'e de Provence
  38 rue Fr\'ed\'eric Joliot-Curie; 13388 Marseille Cedex 13, France 
  \and University College London, Gower Street, London WC1E 6BT, UK
  \and University of Pennsylvania, 4N1 David Rittenhouse Lab 209 S 33rd St, Philadelphia, 
  PA 19104, USA 
  \and Institute of Astronomy, 2680 Woodlawn Drive, Honolulu, HI 96822, USA 
  \and University of California, Space Sciences Laboratory, Berkeley, CA 94720, USA
  \and Centre de Physique de Particule de Marseille,163 av de Luminy, Case 902, 13288 Marseille cedex 09, France 
  \and Space Telescope Science Institute, 3700 San Martin Drive, Baltimore,
  MD 21218, USA 
  \and CFHT, 65-1238 Mamalahoa Hwy, Kamuela, HI 96743, USA 
  }\date{accepted at A\&A}
\authorrunning{St\'ephanie Jouvel et al. }
\titlerunning{Photo-z of space WL optimized survey}
\offprints{St\'ephanie Jouvel, \email{s.jouvel@ucl.ac.uk}}

\abstract{
%context
  With the discovery of the accelerated expansion of the universe,
  different observational probes have been proposed to investigate 
  the presence of dark energy, including possible modifications to 
  the gravitation laws by accurately measuring the expansion of the 
  Universe and the growth of structures. We need to optimize the return 
  from future dark energy surveys to obtain 
  the best results from these probes. 
  }{
%Aims 
  A high precision weak-lensing analysis requires not an only accurate
  measurement of galaxy shapes but also a precise and unbiased
  measurement of galaxy redshifts. % The photometric redshift
  %technique appears as the only possibility to determine the redshift of
  %the background galaxies used in the weak-lensing
  %analysis. 
  The survey strategy has to be defined following both the photometric redshift
  and shape measurement accuracy.
  }{
%Methods 
  We define the key properties of the weak-lensing instrument and compute 
  the effective PSF and the overall throughput and sensitivities. 
  We then investigate the impact of the 
  pixel scale on the sampling of the effective PSF, and place upper limits on
  the pixel scale. We then define the survey strategy computing the survey 
  area including in particular both the Galactic absorption and Zodiacal light 
  variation accross the sky.
  Using the Le Phare photometric redshift code and realistic galaxy mock
  catalog, we investigate the properties of different filter-sets
  and the importance of the u-band photometry quality to optimize the 
  photometric redshift and the dark energy figure of merit (FoM).
  }{
%Results 
  Using the predicted photometric redshift quality, simple shape measurement 
  requirements, and a proper sky model, we explore what could be an optimal 
  weak-lensing dark energy mission based on FoM calculation. We find that we can 
  derive the most accurate the photometric redshifts for the bulk of the faint 
  galaxy population when filters have a resolution $\mathcal{R}\sim3.2$.
  We show that an optimal mission would survey the sky through eight filters using 
  two cameras (visible and near infrared). Assuming a five-year mission duration,
  a mirror size of 1.5m and a 0.5deg$^2$ FOV with a visible pixel scale of 0.15", 
  we found that a homogeneous survey reaching a survey population of I$_{AB}$=25.6 
  (10$\sigma$) with a sky coverage of $\sim$11000deg$^2$ maximizes 
  the weak lensing FoM. The effective number density of galaxies used for WL 
  is then $\sim$45gal/arcmin$^{2}$, which is 
  at least a factor of two higher than ground-based surveys.
  }{
%Conclusions
  This study demonstrates that a full account of the observational strategy is
  required to properly optimize the instrument parameters and maximize the FoM
  of the future weak-lensing space dark energy mission.
  } 

\date{\fbox{\sc Accepted in A\&A \today}}
\maketitle

%an optimal filter resolution, we come up with a
%  final best configuration of 8 filters and a number of 7 filters in a
%  minimum configuration. We argue that the goal of future space
%  mission should have 8 filters as it will minimize the systematic
%  errors which are a key points in achieving precision cosmology.

\section{Introduction}
\label{sec:intro} With the measurement of the accelerated expansion of
the Universe using Type Ia Supernovae
(\citet{Riess98,Wood-Vasey07,Kowalski08,Perlmutter99}), together with the flatness
of the metrics derived from many CMB balloon-borne and space
experiments (WMAP-7 years: \citet{Spergel03,Komatsu09}), cosmology has
entered a new era of precision measurements. The concordance Lambda cold dark matter model of the CMB and SNIa probes is also consistent with other probes 
(Baryonic Accoustic Oscillation (hereafter BAO) eg. \citet{Eisenstein05,Percival10},
cluster counts~ eg. \citet{Takada07}, weak-lensing (hereafter WL)~ eg. \citet{Fu08}).
This successful model has, however,
reintroduced Einstein's controversial cosmological constant, which
remains a mystery for fundamental physics. The contribution of the
cosmological constant could be similar to that of a "dark energy" (hereafter DE) 
that would explain the observation of an accelerating Universe. 
Other theoretical models propose a
change in the laws of gravity instead of adding an unknown "Dark
Energy" component. Discriminating between the several DE
solutions~\citep{Linder08} is the challenge of observational cosmology
over the next decade. It has in particular motivated the preparation of future 
space-based missions such as JDEM, the Joint Dark Energy Mission\footnote{http://jdem.gsfc.nasa.gov/} on the US side (for which 3 concepts were in competition:
SNAP \footnote{http://snap.lbl.gov/} DESTINY: \citet{Morse04}) 
and on the European side the EUCLID mission,
\footnote{http://sci.esa.int/euclid} which represents the ``merging'' of the
DUNE\footnote{http://www.dune-mission.net/} and the SPACE\footnote{http://urania.bo.astro.it/cimatti/space/} concepts.
%or ground based telescope such
%as the Large Synoptic Survey Telescope (LSST; \citet{Ivezic08}),and
%the Panoramic Survey Telescope \& Rapid Response System (pan-STARRS;
%\citep{Magnier07}).

To go beyond our current limited observations of the Universe, we
critically need new experiments that will provide new and numerous
observations of galaxies in the Universe to address the fundamental 
questions of cosmology.
Different cosmology probes have been proposed to measure the DE
equation of state. These include in particular SNIa~\citep{Dawson09}, WL
tomography~\citep{Massey07WL3D,Hu99}, and
3D-WL~\citep{Kitching07,Heavens03,Heavens06},
BAO~\citep{Padmanabhan09}, cluster counts~\citep{Marian09}), cluster strong lensing
\citep{Jullo09}, and Alcock-Pazsinsky test \citep{Marinoni10}.  The best approach 
is most likely to combine different probes, allowing us to minimize possible 
systematic effects.

WL has emerged as one of the most effective cosmological probes (\citet{DETF}
, see also the more recent JDEM FoM working group results
\citep{Albrecht09}) as it is sensitive to both the geometry (through
its dependence on angular-diameter distance ratio) and the growth of
structure. The observed shape of a distant galaxy depends on the
amount of mass distributed along the line of sight. To obtain the highest 
quality cosmological constraints, it is critical to derive
accurate redshift measurements of all the galaxies for which one can
measure their shape (\citet{Massey07STEP}). In other words, any future
WL imaging survey must address the question of the complementary
redshift survey. We are presently unable to measure the galaxy
redshifts of all the galaxies used in the shear estimation using spectroscopic
technique. The only solution is to use photometric redshift. 
Although photometric redshifts have now been used for many
years, the technique has mainly been developed using data available
at various telescopes. However, very rarely has an instrument or a
survey been designed to optimize the photometric redshift
measurement needed to reach a specific goal.

Previous work aimed particularly at optimizing photometric redshifts
for future surveys include e.g., \citet{Benitez09} and
\citet{Dahlen08}, which consider the filter properties, their number and
the photometry efficiency, and also \citet{Bordoloi09},
\citet{Schulz09}, \citet{Quadri09}, and \citet{Sheth10}, which evaluate
the possible improvement of the photometric redshift technique using
respectively, some work on likelihood functions, cross-correlation
methods, close galaxy pairs, convolution, and deconvolution methods
from a subsample of spectroscopic redshifts. 
A detailed study of the impact of photometric redshift errors on dark
energy constraints was performed by \citet{Hearin10} who generalized and 
extended the work of \citet{Bernstein10}. It studies in detail 
the different types of photoz errors, their impact on
dark energy parameters and the tolerances that will be useful in
future survey design. The present paper extends the earlier work of \citet{Dahlen08} 
and places the photometric redshift
determination in the global context of the DE mission optimization. 

As we prepare future cosmological surveys, it is important to develop 
the optimal observational strategy and the photometric 
data of a WL survey to maximize the prime science of the DE mission 
based on the DETF \citep{DETF} figure of merit.
To achieve this goal, we use mock catalogs with realistic galaxy
distributions as described in \citet{Jouvel09} (hereafter Paper I) that is 
specifically designed to address this problem.

This paper is organized as follows. In section \ref{sec:photoz} we
quickly summarize current photometric redshift techniques and characterize 
the likely photometric uncertainties of future WL missions. We develop the WL
requirements for future space DE missions in section \ref{sec:WLrequirements}.
In section \ref{sec:filterset}, we investigate different filter configurations and
underline the key characteristics of favored configurations. 
Section \ref{sec:blue} investigates the impact of the blue-band photometry 
efficiency to help decrease the catastrophic redshift rate.

Finally, in section \ref{sec:surveystrategy} we explore the survey
strategy in terms of a DE figure-of-merit (FoM) by
investigating how the survey efficiency depends on the number of filters,
the area of the sky surveyed, and the total exposure time per
pointing. We discuss the results and possible improvements in section
\ref{sec:conclusion}. 

%COMMENT: YOU MAY WANT TO CHECK THAT I GOT THIS RIGTH.}

Throughout this paper, we assume a flat Lambda-CDM cosmology and use the AB magnitude system.

%%%%%%%%%%%%%%%%%%%%%%%%%%%%%%%%%%%%%%%%%%%%%%%%%%%%%%%%%%%%%%%%%%%
\section{Photometric Redshifts and Photometric Noise}
\label{sec:photoz}

The photomeric redshift technique is to some extent similar to very
low resolution (typically $\mathcal{R}\sim 5$) spectroscopy, but instead of
identifying emission or absorption lines, it relies on the continuum of 
spectra and the detection
of broad spectral features generally strong enough to be detected in 
visible and NIR filters. These features include  ``breaks" or
"bumps" in the galaxy spectral energy distribution (SED)
\citep{Sawicki96,Bolzonella00,Benitez00}. Depending on the filter resolution, 
any spectral features that produce a change in colors can help the photometric 
redshift (hereafter photoz) determination.

There are three or four main spectral features that are particularly
helpful to the photoz procedure of which the most fundamental are the 
Balmer break at $\sim 3700\AA$ and the D4000$\AA$ break. 
Additional useful characteristics are the Lyman break at 912$\AA$ and the Lyman 
forest created by absorbers along the line of sight. 
However, the Lyman break only enters to the U-band filter at $z\approx 2.5$
and therefore only helps in breaking the color-redshift degeneracies 
for high redshift galaxies. In contrast the 1.6$\mu$m bump \citep{Sawicki02} 
might be capable of breaking the color-redshift degeneracies of low redshift
galaxies if a filter with coverage redder than 1.6$\mu$m is added to the filter set.

\subsection{Photometric redshift techniques}
\label{subsec:photoz}
There are two main types of methods that have been used to derive redshifts
based on the photometry of objects: (1.) Empirical methods such as neural
network (NN) techniques \citep{Collister04,Vanzella04} and (2.) template
fitting methods such as the BPZ Bayesian photometric redshift
of \citet{Benitez00}, HyperZ of \citet{Bolzonella00}, and Le Phare
\footnote{http://www.cfht.hawaii.edu/~arnouts/LEPHARE/cfht\_lephare/lephare.html}
used in \citet{Ilbert06,Ilbert09}.  Both methods includes two
steps. The first step is the most critical in ensuring the robustness
of the photometric redshift estimate. For the NN technique, this step
is crucial. It uses a training set of galaxies from which the NN
learns the relation between photometry and redshift. For the template
fitting method, this corresponds to the calibration of the library of 
galaxy templates thereafter used in the
redshift estimation. The template fitting method can work without this
first step but it may then introduce some bias if the templates 
used are not representative 
of the galaxies for which the photometric redshift are measured. 
However, we aim to obtain unbiased photometric redshift 
measurements for many faint galaxies, it is essential that we calibrate
the library of galaxy templates. Indeed, the calibration sample or the training set 
needs to be representative of the galaxy population for which we wish to find a
redshift. The second step in both methods is the photometric redshift computation of
the full galaxy sample from the photometry. The NN uses the complex
function learned from the training set, while the template fitting
method uses the calibrated library with a minimisation procedure to
derive a redshift estimation for each galaxy in a photometric
catalogue.
%%%%%%%%%%%%%%%%%%%%%%%%%%%%%%%%%%%%%%%%%%%%%%%%%%%%%%%%%%%%%%%%%%%
% \begin{figure}
% \caption{Extended CWW library of SED templates in the UV/visible  wavelength range. Templates ranges from star-forming galaxy to old elliptical galaxy.}
% \psfig{figure=/Save/sjouvel/lephare/sed/GAL/CE_NEW/all_sedm.ps,angle=0,width=8cm}
% \label{fig:library}
% \end{figure}
%%%%%%%%%%%%%%%%%%%%%%%%%%%%%%%%%%%%%%%%%%%%%%%%%%%%%%%%%%%%%%%%%%% 

In this investigation, we use the Le Phare photometric redshift code,
which is based on the template fitting method. The code is applied to
galaxies in a mock galaxy catalog that we describe in the next
subsection.  For each galaxy, the code derives a photometric redshift
and a best-fit galaxy template using a $\chi^{2}$~minimisation
defined as
\begin{equation} \label{eq:chi2}
\chi_ {model}^{2}=\sum_{i=1}^{n}([F_{obs}^{i}-\alpha F_{model}^{i}]/\sigma^{i})^{2}
\end{equation}
where $F_{obs}^{i}$~and $F_{model}^{i}$ are the observed and the template
model fluxes inside a filter $i$ and $\sigma^{i}$ is the 
photometric error for this filter in a given survey configuration (as
defined in section \ref{subsec:noise}). Photometric errors play the
role of a weight in the $\chi^2$ minimisation method and $\alpha$ is a
normalisation factor. The photometric redshift and best-fit template
correspond to the minimum value of the $\chi^2$ distribution for a
given simulated galaxy.\\

\subsection{CMC mock catalogue}
In Paper I, we developed  realistic spectro-photometric mock
galaxy catalogs. In this paper, we use one of those catalogs, the
COSMOS mock catalog (hereafter CMC), which was built from the observed
COSMOS data set \citep{Scoville07,Capak08}.  This catalog uses the 
photometric redshift and best-fit template
distribution of \citet{Ilbert09}. Using these two pieces of information, we
calculate the theoretical fluxes of each galaxy in each band of a given survey
configuration. We then draw an observed flux from a Gaussian distribution 
based on the error estimate to simulate the observed galaxy photometric properties. 
The errors depend on the survey configuration, and the method used to calculate 
them is described in Section
\ref{subsec:noise}. We note that the mock galaxy catalog is produced
using the same set of templates utilised by the photometric
redshift code. However, the representativeness of the calibration sample
in the template fitting procedure is not the aim of this paper but
will be studied in a future paper \citet{Jouvel10a}. Thus we assume a ``perfect''
calibration in using the same library of templates for the
development of the mock catalog and in the $\chi^2$
procedure. Despite this being a very optimistic
case, it provides predictions and some results
in the "optimal" case.

The CMC assigns several emission lines to all galaxies in the catalog
based on their $[OII]$ fluxes, using the calibration of
\citet{Kennicutt98}.  The emission line fluxes are added to the flux
derived from the continuum of each galaxy in the mock catalogue. This
creates a natural dispersion in the simulated magnitudes, reflecting
what will be observed in future real data. The bias that the emission lines
will produce in the photometric redshift estimate is one of the justifications for a 
photometric redshift calibration survey (PZCS) ideally covering
the same range in magnitude and redshift as the photometric galaxy
catalogue. A wide and deep PZCS will help us to decrease the bias and
dispersion of the photometric redshift distribution using template 
calibration techniques. In optimizing the
library of templates used in the photometric redshift analysis, we
will be able to reproduce more accurately the diversity of the observed 
galaxy population including the impact of the emission line fluxes as shown in
\citet{Ilbert09}, who found that their results are greatly improved where a
spectroscopic galaxy sample is available. The new version of the Le
Phare code includes the emission line fluxes in the library
of templates as described in \citet{Ilbert09}. This last feature was a major
impact in helping to improve photometric redshift results of \citet{Ilbert09}.
%This is one more free parameter, in addition to
%the dust extinction \citep{Calzetti00}, which increases the photometric
%redshift uncertainty in adding more color degeneracies.
%This will be discussed in section \ref{sec:WLrequirements}.

\subsection{Typical noise properties for a space based survey}
\label{subsec:noise}

In Paper I, we did not discuss in detail the typical photometric uncertainty caused
by the instrument design and survey strategy.
Since we wish to investigate the photometric redshift
quality of future surveys, we now need to produce a realistic noise
distribution for each galaxy in our catalog.
To achieve this, we assign a photometric noise
to each band that depends on the galaxy size and flux.  Since we use
electronic devices, the photometric signal is physically stored as
electrons. Thus we express our formulae in terms of the number of electrons,
which is proportional to the number of photons. We define
$e_{signal}$~ as the number of electrons produced by the galaxy
flux. The photon noise can be described by a Poissonian
statistic. Other sources of uncertainty originate in
the instrument electronic devices and other
astrophysical sources photons detected at the telescope. These studies
are space oriented so the main source of background noise comes from
the Zodiacal light $e_{sky}$ which is true in particular for a mission orbiting L2.
The thermal radiation of the detector
results in a ``dark current" $e_{dark}$, while the reading of
the detectors results in a read-out noise $e_{RON}$ described with a Gaussian 
statistic. We go through each of these four terms contributing to the noise 
in the Appendix.

The signal-to-noise ratio including all the noise contributions is
defined by
%\begin{equation}
%\label{eq:sn}
%S/N = \frac{e_{signal}}{\sqrt{e_{signal}+N_{pix}e_{sky}+N_{pix}N_{expo}e_{RON}^2+N_{pix}N_{expo}T_{obs}e_{dark}}}
%\end{equation}

\begin{eqnarray}
\label{eq:sn}
\lefteqn{S/N = }
\nonumber\\
& & {}\frac{e_{signal}}{\sqrt{e_{signal}+N_{pix}e_{sky}+N_{pix}N_{expo}e_{RON}^2+N_{pix}N_{expo}T_{obs}e_{dark}}},
\nonumber\\
\end{eqnarray}
where $N_{expo}$ is the number of exposures, $T_{obs}$ the exposure time, and 
$N_{pix}$ the number of pixels taken in the flux error calculation. We took the RON
to be $e^{vis}_{RON}=6e^-/pix$ and the dark current $e^{vis}_{dark}=0.03e^-/pix/s$
for the visible detectors and $e^{ir}_{RON}=5e^-/pix$ and $e^{vis}_{dark}=0.05e^-/pix/s$
for the NIR detectors. All parameter values are listed in the Appendix of 
table \ref{tab:jdem}. 
These performances are achieved or expected in the near future
from detectors of future DE missions.
Thus, for each galaxy in each band, we calculate a S/N from 
equation \ref{eq:sn} and compute an observed flux $f^{obs}_{gal}$~
that includes a random noise drawn from a Gaussian distribution whose
characteristics are $(\mu,\sigma)=(f^{theo}_{gal},S/N)$, where
$f^{theo}_{gal}$ is the noiseless or theoretical flux value given by
the CMC mock catalog. Thereby, using the mock catalogs of
\citet{Jouvel09} and characteristics of future surveys, we compute
realistic mock galaxy catalogs for future WL DE surveys
including a redshift $z_s$, template model, galaxy fluxes, and uncertainties
in each photometric band, in addition to a galaxy size. More details about the
calculation of the S/N are given in the Appendix.
%%%%%%%%%%%%%%%%%%%%%%%%%%%%%%%%%%%%%%%%%%%%%%%%%%%%%%%%%%%%%%%%%%%
\begin{figure}[!ht]
\resizebox{\hsize}{!}{\includegraphics{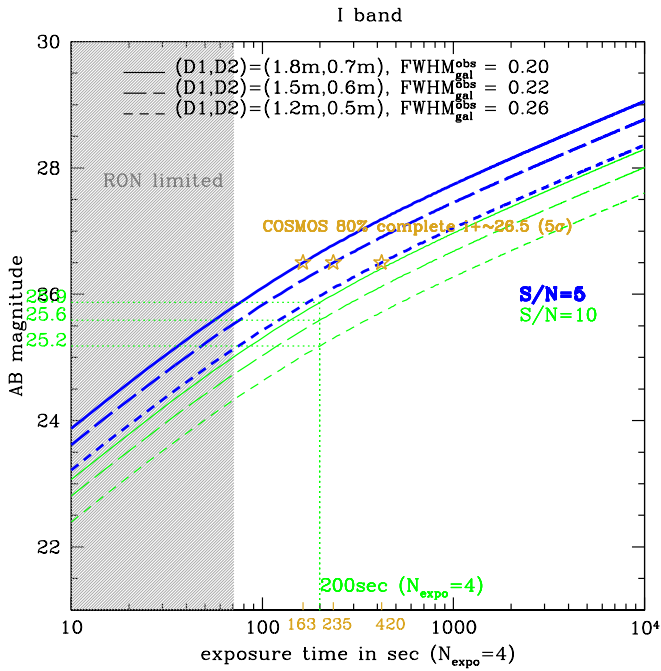}}
  \caption{$I_{AB}$ magnitude as a function of exposure time for
    mirror diameters of 1.2m, 1.5m, and 1.8m using a pixel size of
  respectively 0.19",0.15", and 0.12" and a filter resolution of 3.2. 
  The magnitude is calculated for four exposures ($N_{expo}=4$) assuming an   
  exposure time by exposure written on the axis, a RON of $e^{vis}_{RON}=6e^-/pix$ 
   and a dark current of $e^{vis}_{dark}=0.05e^-/pix/s$. 
   All parameters values are listed in the Appendix table \ref{tab:jdem}. 
  Magnitudes are calculated inside a circular aperture of 1.4$\times$FWHM.}
\label{fig:expomag}
\end{figure}
%%%%%%%%%%%%%%%%%%%%%%%%%%%%%%%%%%%%%%%%%%%%%%%%%%%%%%%%%%%%%%%%%%%
Following this noise prescription,  {\bf Figure~\ref{fig:expomag}}
shows the I-band magnitude as a function of exposure time for a given
$S/N\approx 5$ and 10 in blue and green, respectively, and for mirror
sizes of 1.2m (small-dashed lines), 1.5m (large-dashed lines), and
1.8m (solid lines).  These values are derived assuming an obstructed 
telescope design with
a mirror size for the secondary of 60\% of the primary mirror. 
This shows for example that a 1.5m telescope and a survey strategy of four
exposures of 200s (800s of total integration time) reaches a
magnitude of $I_{AB}$=25.8 ($S/N\approx10$) for a galaxy source of
$FWHM^{obs}_{gal}=0.20$, where $FWHM^{obs}_{gal}$ is the observed FWHM of a galaxy. 
Magnitudes are computed inside a circular aperture 
of 1.4$\times$FWHM. The stars in gold represents the exposure time needed to reach
the COSMOS completeness for different telescope diameters calculated using our
noise prescription.  
The magnitude calculation is described in the Appendix.

To obtain an accurate WL measurement, it is safe to use the galaxies
whose FWHM are larger than 1.25$\times$[FWHM(ePSF)] and $S/N>10$, where the ePSF is
the effective PSF of the telescope defined in section
\ref{subsec:ePSF}. The COSMOS WL analysis used a criterion
of 1.6x[FWHM(ePSF)] and a S/N$>$10, but we hope that an image 
analysis technique of higher quality will improve the COSMOS limit 
in the future. The choice 
of a factor of 1.25 although an arbitrary criterion, allows us to easily
compare different survey designs by using a simple size cut as a quality cut.

\begin{figure}[!ht]
\hbox{
\resizebox{\hsize}{!}{\includegraphics{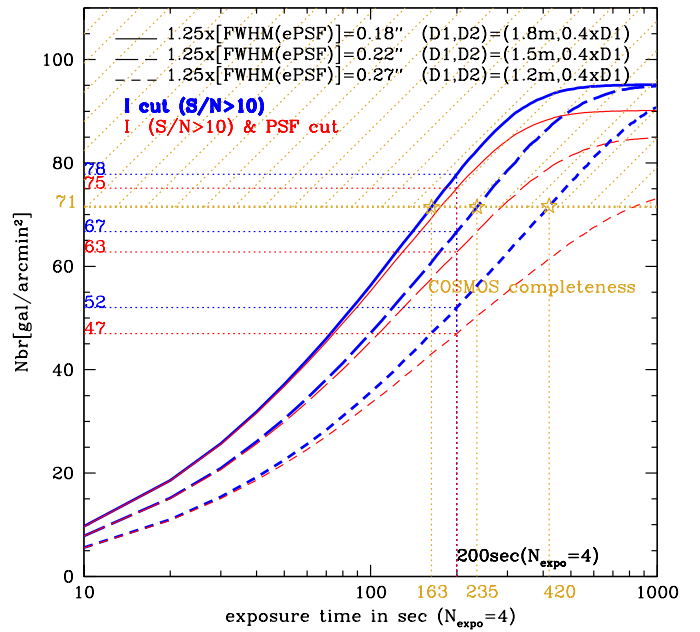}}
}
\caption{Effective number of galaxies as a function of exposure time
  for a mirror diameter of 1.2m, 1.5m , and 1.8m using a pixel size of
  respectively 0.19",0.15", and 0.12" and a filter resolution of 3.2.
   All parameters used to produce this figure are listed in the Appendix Table \ref{tab:jdem}.}
\label{fig:expodensity}
\end{figure}

{\bf Figure~\ref{fig:expodensity}} shows the
number density of galaxies reached using these criteria for a primary
mirror size of 1.2m, 1.5m, and 1.8m. In decreasing the primary mirror
diameter by 0.3m, the galaxy number density is reduced by 13gal/arcmin$^2$ 
when going from 1.8m to 1.5m,
and 18gal/arcmin$^2$ when going from 1.5m to 1.2m. We choose to use a
pixel scale that varies with the mirror size to ensure an equal sampling
of the effective PSF. We choose, respectively, a pixel scale of 0.19" for
a mirror size of 1.2m, and 0.15" for 1.5m and 0.12" for 1.8m.
%%%%%%%%%%%%%%%%%%%%%%%%%%%%%%%%%%%%%%%%%%%%%%%%%%%%%%%%%%%%%%%%%%%
%%%%%%%%%%%%%%%%%%%%%%%%%%%%%%%%%%%%%%%%%%%%%%%%%%%%%%%%%%%%%%%%%%%
Figure~\ref{fig:expodensity} also shows that the quality cut based on galaxy size
produces a loss of 3, 5, and 6gal/arcmin$^2$, respectively for a primary 
mirror diameter of 1.8m ,1.5m, and 1.2m.  We note that the loss is more
significant for smaller mirror sizes. This is due to the relationship 
between the mirror size and the pixel scale. Smaller mirror have larger pixel 
scales, which makes the quality cut on galaxy size
more stringent. However, this is a small loss compared to the
31gal/arcmin$^2$ that one loses when going from a mirror size of 1.8m to
1.2m based on 4 exposures of 200s. Using the exposure time needed 
to reach the COSMOS completeness (shown in Figure~\ref{fig:expodensity}), 
we have a galaxy density of 71gal/arcmin2. This defines an exposure time-density 
domain in which the COSMOS catalog and the CMC are complete. The dashed gold region
corresponds to areas where the CMC catalogues produced are incomplete. 
In these areas, conclusions may be affected by the incompleteness of the CMC. 

% We need to say what weak lensing wants from photo-z:\\
%- unbiased measurement \\
%- maximise the number of galaxies with photo-z: define the fraction of
%galaxies with trustable photometric redshift.  To acheive
%that goal we have to optimise our templates. There is differents
%technique like the one implemented in ZEBRA \citet{Feldmann06} which
%uses a mathematical decription to create optimised template that they
%add in their library. An other technique is developped in
%\citet{Ilbert08} based on the emission line galaxies on the input
%library.

\subsection{How to characterize the photometric-redshift quality}
\label{subsec:photozq}

We call the redshift coming from the input mock catalog the
spectroscopic redshift $z_s$, while the photometric redshift $z_p$
corresponds to the redshift calculated by the photometric redshift
technique. Considering a photometric redshift distribution, we define
the ``core of the distribution" as the galaxies for which
$|zp-zs|<0.3$ and the ``catastrophic redshift" as the galaxies outside
the core. We did not include the division by $1+zs$ since we had not
intended to produce results to be used in WL analyses, but to instead 
assess the photometric redshift quality.
 %%%%%%%%%%%%%%%%%%%%%%%%%%%%%%%%%%%%%%%%%%%%%%%%%%%%%%%%%%%%%%%%%%%
\begin{figure}[!h]
%\resizebox{\hsize}{!}{\includegraphics{zpzs_def_Qjpeg.ps}}
%\resizebox{\hsize}{!}{\includegraphics{zpzs2D_bin0.03_col500_6QE_Qjpg.ps}}
%\resizebox{\hsize}{!}{\includegraphics{zpzs2D_bin0.05_col300_6QE_Q50.ps}}
\resizebox{\hsize}{!}{\includegraphics{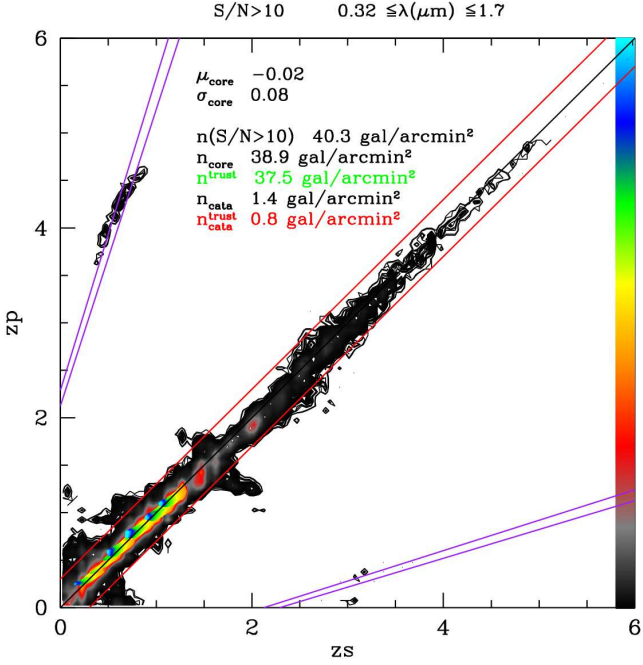}}
\caption{Illustrative diagram of photometric versus spectroscopic
  redshift, where we identify the quantities assessing the photometric redshift
  quality. }
\label{fig:photoz}
\end{figure}
%%%%%%%%%%%%%%%%%%%%%%%%%%%%%%%%%%%%%%%%%%%%%%%%%%%%%%%%%%%%%%%%%%%
We define some characteristic numbers that we use to quantify the
quality of a photometric redshift distribution:
\begin{itemize}
\item $\sigma_{core}$, the dispersion of the core distribution defined as $\sigma(|z_p-z_s|<0.3)$.
\item $\mu_{core}$, the bias measured from the mean or median of the core
  distribution defined as $\mu(|z_p-z_s|<0.3)$.
\item $n_{core}$, the number density of galaxies inside the core distribution. 
\item $n^{trust}$, the number density of galaxies with a photoz of high confidence that we defined as $\Delta^{68\%} z<0.5$.
\item $n_{cata}$, the number density of galaxies with a catastrophic redshift defined as $|z_p-z_s|>0.3$.
\item $n^{trust}_{cata}$, the number density of galaxies with a photoz of high 
confidence being catastrophic redshifts.
\end{itemize}

{\bf Figure~\ref{fig:photoz}} is an illustrative density diagram
showing photometric versus (vs.) spectroscopic redshift with the core of the
distribution being located inside the two red lines, and the catastrophic
redshifts outside these red lines. Following this definition
of catastrophic redshift, there are two kinds of "bad" redshift. The
galaxies surrounding the red lines and the galaxies situated close to
and within the two purple lines. There are two main
reasons for the redshift procedure to fail, which we discuss below. \\

\subsubsection{A confusion between the Lyman and Balmer breaks}  
This break confusion is represented by the four purple lines. In an
ideal case, if the photometric redshift procedure identifies a 
higly accurate redshift $zp=zs$, we find that
\begin{equation}
\lambda_{break-obs}=(1+zs)\lambda_{break-rf},
\end{equation}
where $\lambda_{break-obs}$ is the observed wavelength of one of the
breaks used in the photometric redshift procedure, $\lambda_{break-rf}$ is the
rest-frame wavelength of the same break, and $zs$ the spectroscopic
redshift of the galaxy. In the case of a catastrophic redshift
$zp=z_{cata}$, this equation becomes
\begin{equation}
\lambda_{break-obs}=(1+z_{cata})\lambda_{break-cata}.
\end{equation}
We can then write:
\begin{equation}\label{eq:zcata}.
z_{cata}=(1+zs)\frac{\lambda_{break-rf}}{\lambda_{break-cata}}-1
\end{equation}
We define four line couples $(break-rf,break-cata)$ that are sources of the 
color degeneracy producing the catastrophic redshifts, where $break-rf$ is 
the real feature and $break-cata$ is the wrong feature found by a photoz code :
\begin{equation}
(break-rf,break-cata)=
\left( \begin{array}{cc}
Ly\ \alpha & Balmer \\
Ly\ \alpha & D4000 \\
Ly\ break & Balmer\\
Ly\ break & D4000
\end{array} \right)
\end{equation}
These couples used in equation \ref{eq:zcata} define the four
purple lines in the zp-zs plane where the catastrophic redshift
 happens with the highest probability.

This confusion occurs for both low and high redshift galaxies,
generally at $z<0.5$ and $z\geq2.5$, depending on the wavelength
range available to the instrument. A wide wavelength range
going from U to K band would avoid most catastrophic redshifts by using both
the U-band and NIR photometry. The Balmer break can be followed at all
redshifts from the V-band photometry ($z\sim0$ Balmer break$\sim$ 4000
\AA) to H-band photometry ($z\sim3$ Balmer break$\sim$ 16000
\AA). However, it can be misidentified as the Lyman break leading to
the creation of catastrophic redshifts. This misidentification can be
avoided by using deep U-band photometry. 

% The Lyman break is followed by a
% strong Lyman forest due to the absorbers along the line of sight. This
% information helps to distinguish between high and low reshift galaxies
% thanks to the U-band photometry as can be seen in {\bf Figure
%   \ref{fig:ubandpower}}. The Madau extinction \citep{Madau95} has not
% been applied in this figure. This is just an illustrative diagram
% aiming to show that the U-band photometry can be useful.
%%%%%%%%%%%%%%%%%%%%%%%%%%%%%%%%%%%%%%%%%%%%%%%%%%%%%%%%%%%%%%%%%%%
% \begin{figure}[!h]
% \resizebox{\hsize}{!}{\includegraphics{uband_power.ps}}
% \caption{Illustrative diagram showing the importance of the U-band
%  photometry.}
%{\bf COMMENT: IS THE MADAU ATTENUATION APPLIED TO THE
%SEDs PLOTTED?, IT DOESN'T LOOK LIKE IT}}
% \label{fig:ubandpower}
% \end{figure}
%%%%%%%%%%%%%%%%%%%%%%%%%%%%%%%%%%%%%%%%%%%%%%%%%%%%%%%%%%%%%%%%%%%

The break confusion will generally produce a double peak in the
redshift probability distribution of low-redshift galaxies $0<z_s<0.5$, one 
at the correct redshift, and one at higher redshift $3.5<z_p<4$, 
which corresponds to a ``catastrophic redshift''. Hence, the derived photometric 
redshift distribution can be biased having an excess of galaxies 
with $3.5<z_p<4$, which will strongly perturb the
DE parameter estimation~\citep{Huterer06}. In
section \ref{sec:blue}, we investigate more quantitatively the gain
of an efficient U-band in minimizing the break confusion.

\subsubsection{An inaccurate template fitting}
The photometric redshift
dispersion and biases depend on the quality of the photometry of
galaxies (which can be affected by instrumental defects or crowded fields). 
Deeper photometry helps to provide higher accuracy photometric redshifts
at a given magnitude.  The galaxy color
accuracy is higher with deeper photometry and the weight of the fit
given by the S/N is higher, which both decrease the
dispersion and possible biases in the photometric redshift
estimate. In addition, a slight filter calibration error is enough to bias
the photometric redshift distribution. 
A way in minimizing the dispersion
and biases of the photometric redshift estimate is to optimize the
resolution of the photometric bands. We explore this solution in section
\ref{subsec:res}.

\section{Weak lensing survey key parameters and definitions}
\label{sec:WLrequirements}
To reach the goal of precision cosmology, it is essential to optimize
the instrument design and survey strategy, which both impact the
quality of the WL results. The present section aims
to introduce the quantities used in the DE parameter
estimation such as: 
(1) the galaxy number density which is a function of the exposure time and the photometric redshift quality 
(2) the survey area including the impact of the Galactic absorption
(3) the pixel size which impact the quality of the photometry and the shape measurement
(4) the minimum exposure time to be in the photon noise regime.  

%While we define all the WL survey key parameters here, it is only in section \ref{sec:surveystrategy}, that we will compare the different survey strategies.

\subsection{Weak lensing dark energy parameter list}

%To estimate the dark energy parameters, the cosmological WL analyses
%aims to measure the fluctuations of matter density on large scales
%using the distorted galaxy shapes due to the density fluctuations at
%different angular distance scales. These deformations are very small
%typically of the order or less than the deformations introduced by the
%PSF and one has to average the galaxy shapes in order to measure the
%WL signal. We thus need a large number of galaxies to reduce the
%statistical noise on the measured signal.

One of the possible way of constraining DE is
the WL tomography described in either \cite{Hu04} or \cite{Amara07}. 
This method divides the source distribution in redshift
slices, thus requires that accurate and unbiased photometric redshifts 
be available for most galaxies. A number of factors affect the FoM of 
this technique including, (1) the number of galaxies useful to the WL
measurement, (2) the systematic errors in the shape measurement, and
(3) the errors and biases in the photometric redshift distribution.

The FoM formalism of the iCosmo package
\citep{icosmo} is based on the WL tomography method. Using the galaxy
densities defined from the photometric redshift results from our mock
catalogs and the FoM from iCosmo, we look at the impact of the
photometric redshift quality on the DE parameter
estimation. We assume a flat cosmology where the fiducial values of
cosmological parameters are
($\Omega_m,w_0,w_a,h,\Omega_b,\sigma_8,\Omega_{\Lambda})=[0.3,-0.95,0,0.7,0.045,0.8,1,0.7]$.
We compute the FoM of the $(w_0,w_a)$ DE
parameters in marginalising over the other cosmological parameters and
using five tomographic redshift bins that have been found to provide the
most accurate FoM \citep{Sun09}. The redshift distribution follows a
distribution described in \citet{Smail95} and \citet{Efstathiou90}
\begin{equation}
N(z)\propto z^\alpha\exp{-(1.41z/z_{med})^\beta}
\end{equation}
with parameters $\alpha,\beta=[2,1.5]$ following the COSMOS redshift
distribution fit of \citet{Massey07WL3D}. The boundaries of the tomographic
redshift bins are calculated to produce an equal repartition of the
number of galaxies in each of the five redshift bins. To calculate the
FoM, the key numbers that we derive from our mock catalogs are
\begin{itemize} 
\item $N_{gal}$, the galaxy number density of galaxies that satisfy
\begin{eqnarray}
  \frac{\Delta^{68\%}z}{1+zp}<\epsilon \quad \textrm{S/N}_{I-band}\ge10 \nonumber\\
\textrm{FWHM}_{gal}>1.25\times \textrm{FWHM}_{ePSF},
\end{eqnarray}
where $ePSF$ is defined in section \ref{subsec:ePSF} and $\epsilon=(0.1,0.5)$
is a parameter that defines the quality of the photometric redshift.
\item $z_{med}$ is the median of the photometric redshift
  distribution of $N_{gal}$.
\item $\mathcal{A}$ is the survey area derived from the instrument
  field-of-view and the survey strategy, explained in section \ref{subsec:area}.
%\begin{equation}
%\mathcal{A}=\frac{year_{sec}\ FOV\ \zeta\ n_{cam}}{T_{obs}\ N_{expo}\ n_f}
%\end{equation}
%  year_in_sec=365.25*24*3600 expotime_ron=expotime[il]+30
%  area=year_in_sec*fov*survey_eff/expotime_ron/nb_expo*fov/nfilt_by_cam*(1.-loss_stars)
\end{itemize} 
The number of objects $N_{gal}$ depends on the photometric redshift
error criteria that we assume, which are parametrized by $\epsilon$ (studied in
section \ref{sec:surveystrategy}), the primary mirror size ($D_{1}$), 
and the pixel scale ($p^{vis}$), which enter in the definition of the 
effective PSF (ePSF) discussed in section \ref{subsec:ePSF} and  
in the photometric uncertainties described in section \ref{subsec:noise}.
In section \ref{subsec:ePSF}, we define a maximal and an optimal pixel size 
by means of their impact on the size of the effective PSF, which determines 
the useful number of galaxies: $N_{gal}$. In section \ref{subsec:tobs}, 
we define the minimum exposure time in the photon noise regime depending 
on the instrument parameters.
We then study in section \ref{subsec:area} the survey area $\mathcal{A}$ 
taking into account the Zodiacal light and Galactic absorption, which 
both depend on the sky position.

\subsection{ePSF: Effective PSF of the telescope and optimal pixel scale}
\label{subsec:ePSF}
The future observation strategy is to survey a large fraction of the sky. 
This would be easier using large pixels typically of the order of the PSF size,
which is a function of the mirror size (see Table \ref{tab:jdem}); this would 
help to optimize the area versus observation time without under-sampling 
the PSF too much, which would affect the quality of the WL measurement. In
this section, we define the pixel scale to be used in the
calculation of the noise properties.
\subsubsection{Formalism}
Using the formalism of \citet{High07}, the observed galaxy shape 
$I_{obs}(\theta)$ is expressed as the convolution of three components of 
the intensity profile of galaxies, the pixel response $p(\theta)$ and the 
PSF of the telescope
\begin{equation}
I^{obs}(\theta)=I^{galaxy}(\theta)*PSF(\theta)*p(\theta).
\end{equation}
The theoretical PSF size $PSF(\theta)$ corresponds to the size of the Airy
disk. Its size is a function of the wavelength $\lambda$ and the
mirror size on the basis of the relation
\begin{eqnarray}\label{eq:airy_disk}
PSF(\theta)=2.44\frac{\lambda}{D_1}.\\
\end{eqnarray}
%Airy\ disk\ radius &=& 1.22. \frac{\lambda}{D_1}\\
Similarly the full width half maximum of this PSF is defined by 
\begin{eqnarray}\label{eq:airy_fwhm}
\textrm{FWHM}[Airy\ disk]  &=& 1.02. \frac{\lambda}{D_1},
\end{eqnarray}
where $D_1$ is the diameter of the primary mirror. The PSF and pixel
response introduce systematics in the WL measurement and
need to be extracted from the galaxy shape before doing any lensing
calculations. For this purpose, we use point-like sources such as
stars to correct for both PSF circularization and anisotropic deformation. 
Thus, we define the effective PSF
($ePSF$) corresponding to the star intensity profile, which is the
convolution of the PSF and the pixel response that will be observed on
telescope images
\begin{equation}
ePSF(\theta)=\delta(\theta)*PSF(\theta)*p(\theta).
\end{equation}
This ePSF corresponds to the resolution of the instrument or the
smallest size resolved by the telescope. To obtain a rough estimate of
the size of the $ePSF$, we assume Gaussian distributions
for the pixel response, the PSF of the telescope, the jitter, and the pixel
diffusion. The jitter and pixel diffusion also affect
to the size of the observed PSF and need to be taken into account
\citep{Ma08b}. 
Thus we define the effective PSF expressed in arcsec as
%% JPK - I changed the 0.5 p^2 into a 0.2 p^2 to math the equation in the appendix because FWHM=2.36\sigma
\begin{equation}\label{eq:ePSF}
ePSF=\sqrt{\left(PSF(\theta)\right)^2+0.2\ p^2+\sigma_j^2+\left(\frac{\sigma_d}{0.1}.p\right)^2}
\end{equation}
where $p=p^{vis/ir}$ is the pixel scale (visible or IR camera),
$\sigma_j$ represents the jitter of the telescope and $\sigma_d$ the
diffusion of the pixel. The pixel diffusion varies as a function of the
pixel size ($p$) and has a typical value of $\sigma_{d}=$0.04" for a 
pixel size of 0.1",
which is equal to a diffusion of 0.4 pixel (see Table \ref{tab:jdem}).

\subsubsection{Maximal and optimal pixel scale}
Extrapolating from the results of \citet{High07}, we define the maximal
pixel scale. In the context of a DE WL survey, \citet{High07} defined 
an optimal pixel size for a primary mirror size of 2m to be 0.09" 
with one exposure at 0.8$\mu$m. This pixel scale slightly undersamples the ePSF. 
However, the pixel scale can be increased if a combination of 
sub-pixel dithered images are used. Different techniques can be used to 
combine the sub-sampled images such as the drizzling technique \citep{Fruchter02} 
working in real space or the method
proposed by \citet{Lauer99} which works in Fourier space. To
recover the loss of information caused by the ePSF under-sampling,
the minimum number of exposure has to be
$N_{min}=\Big(\frac{p^{us}}{p^{vis}}\Big)^2$, where $p^{us}$ is the
under-sampled pixel scale and $p^{vis}$ the optimal pixel scale for
one exposure. \citet{High07} showed that for a primary mirror size of 2m,
a pixel scale of 0.16" using four perfectly interlaced images is 
a good alternative to one exposure with 0.09", if assuming a perfect 
image reconstruction from the four dithered exposures. 
In terms of PSF sampling, this would allow us to
undersample the PSF by a factor $\xi$ of
\begin{equation}\label{eq:NIRpix}
\xi= \frac{PSF\ size\ (D_1=2m)}{p^{us}}=\frac{0.14}{0.16}\approx0.87. 
\end{equation}
Using the under-sampling factor $\xi$~and a 1.5m telescope, we find a 
pixel scale of $p^{us}\approx PSF\ size\ (D_1=1.5m)/\xi\approx0.2$", and
for a 1.2m telescope $p^{us}\approx 0.25$".

\subsubsection{Pixel scale estimated with the MTF (modulation transfer function)}
%%%%%%%%%%%%%%%%%%%%%%%%%%%%%%%%%%%%%%%%%%%%%%%%%%%%%%%%%%%%%%%%%%% 
\begin{figure}[!h]
  \resizebox{\hsize}{!}{\includegraphics{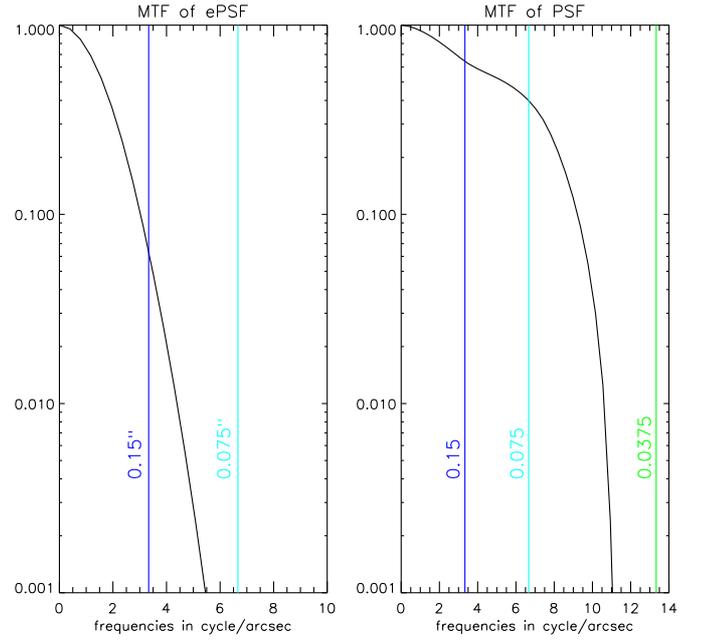}}
  \caption{(right) MTF[PSF] and (left) MTF[ePSF] at $\lambda\approx
    8000$\AA, for a mirror size of 2m. The green line stands for a
    pixel scale of $p^{vis}=0.0375$", cyan $p^{vis}=0.075$", and blue
    $p^{vis}=0.15$". The ePSF comes from equation with a pixel scale
    of $p^{vis}=0.075$", a jitter $\sigma_j\approx0.04$ and a
    diffusion $\sigma_d=0.04$. The colors represents
    different pixel scales studied for the optical camera in arcsec. 
    The pixel scale of the NIR camera is explained in equation \ref{eq:NIRpix}.}
  \label{fig:MTF}
\end{figure}  
%%%%%%%%%%%%%%%%%%%%%%%%%%%%%%%%%%%%%%%%%%%%%%%%%%%%%%%%%%%%%%%%%%% 

\begin{figure}[!h]
  \resizebox{\hsize}{!}{\includegraphics{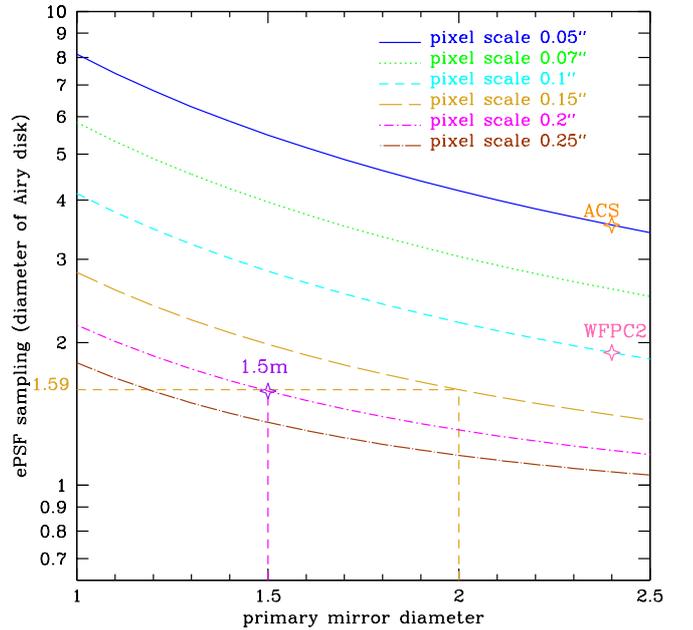}}
  \caption{ePSF sampling as a function of primary mirror diameter for
    different pixel scale at 800nm. The pink and orange points
    represent respectively the WFPC2, and ACS camera of the Hubble
    Space Telescope. The purple point represents a 
    mirror size of 1.5m and a pixel scale of 0.2".}
  \label{fig:ePSFmir}
\end{figure}

Another way to define the optimal pixel scale of our configuration
($D_1=1.5m$ and $N_{expo}=4$) is to apply the
Nyquist-Shannon theorem. This theorem says that a function is
completely determined if it is sampled at $\frac{1}{2f_{max}}$, where
$f_{max}$ is the highest frequency of the given function. We thus
trace the Fourier transform coefficients of the ePSF and check that a
pixel scale of $p^{vis}$ is given by $p^{vis}\geq\frac{1}{2f_{max}}$. {\bf Figure
\ref{fig:MTF}} shows the modulation transfer function $MTF$ of the
PSF and ePSF of a 2m mirror, where ePSF is defined in equation
\ref{eq:ePSF}. The MTF corresponds to the coefficients of the Fourier
transform, that we trace as a function of the frequency. This shows
that using the instrumental characteristics we defined, a pixel
scale of 0.15" is a good choice, if we assume a perfect information
recovery using the dithering technique. A pixel scale of
0.075" samples the ePSF well and allows us not to lose any information
at any frequency. We translate this information to a mirror of 1.5m
using {\bf Figure \ref{fig:ePSFmir}}.
%%%%%%%%%%%%%%%%%%%%%%%%%%%%%%%%%%%%%%%%%%%%%%%%%%%%%%%%%%%%%%%%%%% 
%%%%%%%%%%%%%%%%%%%%%%%%%%%%%%%%%%%%%%%%%%%%%%%%%%%%%%%%%%%%%%%%%%%
This figure shows the ePSF sampling in an I-band filter as a function
of the primary mirror diameter for different pixel scales. The ePSF
diameter is defined as two times the ePSF radius in equations
\ref{eq:airy_fwhm} and \ref{eq:ePSF} in section
\ref{sec:WLrequirements}. This figure shows that a pixel scale of
0.15" for a primary mirror size of 2m is equivalent to a pixel scale
of 0.2" for a primary mirror size of 1.5m in terms of ePSF
sampling. We thus define 0.2" pixel as the maximal choice for future WL
surveys, assuming a perfect information recovery with a perfect half-pixel 
dithering. %With a pixel scale of 0.2", the ePSF is thus sampled with 1.5 pixels. 
With the perfect sub-pixel dithering, we can hope to recover the information to reach
0.1"/pixel with a sampling of 3 pixels for the full ePSF equivalent to
1.5 pixel over the FWHM[ePSF].
Such a configuration will be discussed in more detail in a separate paper \citep{Jouvel10b}.

The maximum pixel scale of 0.2" for a 1.5m telescope 
should be considered as the upper limit while a safer solution, 
which we suggest is ``optimal'', has a pixel scale of 0.15" 
(two pixels sampling of ePSF), which is comparable in terms of the 
sampling of the ePSF to the sampling of the WFPC-2 camera of the 
{\it Hubble Space Telescope}. 

A higher sampling rate was proposed by \citet{Paulin-Henriksson08} to reach 
a good ePSF sampling and measure accurately this ePSF, allowing excellent 
WL measurement.
However, enlarging the pixel size also provides a larger field of view 
(for a given number of detectors), which is a key parameter in the FoM determination. 
As we expect the ePSF of a space mission at L2 to be extremely stable 
\citep{Bernstein09pro} thus well constrained, a full optimization of the pixel 
size that takes account of full observation strategy and the final WL FoM 
must be investigated before committing to a final design.
 
%%%%%%%%%%%%%%%%%%%%%%%%%%%%%%%%%%%%%%%%%%%%%%%%%%%%%%%%%%%%%%%%%%%
\begin{figure}[!h]
\resizebox{\hsize}{!}{\includegraphics{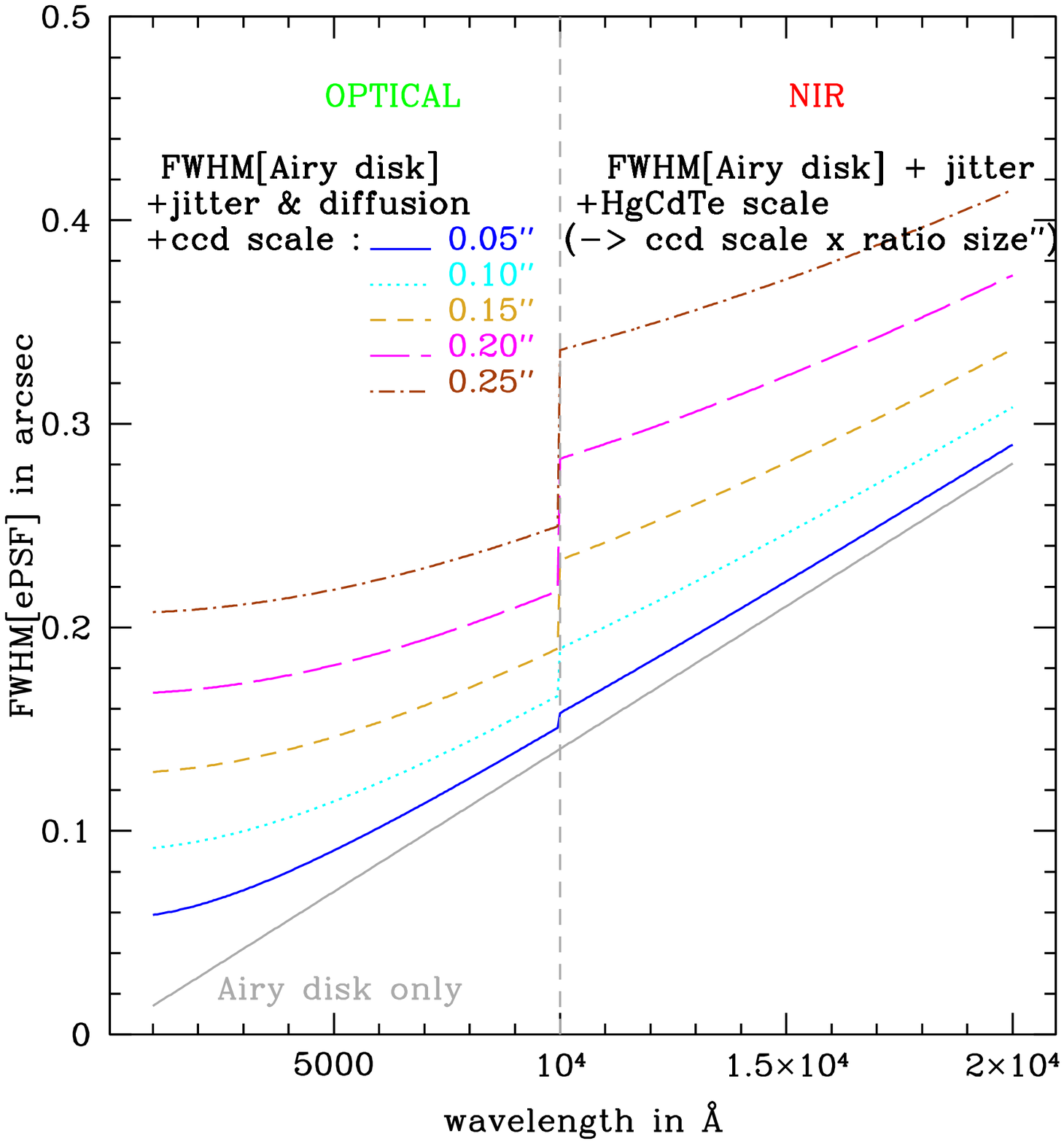}}
\resizebox{\hsize}{!}{\includegraphics{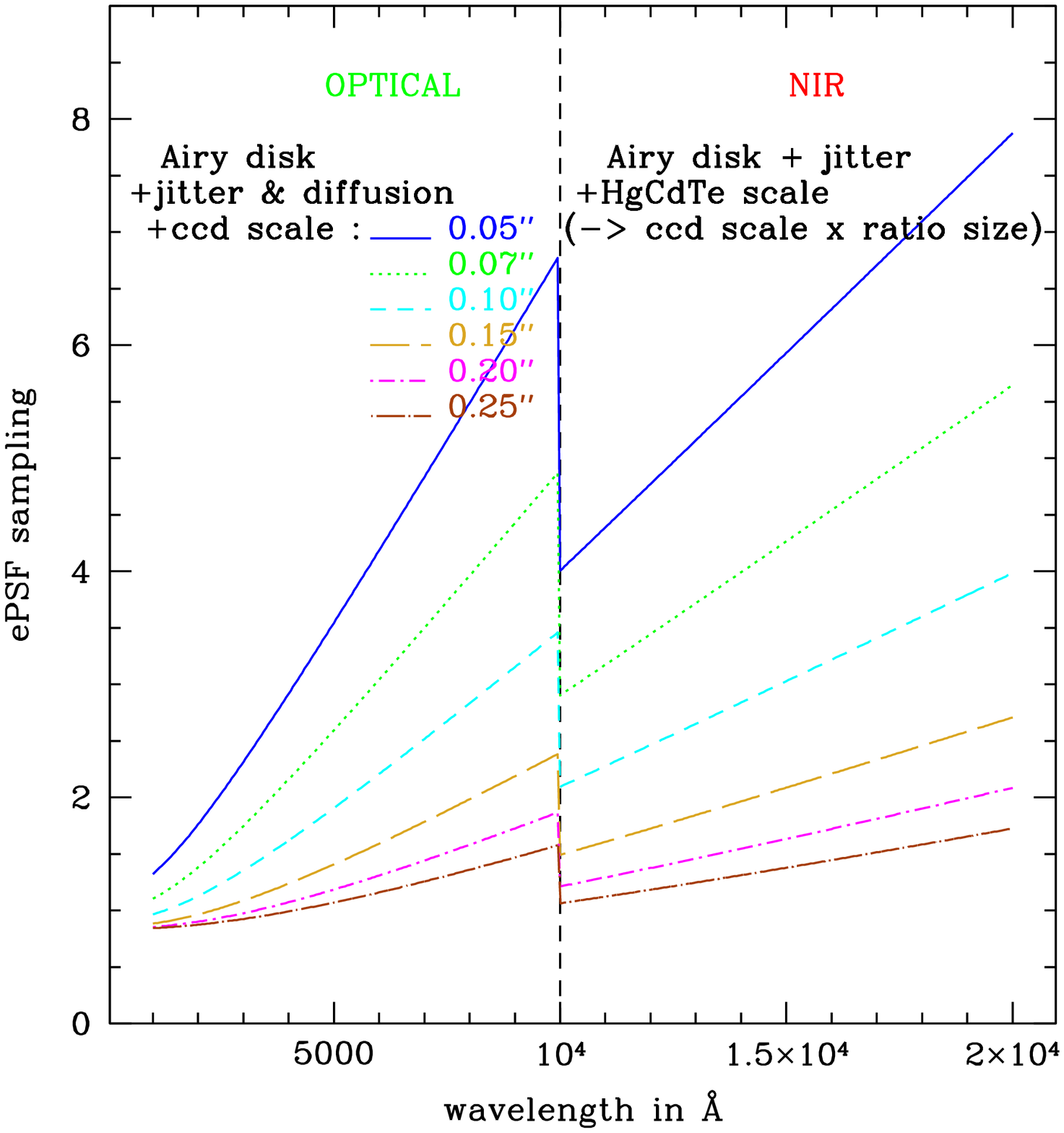}}
\caption{(Top) FWHM of the ePSF in arcsec and (Bottom) ePSF sampling
  as a function of wavelength in \AA~for a mirror size of 1.5m.}
\label{fig:fwhmwvl}
\end{figure}
%%%%%%%%%%%%%%%%%%%%%%%%%%%%%%%%%%%%%%%%%%%%%%%%%%%%%%%%%%%%%%%%%%% 

\subsubsection{Discussion}
{\bf Figure~\ref{fig:fwhmwvl}} shows the sampling of the full ePSF in
the top figure and the FWHM of this ePSF as a function of
wavelength using pixel scales of 0.05",0.07",0.15",0.2", and 0.25"
respectively in blue, green, cyan, gold, magenta, and brown. These
pixel scales correspond to the  visible CCD detectors. For simplicity,
 we assume that the NIR detectors share the same focal plane so that the pixel ratio
 is just the ratio of the pixel physical size
\begin{equation}
p^{ir} = [ratio\ size]\times p^{vis}= \frac{18\mu m}{10.5\mu m}\times p^{vis}=1.71\times p^{vis},
\end{equation}
where $p^{ir}$ and $p^{vis}$ are, respectively, the pixel scale of the
NIR detector and the visible  detector. We consider for the NIR detectors the
physical size of 18$\mu$m  and
10.5$\mu$m for visible  detectors (LBNL CCD). 

We note that the ePSF is similarly sampled in the NIR wavelength
range. The PSF size is proportional to the wavelength $\lambda$ such that
$PSF\propto\lambda/D_{1}$, where $D_{1}$ is the primary mirror size, that
allows a higher sampling. However, a large PSF causes a substantial decrease in
the galaxy number density. The WL analysis makes use of the shape of galaxies. 
Thus, one has to make a cut in the galaxy size to use only the galaxies 
whose shapes are not contaminated by the instrumental PSF. 
As an illustrative example the PSF size at $\lambda=1200$nm with a mirror 
size of 1.5m is equivalent to that at $\lambda=800$nm for a mirror size of 1m. 
Even if the count slope differes between J band and I band, it will decrease 
the galaxy number density significantly, suggesting that the WL measurement 
should be conducted more efficiently in the visible bands.

\subsection{Exposure time}
\label{subsec:tobs}
To establish the optimal DE FoM, we need to define a
``minimal'' exposure time for WL, which is a combination of three
dependent factors: 
1) the photon noise, which must dominate the detector noise for typical galaxy photometry and shape measurements; 
2) the exposure time should be small
enough to cover the largest possible area of available sky, but long
enough to obtain a good photometric redshift distribution;
3) To reach a homogeneous survey across the sky, the exposure time
should be adjusted depending on the Zodiacal light and Galactic absorption. 
This is particularly important at visible wavelengths where the WL measurement will be conducted.

%%%%%%%%%%%%%%%%%%%%%%%%%%%%%%%%%%%%%%%%%%%%%%%%%%%%%%%%%%%%%%%%%%% 
\begin{figure}[!ht]
  \resizebox{\hsize}{!}{\includegraphics{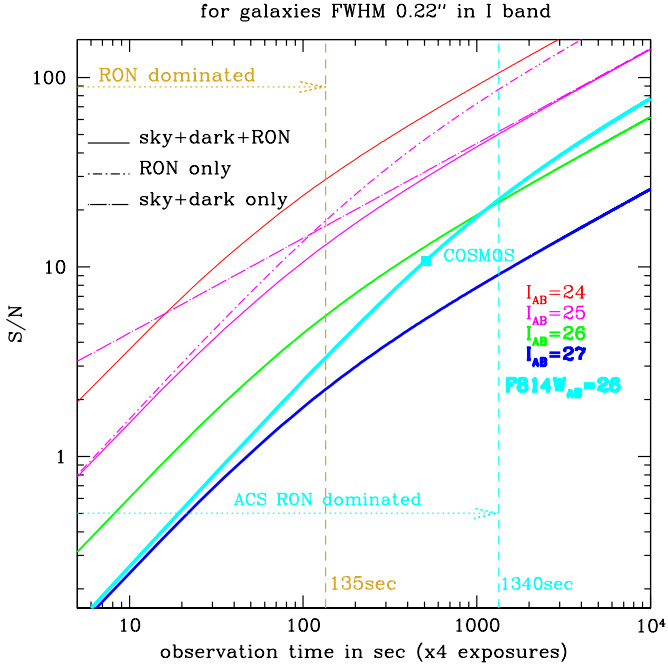}}
  \caption{Signal-to-noise ratio as a function of observation time for
    different I-band magnitudes. The telescope characteristics are listed in Table~\ref{tab:jdem}.
    We use the I-band filter for the blue, green, magenta, and red curves. 
    The cyan curve uses the properties of the ACS camera in the F814W filter. 
    The square cyan dot on this last curve
    represents the COSMOS survey with an observation time of 507s by exposure.
    The dotted gold line separates the RON (read-out noise) dominated regime (exposure time less than
    135s) from the photon noise dominated regime at longer exposure
    times. The dotted cyan line represents the same as the dotted
    gold line for the ACS camera.}
  \label{fig:sntobs}
\end{figure}  
%%%%%%%%%%%%%%%%%%%%%%%%%%%%%%%%%%%%%%%%%%%%%%%%%%%%%%%%%%%%%%%%%%% 

To study the impact of the filter set properties on the photometric
redshift quality, we first have to define a minimal exposure time
beyond which the photon noise dominates the detector
noise. Thus, for a given exposure, we extract the minimum
exposure time $T^{min}_{obs}$ from which the read-out noise becomes
sub-dominant in using the denominator of equation \ref{eq:sn}

\begin{equation}\label{eq:tobsmin}
e_{signal}+N_{pix}T^{min}_{obs}e_{sky/sec}=N_{pix}e_{RON}^2+N_{pix}T^{min}_{obs}e_{dark},
\end{equation}
where the left-hand term is the photon noise contribution from the Zodiacal
light and a galaxy, respectively, $e_{signal}$ and $e_{sky}$ and the
right-hand term is the detector noise with the read-out noise (RON) and
the dark current (for more details, see the Appendix).

{\bf Figure~\ref{fig:sntobs}} shows the S/N as a function
of the observing time at different I-band magnitudes of 24,25,26, and
27 (red, magenta, green, blue respectively) using a filter resolution
of $\mathcal{R}=3.2$, a pixel size of 0.15", and a mirror diameter of 1.5m. 
The cyan curve corresponds to the ACS-COSMOS expectations for
this noise simulation at a magnitude of 26 in the F814W filter with
the ACS pixel scale of 0.05" and the HST mirror size of 2.4m. The
square point represents the COSMOS survey (with four exposures of 507s) 
performed using the ACS camera, which has a RON of $e^{vis}_{RON}=7e^-/pixel$ 
and a dark current of $e^{vis}_{dark}=13e^-/pixel/hour$ as stated in 
\citet{Koekemoer07}. We find a limiting magnitude of AB(F814W)=26.6 ($5.2\sigma$) 
for a galaxy size of 0.21'' effective radius. This estimate is very 
close to that of \citet{Leauthaud07}, who find a limiting magnitude of 
AB(F814W)=26.6 ($5\sigma$) for a galaxy effective radius of 0.2". 

We note a change of slope for the solid line curves denoting the detector
and photon noise regime. For short exposures,
the read-out noise dominates the denominator term and the
S/N grows proportionally to the exposure time as shown by
the magenta dot-little-dashed line.  Thus, $e_{signal} \propto T_{obs}$
and the S/N is
\begin{equation}
S/N \sim \frac{e_{signal}}{\sqrt{e^2_{RON}}}\propto T_{obs},
\end{equation}
where $e_{RON}=6e^-/pix$ is not time-dependent. For long exposures,
the S/N varies as the square root of the observing time
as shown by the magenta dot-long-dashed line
\begin{equation}
S/N \sim \frac{e_{signal}}{\sqrt{\alpha\ T_{obs}}} \sim \frac{\beta\ T_{obs}}{\sqrt{\alpha\ T_{obs}}} \propto \sqrt{T_{obs}},
\end{equation}
where $\alpha$ holds for the galaxy and sky photons, whose fluxes are a
function of $T_{obs}$, and $\beta$ can be deduced from
equation \ref{eq:esignal} in the Appendix. \\

Equation \ref{eq:tobsmin} can be simplified to define a minimum
exposure time at which the sky noise equals the collective detector noises (the
dark current and read-out noise)
\begin{equation}\label{eq:Tobsmin}
T^{min}_{obs}\ e_{sky/sec}=e^2_{RON}+T^{min}_{obs}\ e_{dark},
\end{equation}
where $e_{sky}=e_{sky/sec}\ T_{obs}$. We find a minimum exposure time
of 135 sec for a 1.5m mirror diameter with a 0.15" pixel scale. 
This number defines the minimum exposure time for which a WL survey is optimal.
It is interesting to raise the S/N as long as we are in
the detector noise regime and $S/N\propto T_{obs}$. We also 
note that current software assumes that the noise properties follow Poisson statistics,
which is true in the photon noise regime. However, in the detector noise regime, 
the noise properties follow Gaussian statistics. If not taken into account, this 
will impact the galaxy properties calculated from the software as well as the 
galaxy extraction. We return to the observation time in section 
\ref{sec:surveystrategy}, where we study its impact on DE parameter estimations.

%%%%%%%%%%%%%%%%%%%%%%%%%%%%%%%%%%%%%%%%%%%%%%%%%%%%%%%%%%%%%%%%%%% 
\begin{figure}[!ht]
  \resizebox{\hsize}{!}{\includegraphics{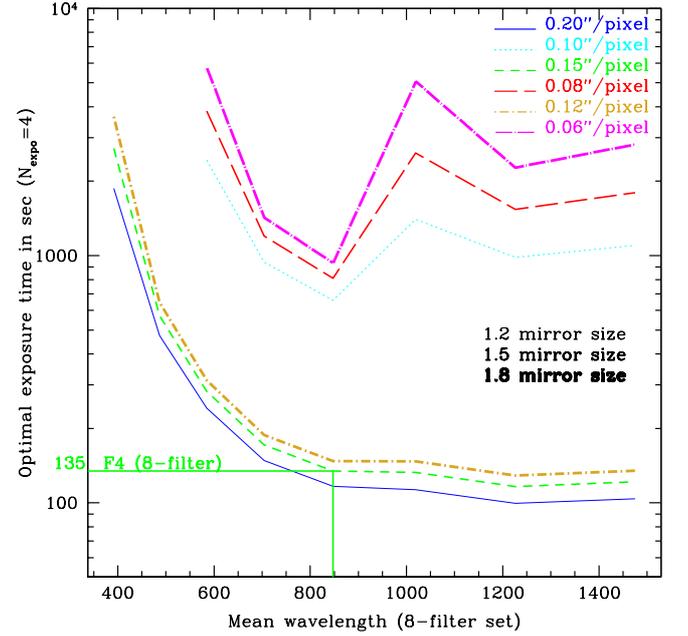}}
  \caption{Minimum exposure time for WL  surveys as a
    function of wavelength integrated in filters of resolution
    $\mathcal{R}=3.2$. The minimal exposure time is defined in 
    equation \ref{eq:Tobsmin} and corresponds to the photon noise
    dominating detector noises. It takes into account the mirror size,
    the pixel scale, and the filter efficiency. The thickness of the lines
    grows with the mirror size.}
  \label{fig:expotime}
\end{figure}  
%%%%%%%%%%%%%%%%%%%%%%%%%%%%%%%%%%%%%%%%%%%%%%%%%%%%%%%%%%%%%%%%%%% 

{\bf Figure~\ref{fig:expotime}} shows the two dependences on both the mirror
and pixel scales of the minimum exposure time as a function of
wavelength integrated in filters of resolution $\mathcal{R}=3.2$. This
figure shows that a smaller mirror and pixel scale requires a longer
minimum exposure time for the photon noise to dominate the
detector noise. We note that a 1.2m mirror diameter with 0.1" by
pixel requires a minimum of 700s exposure to be photon-noise-dominated. 
It is possible to decrease the minimum exposure time needed by using 
filters that are broader than our optimal resolution of $\mathcal{R}=3.2$.
However, this would reduce the photoz quality (as shown 
in section \ref{sec:filterset}) and may jeopardize the PSF color correction. 
The shape of curves reflect the logarithmic width of the filter set 
configuration (shown in Figure \ref{fig:filtersets})
and the drop in the detector efficiency at blue wavelengths 
(as studied in Figure \ref{fig:eper}).
This also explains the decrease in the sky background magnitude at blue wavelengths
shown in {\bf Table~\ref{tab:mag_noise}}. This noise magnitude is
the Zodiacal light flux (explained in the Appendix) integrated within the
photometric bands of the eight-filter set without taking into account any 
instrument characteristic other than the filter efficiency. In the table,
$\eta$ represents the whole transmission including filter
transmission, mirror reflectivity, and detector efficiency.

We note that a fixed exposure time of 200s allows us to use optimally the
information contained in almost all bands, except the two bluest
bands (which would ideally require longer exposure times). 
For simplicity, we thus choose to use this exposure time to study
the photometric redshift quality as a function of the resolution of filters 
in section \ref{sec:filterset}.

%%%%%%%%%%%%%%%%%%%%%%%%%%%%%%%%%%%%%%%%%%%%%%%%%%%%%%%%%%%%%%%%%%%
\begin{table}[!h]
\caption{Noise magnitude (Zodiacal light) in mag/arcsec$^2$ for the
  eight-filter set configuration and the total telescope throughput $\eta$ in each band.}
\begin{tabular}{cccccccc} \hline \hline 
Camera & Filters & Noise mag & $\Delta\lambda$ & $\lambda_{central}$ & $\eta$\\
\hline \hline 
Visible & F0 & 24.01  & 149nm &  392nm &  0.25\\
& F1 & 23.54  & 150nm &  487nm &  0.43\\ 
& F2 & 23.16  & 181nm &  585nm &  0.51\\                                         
& F3 & 22.91  & 218nm &  704nm &  0.61\\                                    
& F4 & 22.76  & 262nm &  847nm &  0.67\\                                          
NIR & F5 & 22.67  & 315nm & 1019nm &  0.61\\                                          
& F6 & 22.64  & 379nm & 1226nm &  0.66\\                                        
& F7 & 22.68  & 456nm & 1475nm &  0.66\\
\hline \hline
\label{tab:mag_noise}
\end{tabular}
\end{table}
%%%%%%%%%%%%%%%%%%%%%%%%%%%%%%%%%%%%%%%%%%%%%%%%%%%%%%%%%%%%%%%%%%%

\subsection{Galactic absorption, Zodiacal light, and survey area}
\label{subsec:area}
For most current surveys (COSMOS, CFHT-LS, RCS2), the corrections for Galactic
absorption and Zodiacal light are not difficult to make. These surveys cover relatively
small fields and are generally located at high Galactic latitudes. 
This will not be the case for the next generation of weak lensing surveys, 
which will be limited by both Galactic absorption and Zodiacal light variation.

In general, the overall number of galaxies grows faster when surveying wider fields
rather than going deeper in smaller fields (which reflects the small
gradient of the galaxy count slope). 
Future cosmological surveys should cover more than ten thousand  
square degrees as advocated by \citet{Amara07}, who demonstrated that DE 
constraints grow proportionally to the number of galaxies. However, when reaching such 
wide areas, the impact of Galactic absorption and Zodiacal light variation
has to be accounted for in the survey strategy or it will otherwise 
severely affect the photometry quality, leading to degradation of the DE constraints.
This was not addressed in \citet{Amara07}.

\begin{figure}[!h]
  \resizebox{\hsize}{!}{\includegraphics{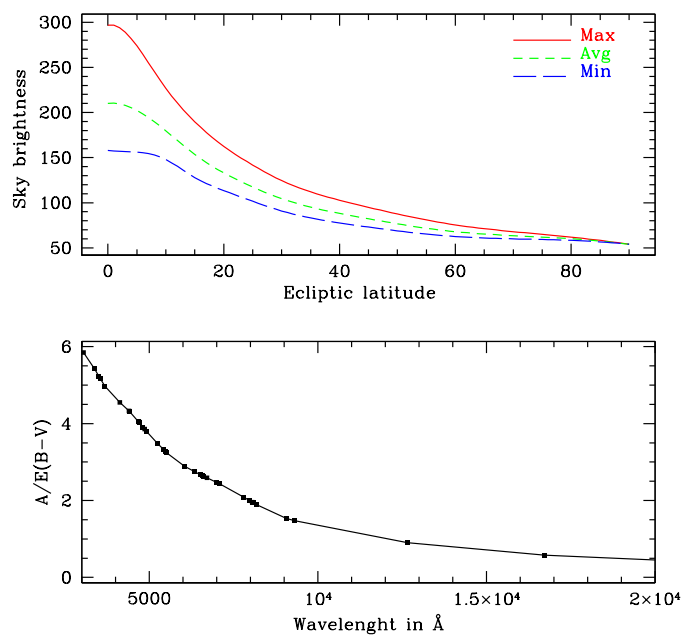}}
  \caption{(Top) Zodiacal background level as a function of ecliptic
    latitude from \citet{Leinert02}, assuming the telescope viewing
    angle is between 70 and 110 degrees from the Sun. (Bottom) Galactic
    absorption as a function of wavelength from \citet{Schlegel98}.}
  \label{fig:zodigalac}
\end{figure}

\begin{figure}[!h]
  \resizebox{\hsize}{!}{\includegraphics{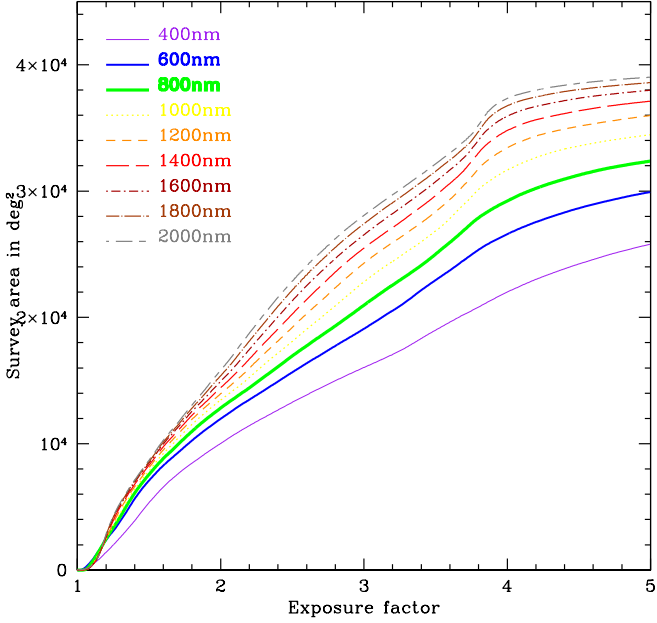}}
  \caption{Cumulative distribution of the sky coverage for a given
    exposure time factor, i.e. sky coverage with exposure factor less
    than the given exposure factor, at several wavelengths.%Exposure
    %time factor as a function of sky coverage at wavelengths of 400nm
    %to 2000nm by bin of 200nm.
  }
  \label{fig:expfactor}
\end{figure}

To reach a homogeneous data quality, we need to adjust the exposure time
of the survey as a function of the pointing position on the sky ($\alpha$,$\delta$). 
We can define the exposure time factor
needed to reach the intrinsic magnitude limit as a function of the
coordinates (assuming that the survey is photon-noise-limited)
\begin{equation}
t_f(\alpha,\delta)=\frac{zodi(\alpha,\delta)}{trans(\alpha,\delta)^2},
\end{equation}
where $t_f$ is the exposure time required to reach the desired limit
displayed in {\bf Figure \ref{fig:expfactor}}, $zodi$ is the Zodiacal
background light level plotted in  {\bf Figure \ref{fig:zodigalac}}, and
$trans$ is the Galaxy absorption defined by
\begin{equation}
trans(\alpha,\delta)=10^{-0.4\ A_{\lambda}(\alpha,\delta)},
\end{equation}
where $A_{\lambda}$ is the extinction map at wavelength $\lambda$ due to Galactic
dust from \citet{Schlegel98} shown in Figure \ref{fig:zodigalac}.

%%%%%%%%%%%%%%%%%%%%%%%%%%%%%%%%%%%%%%%%%%%%%%%%%%%%%%%%%%%%%%%%%%% 
%%%%%%%%%%%%%%%%%%%%%%%%%%%%%%%%%%%%%%%%%%%%%%%%%%%%%%%%%%%%%%%%%%% 
%%%%%%%%%%%%%%%%%%%%%%%%%%%%%%%%%%%%%%%%%%%%%%%%%%%%%%%%%%%%%%%%%%% 
%%%%%%%%%%%%%%%%%%%%%%%%%%%%%%%%%%%%%%%%%%%%%%%%%%%%%%%%%%%%%%%%%%% 

In section \ref{sec:surveystrategy}, we use this model to investigate
the fraction of the sky a telescope should survey to optimize the cosmological 
constraints. To do this, we need to define the
telescope characteristics, such as the number of filters $n_f$ for each of the
two cameras ($n_{cam}=2$: one infrared with HgCdTe detectors
and one visible with CCDs - assumed here to have exactly the same field of view),
the number of exposures $N_{expo}=4$ per filter, and
the observation time $T_{obs}$ per exposure. These parameters define 
the survey configuration.

Using the characteristics of the survey configuration, we define the minimum
 exposure time for each pointing as
\begin{equation}\label{eq:tobs_pting_min}
T_{obs/pting}^{min}=T_{obs/cam}\ N_{expo}\ n_{f/cam},
\end{equation}
where $N_{expo}$ is the number of exposures (see Table \ref{tab:jdem}),
$n_{f/cam}$ the number of filters by camera, and $T_{obs/cam}$ the
individual image exposure per filter for a given camera.

The exposure time for a given position on the sky $(\alpha,\delta)$ is then defined as
\begin{equation}\label{eq:tobs_pting}
T_{obs/pting}(\alpha,\delta)=t_{f}(\alpha,\delta) \times T_{obs/pting}^{min},
\end{equation}
where, for simplicity, the exposure time factor is computed
at a wavelength of 800nm, which corresponds to the wavelength 
of the weak lensing measurement, and is applied globally.

\begin{figure}[!h]
  \resizebox{\hsize}{!}{\includegraphics{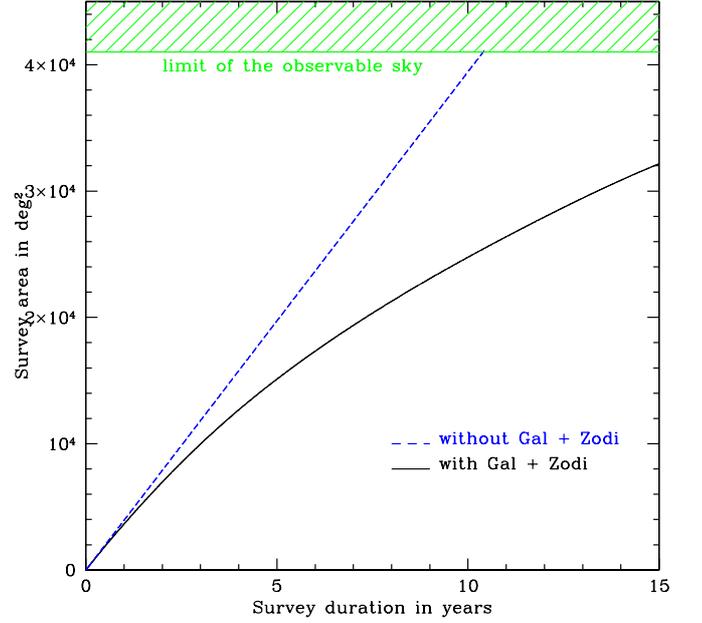}}
  \caption{Cumulative distribution of the sky coverage as a function of
    the mission duration in years using the survey characteristics described in Table \ref{tab:jdem}
    at 8000 \AA. The black curve includes Galactic absorption and Zodiacal light 
    variation, while the blue curve does not.
  }
  \label{fig:year_area}
\end{figure}

We define the survey area for a given camera field-of-view ($FOV$) and a total 
mission time $T_{mission}$ as
\begin{equation}\label{eq:area}
\mathcal{A}= \frac{ FOV \times T_{mission}}{\sum_{p=1}^{N_p}\frac{T_{obs/pting}(p)}{N_p(T_{mission})}}.
\end{equation}
We note that $T_{mission}$ includes a survey efficiency $\zeta=0.7$, which accounts for 
observation overheads (telescope slewing, guide star acquisition, read out time, ...) 
and data transmission.
The denominator is the mean observation time over the whole field surveyed and is calculated iteratively.  
{\bf Figure \ref{fig:year_area}} shows a cumulative distribution of the sky coverage as a function
of the total mission time of a survey in years. This figure uses a survey strategy as defined above
consisting of changing the exposure time as a function of the Galactic absorption strength on the sky area observed.
The black line includes the Galactic absorption and Zodacal light variation, while the blue line does not. 

%%%%%%%%%%%%%%%%%%%%%%%%%%%%%%%%%%%%%%%%%%%%%%%%%%%%%%%%%%%%%%%%%%% 
%%%%%%%%%%%%%%%%%%%%%%%%%%%%%%%%%%%%%%%%%%%%%%%%%%%%%%%%%%%%%%%%%%% 

%%%%%%%%%%%%%%%%%%%%%%%%%%%%%%%%%%%%%%%%%%%%%%%%%%%%%%%%%%%%%%%%%%% 
\begin{figure}[!h]
  \resizebox{\hsize}{!}{\includegraphics{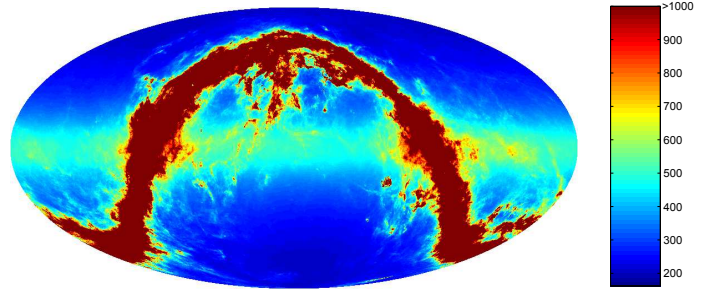}}
  \caption{Exposure time map of the sky (in seconds) needed to reach 
   S/N=10 for a $I_{AB}$ = 25.6 galaxy at a wavelength of 8000 \AA  
   (using the survey characteristics described in Table \ref{tab:jdem}). 
  }
  \label{fig:sky_map}
\end{figure}  
%%%%%%%%%%%%%%%%%%%%%%%%%%%%%%%%%%%%%%%%%%%%%%%%%%%%%%%%%%%%%%%%%%% 

This observation strategy ensures a uniform photometry quality 
over the whole survey by spending more time on pointings closer to the Galactic plane,
as shown in {\bf Figure \ref{fig:sky_map}}. This figure is a version of the sky map 
showing the exposure time required to achieve a S/N of 10 for an $I_{AB}$ = 25.6 galaxy.
The scale is in seconds. This assumes that four exposures of the time listed were taken. 
The time spent then depends on the field location on the sky and is determined with
the exposure factor.

We note that the survey area $\mathcal{A}$ scales as $FOV\times T_{mission}$, hence
a reduction in the camera FOV thus reduction in the number of detectors can be 
compensated for by a longer mission time.

\section{Filter resolution studies}
\label{sec:filterset}
To optimize the survey strategy, we must carefully study all parameters
that affect the galaxy photometry. Using the noise properties defined in
section \ref{subsec:noise}, the optimal pixel scale, and the exposure time
 defined in section \ref{sec:WLrequirements}, we study in
this section the whole telescope transmission i.e. detector sensitivity
and filter transmission assuming mirror reflectivity of bare
silver. In section \ref{subsec:filterset}, we define the filter
properties. In sections \ref{subsec:res}, \ref{subsec:res_sig}, and
\ref{subsec:res_cata}, we study the filter resolution to improve the
photometric redshift accuracy and decrease the number of catastrophic
redshifts. 

To develop a survey strategy we need to define the number of filters
and their shapes, since this will affect the survey speed in terms of the
required exposure time per filter. In this paper, we use square shaped 
filters and vary the filter shape in changing their width.

\subsection{Properties of filter set}
\label{subsec:filterset} 

To optimize the design of the filter set for photometric redshift
quality, we choose logarithmically spaced filters within a given
wavelength range \citep{Davis06}. The wavelength range is chosen so as to
use the full capacity of the detectors. This log-spaced repartition of
filters mimics the wavelength shift-dilation of galaxy spectra as a
function of redshift $z$, expressed by the formula
$\lambda_{obs}=\lambda_{rest}(1+z)$.
The useful spectral features
for the photometric redshift are shift-dilated as a function of
redshift and the filter set is designed to follow this evolution
allowing a direct comparison of galaxy luminosity as a function of
redshift. Thus, each filter is a redshifted copy of the previous one and
its width is multiplied by a factor of $\alpha$ (explained below). 
This filter design was
first developed for the SN probe to improve the K-correction \citep{Davis06}. 
However, this is also relevant for photometric redshifts and galaxy evolution studies. 
We construct the first filter using the detector cut-off in the near-UV of the visible  CCDs
$(\lambda_{min}^{vis}\sim3200\AA)$ and a width $w$:
$(\lambda_{min}^{vis},\lambda_{min}^{vis}+w)$.
 
%%%%%%%%%%%%%%%%%%%%%%%%%%%%%%%%%%%%%%%%%%%%%%%%%%%%%%%%%%%%%%%%%%%
%\begin{table}[!ht]
%\caption{Detectors characteristics.}
%\begin{tabular}{cccccccc} \hline \hline 
%Detectors & $\lambda_{min}$ & $\lambda_{max}$ \\
%\hline \hline 
%CCD &  3200 \AA & 10000 \AA \\ 
%HgCdTe &  8000 \AA & 17000 \AA\\ 
%\hline \hline
%\label{tab:detectors}
%\end{tabular}
%\end{table}
%%%%%%%%%%%%%%%%%%%%%%%%%%%%%%%%%%%%%%%%%%%%%%%%%%%%%%%%%%%%%%%%%%%

The subsequent filters are based on the first filter multiplied by the
factor $\alpha$.  This replication factor is defined as a function of
the wavelength range available to the instrument 
$(\lambda^{detector}_{min},\lambda^{detector}_{max})=(\lambda^{ccd}_{min},\lambda^{HgCdTe}_{max})=(3200\AA,17000\AA)$,
the width of the first filter $w_0$, and the number of filters $n$
\begin{eqnarray}
\lefteqn{\lambda^{detector}_{max}=\alpha^{n} \lambda^0_{max}=\alpha^{n} (\lambda^{detector}_{min}+w_0)}
\nonumber\\
& & {} \Rightarrow \alpha=\Big(\frac{\lambda_{max}^{detector}}{\lambda_{min}^{detector}+w_0}\Big)^{1/n},
\end{eqnarray}
where $\lambda^0_{max}$ is the maximum wavelength of the $i=0$ filter.
Following these properties of filter sets, we define the resolution
of a filter or a filter set as
\begin{equation}
\mathcal{R}_{(i=filter)}=\frac{\lambda^{i}_{mean}}{\Delta \lambda^{i}}=\frac{\alpha^{i} \lambda^{0}_{mean}}{\alpha^{i} w_0}=\mathcal{R}_{(filter\ set)}.
\end{equation}
Using this definition of filter sets, we attempt in Section
\ref{subsec:res} to find a filter set resolution that gives the
best results in term of photometric redshift quality for WL studies
using the telescope design and noise prescription that we defined in
Section \ref{subsec:noise}.

\subsection{Filter resolution studies: methods and hypothesis} 
\label{subsec:res}
The filter set properties defined in Section \ref{subsec:filterset}
are determined by the width $w_0$ of the first filter and the total number of filters $n$.
This also defines the resolution of the filter set.

In this section, we study the impact of the resolution on the
photometric redshift quality and WL analysis. Following the studies in
\citet{Benitez09}, we choose to use eight filters that they proved to
be the optimal number of filters to reach the highest completeness in
depth and quality of the photometric redshift distribution. 
Their studies are based on mock
catalogs derived from the HDF catalog artificially extended to 5000
objects. Their photometric redshift distribution was computed using the BPZ code
\citep{Benitez00}. We also study the minimum number of filters
required using our optimal filter resolution in Section \ref{sec:surveystrategy}.

To test the impact of the filter resolution, we made a grid in
resolution by raising $w_0$ -- the width of the first filter -- in covering steps
of $100\AA$. Using 16 configurations of filter set
$w_{0}=[600\AA,2000\AA]$, we study the evolution of the scatter and
the number of catastrophic redshifts as a function of the filter resolution.\\
%%%%%%%%%%%%%%%%%%%%%%%%%%%%%%%%%%%%%%%%%%%%%%%%%%%%%%%%%%%%%%%%%%%
\begin{figure}[!h]
  \resizebox{\hsize}{!}{\includegraphics{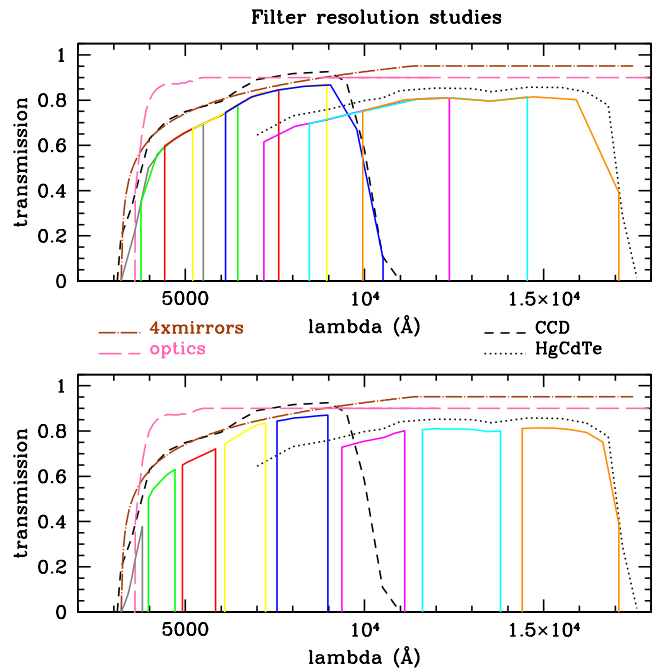}} 
  \caption{Transmission as a function of wavelength in  
  in the extreme cases of the filter sets tested: $\mathcal{R}=6$
    (top panel) and $\mathcal{R}=2$ (bottom panel). We include the
   transmission of optics (long-dashed pink), CCD (small-dashed black),
  NIR detector (dotted black), and the four-mirror reflectivities (solid brown). } 
  \label{fig:extremeres}
\end{figure}
%%%%%%%%%%%%%%%%%%%%%%%%%%%%%%%%%%%%%%%%%%%%%%%%%%%%%%%%%%%%%%%%%%%

{\bf Figure ~\ref{fig:extremeres}} shows the two
extreme cases of filter resolution we tested. The upper panel
contains our results for  a filter set with the highest filter resolution, 
$\mathcal{R}=6$. This high filter resolution
makes the filters very narrow: indeed gaps appear in
the wavelength coverage, which is something we wish to
avoid. We note that $\mathcal{R}=6$ is the only filter resolution studied here 
that has wavelength gaps in its transmission. This is a filter configuration
that is not desired unless complemented with broader band observations. 
The bottom panel shows the lowest filter resolution
$\mathcal{R}=2$ with extremely broad filters. The multicolor lines are
the filter transmission curves. The dashed and dotted lines are,
respectively, the assumed CCD and NIR detector transmission curves. The
dot-dashed line is the four-mirror reflectivity curve using bare
silver reflectivity as described for the SNAP/JDEM mission
\citep{Levi07}. We assume a survey configuration of four bare
silver mirrors to focus the light on the focal plane using a
quantum efficiency (QE) similar to the LBNL detector transmission
properties shown Figure \ref{fig:extremeres}. These characteristics
have an impact on the photometric redshift accuracy that depends on the
photometric errors calculated using equations in section
\ref{subsec:noise}. For each filter set created, we compute
photometric redshifts using the Le Phare photometric redshift code
briefly described in section ~\ref{subsec:photoz}.

\subsection{Filter resolution studies: Photometric redshift quality}
\label{subsec:res_sig}
%%%%%%%%%%%%%%%%%%%%%%%%%%%%%%%%%%%%%%%%%%%%%%%%%%%%%%%%%%%%%%%%%%%
\begin{figure}[!h]
  \vbox{
    %\resizebox{\hsize}{!}{\includegraphics{dispm_9sqr_0_6_6sig.ps}}
    %\resizebox{\hsize}{!}{\includegraphics{dispz_9sqr_20_26_6sig.ps}}
    \resizebox{\hsize}{!}{\includegraphics{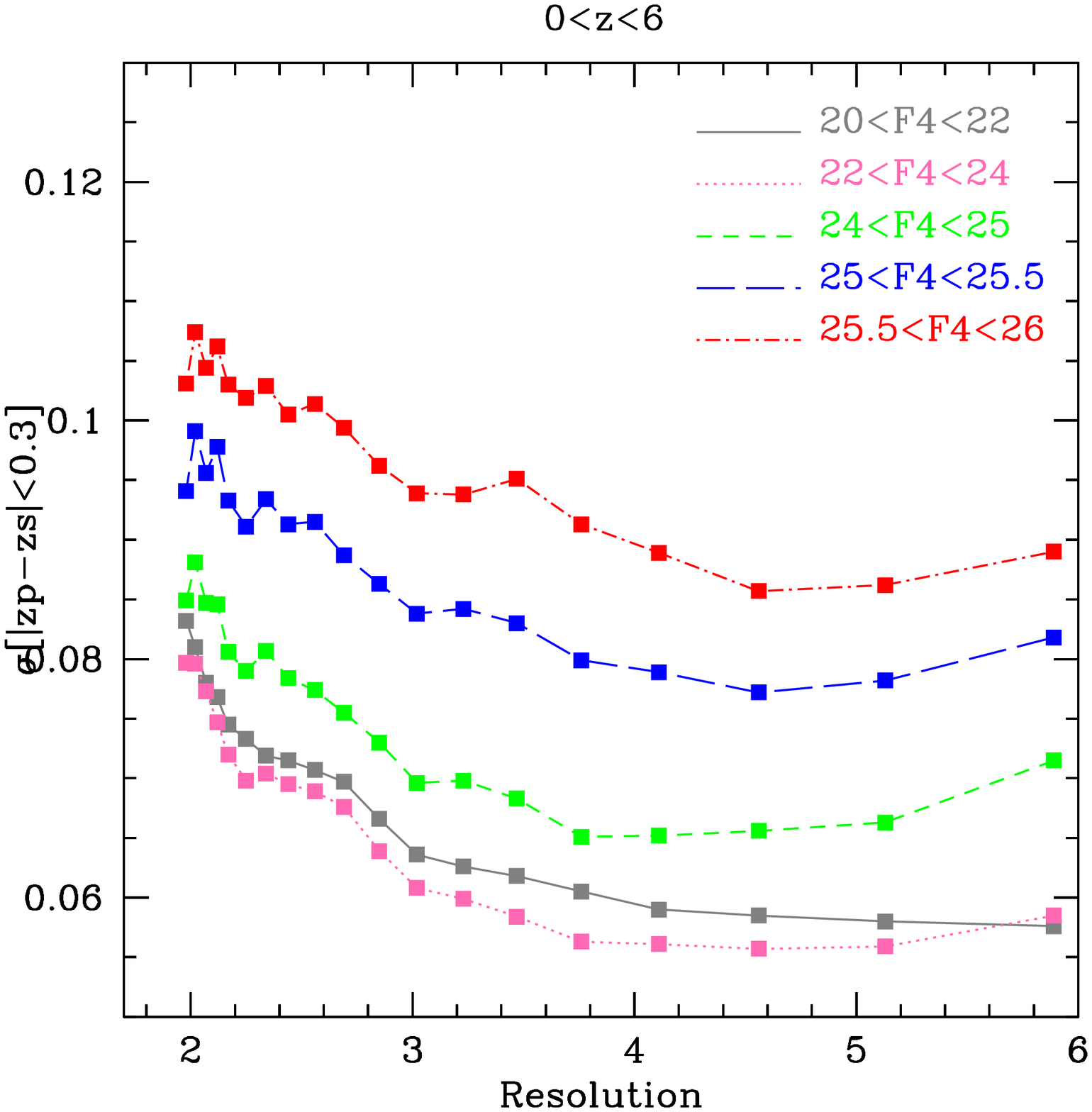}}
    \resizebox{\hsize}{!}{\includegraphics{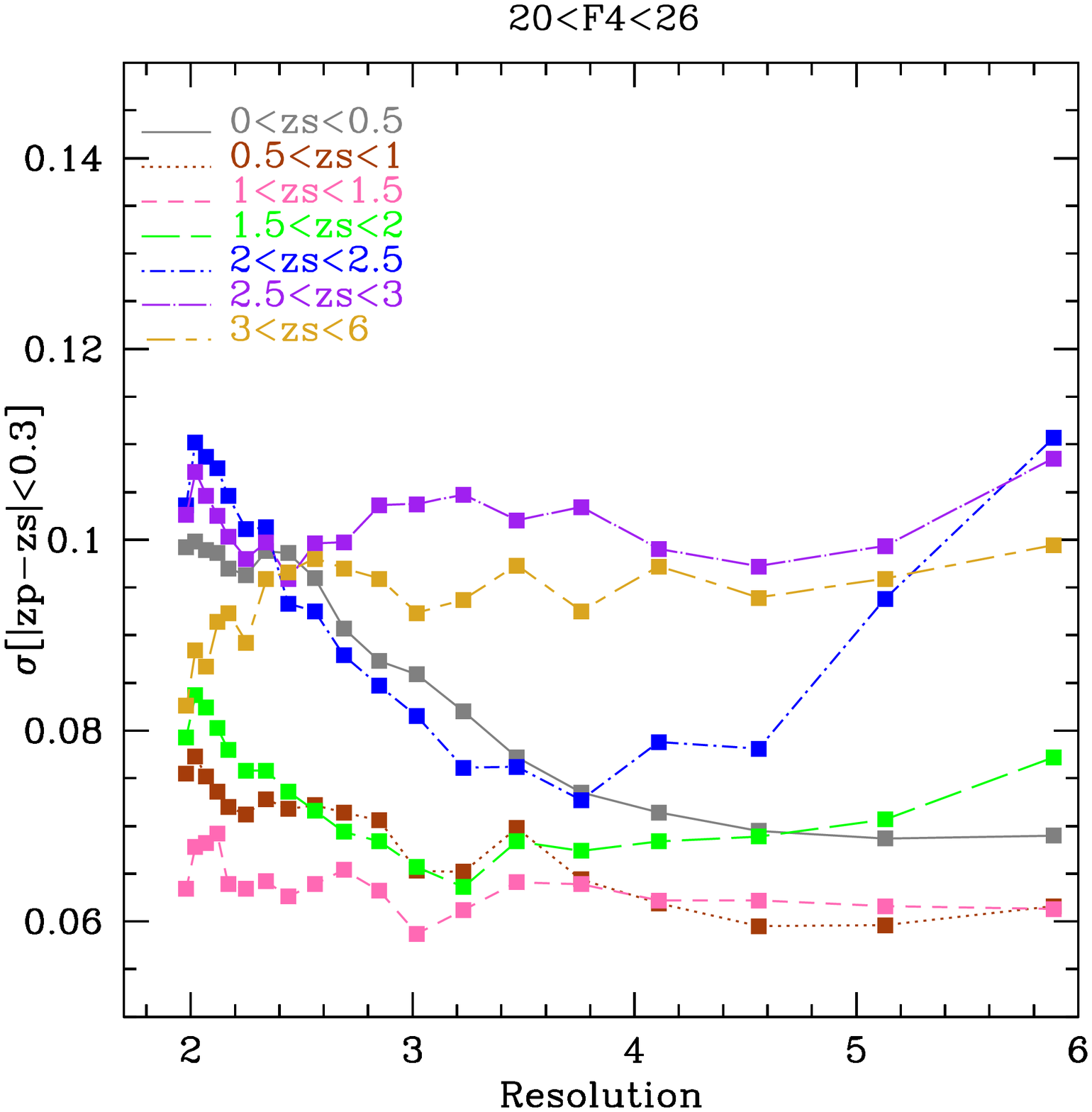}}
  }
  \caption{Photometric redshift scatter $\sigma_{core}$ as a function
    of filter set resolution binned by magnitude up to $z~6$ (top panel) 
    and by redshift (bottom panel) up to $I_{AB}=26$ mag. 
    $F_4$ represents the I-band filter and $z_s$ the
    spectroscopic redshift.} 
  \label{fig:sigres}
\end{figure}
{\bf Figure ~\ref{fig:sigres}} shows the photometric redshift
scatter $\sigma(|zp-zs|)$ as a function of filter set resolution, binned by
magnitude in the upper panel and by redshift in the bottom
panel. Each square point of these curves shows the result for a
particular filter set configuration with a resolution
$2<\mathcal{R}<6$. We use the I-band like F4 filter for the magnitude binning 
(see Table~\ref{tab:mag_noise}). 
To minimize the photoz scatter a filter resolution of $\mathcal{R}>3$ is preferred 
when looking at the top panel of Figure \ref{fig:sigres}.
The bottom panel of Figure \ref{fig:sigres} suggests a preferred filter 
resolution of $\mathcal{R}\approx3-4$.

The accuracy of the photometric redshifts depends on the color
gradients of galaxy templates. It also depends
on the photometric errors that are used as a weight in the template
fitting procedure (see Equation \ref{eq:chi2}). A high filter
resolution ($\mathcal{R}>5$) lowers the S/N in each filter and the weight 
derived from it do not place sufficient constraints to ensure an accurate photoz
estimation. In the case of a low filter resolution 
($\mathcal{R}<3$), the overlap between
filters lowers the galaxy color gradient degrading the quality of the photoz results. 
Figure~\ref{fig:sigres} shows that an optimal filter resolution
is around $\mathcal{R}\approx3-4$.

\subsection{Filter resolution studies: Catastrophic redshift rate} 
\label{subsec:res_cata}
%%%%%%%%%%%%%%%%%%%%%%%%%%%%%%%%%%%%%%%%%%%%%%%%%%%%%%%%%%%%%%%%%%%
\begin{figure}[!ht]
  \vbox{
    \resizebox{\hsize}{!}{\includegraphics{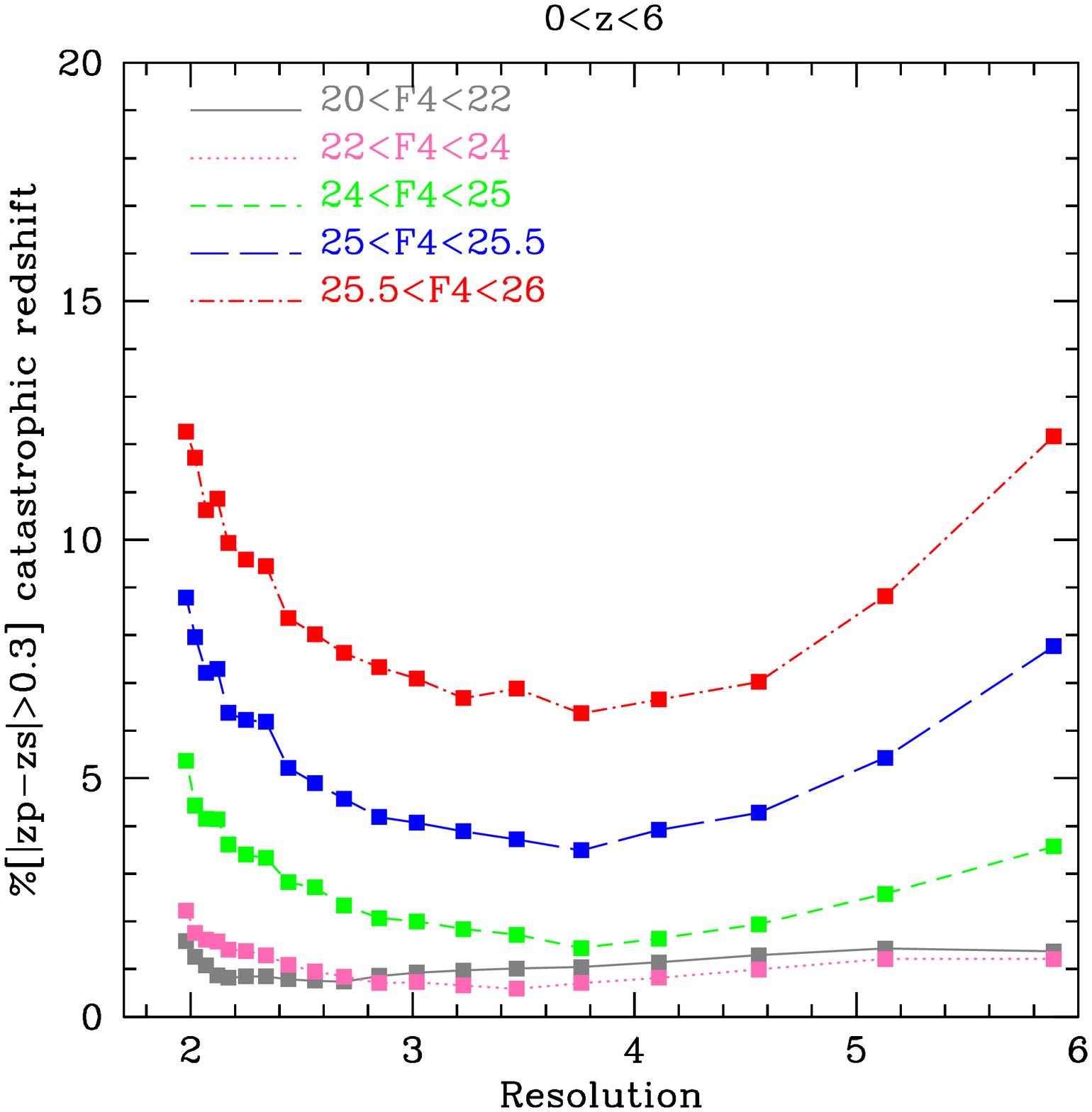}}
    \resizebox{\hsize}{!}{\includegraphics{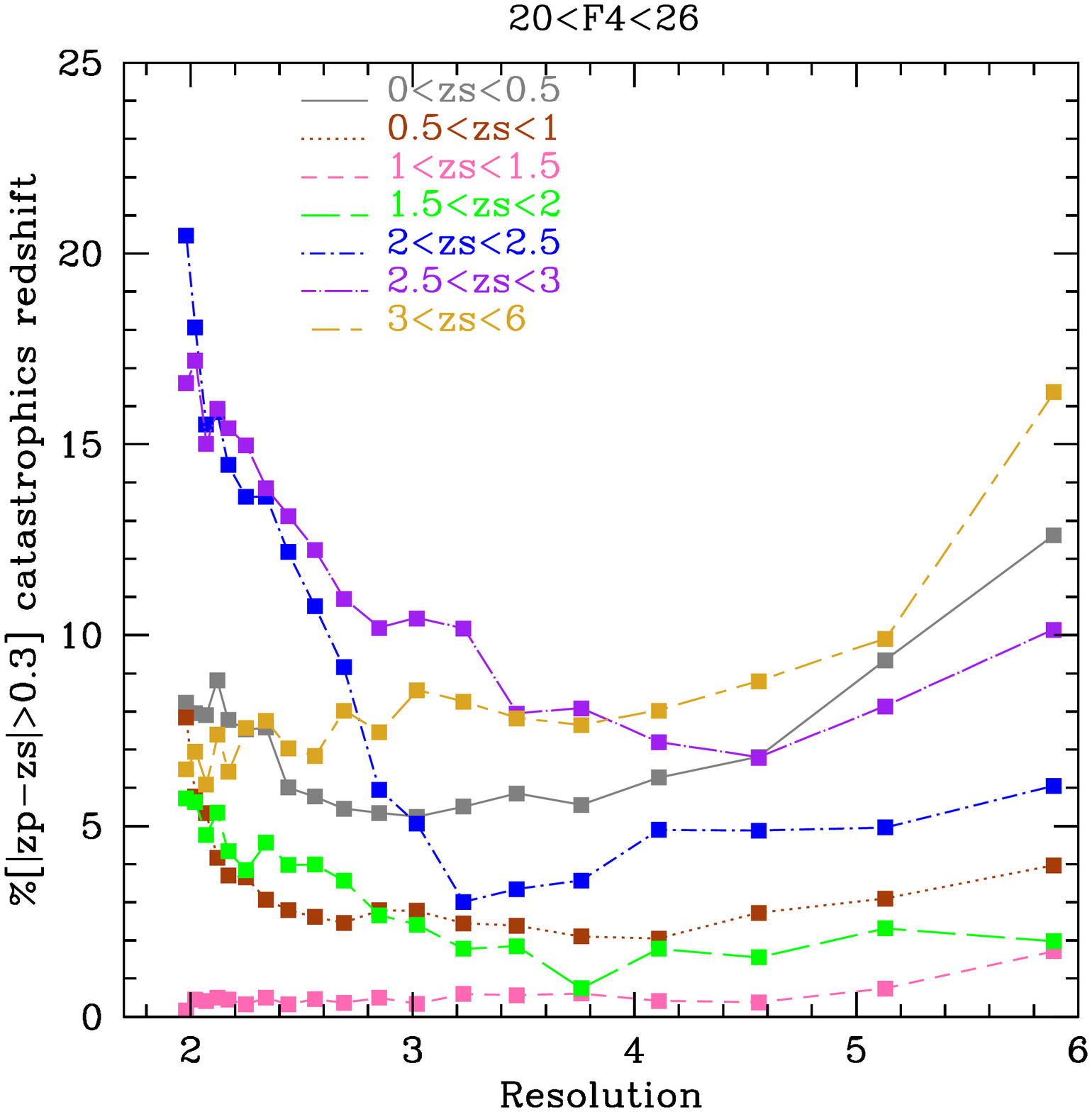}}
  }
  \caption{Percentage of catastrophic redshifts as function of the filter set
    resolution binned by magnitude up to $z~6$ (top panel) and by redshift
    (bottom panel) up to $I_{AB}=26$ mag. 
    $F_4$ represents the I-band filter and $z_s$ the spectroscopic redshift.} 
  \label{fig:catares}
\end{figure}
%%%%%%%%%%%%%%%%%%%%%%%%%%%%%%%%%%%%%%%%%%%%%%%%%%%%%%%%%%%%%%%%%%%

{\bf Figure~\ref{fig:catares}} shows the percentage of catastrophic
redshifts as a function of the filter set resolution binned by magnitude 
(top panel) and by redshift (bottom panel). Each square point corresponds 
to a filter set configuration. Catastrophic redshifts are defined as
$|z_{phot}-z_{spectro}|>0.3$. 
%We observe a similar behavior for the
%number of catastrophic redshifts as a function of the filter
%resolution as for the dispersion figure ~\ref{fig:sigres}. \\

To constrain the DE parameters, one of the WL techniques consists of
dividing the galaxy distribution into redshift slices
\citep{Bernstein10,Sun09}. We thus need accurate
photometric redshifts to avoid a contamination between slices. 
For the redshift range $1<z<3$, the Balmer
break or the D4000 is in the wavelength range fully covered by the
filter set $3200\AA<\lambda<17000\AA$. The color gradient produced will thus
ensure a robust photoz estimation. 
In the $2<z<3$ redshift range, the galaxies are fainter increasing 
the probability of color confusion and resulting in a higher catastrophic redshift rate.
Consequently, in this redshift range, a higher filter resolution increases the
color gradient accuracy which improves the photoz accuracy as shown 
by the blue and violet curves in the bottom panel of Figure~\ref{fig:catares}.

%The Lyman break enters the bluest filter
%at$z \approx 2.5$, but 
High redshift galaxies ($3<zs<6$) usually have faint apparent magnitudes. 
Broader filters are then more suitable to maximize the S/N as shown in terms of 
the percentage of catastrophic redshifts binned by magnitude 
(top panel of Figure~\ref{fig:catares}). The top panel of Figure~\ref{fig:catares} 
shows a preferred resolution range of $3<\mathcal{R}<4$
with a significant decrease in the fraction of outliers for galaxies $F4>24$. 
It reduces the outlier rate by 3\% to 10\% depending on the magnitude range considered.

For future dark energy surveys, the WL analysis is based on the statistics
of faint and numerous galaxies. Figures~\ref{fig:sigres} and ~\ref{fig:catares} show 
that the optimal WL choice uses broad filters and has a resolution in the
range $\mathcal{R}=3-4$.
%%%%%%%%%%%%%%%%%%%%%%%%%%%%%%%%%%%%%%%%%%%%%%%%%%%%%%%%%%%%%%%%%%%
\begin{figure}[!ht]
\resizebox{\hsize}{!}{\includegraphics{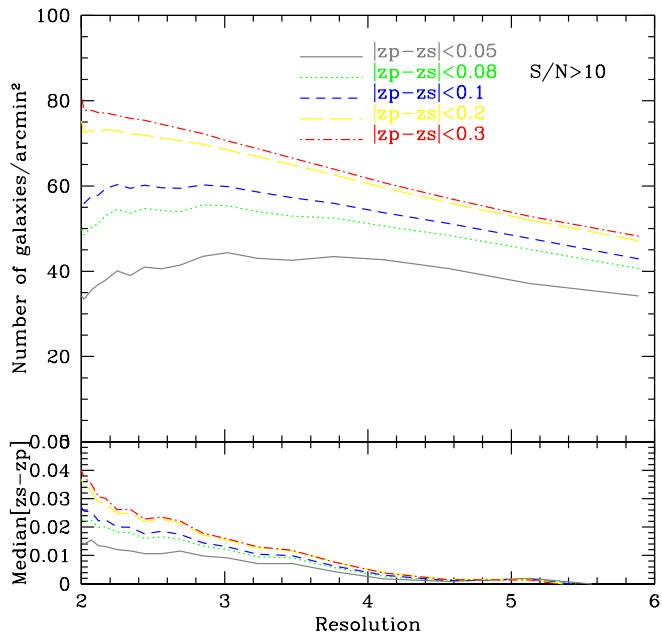}}
\caption{Number of galaxies and median($|zp-zs|$) as a function of the
  filter set resolution for different photometric redshift quality selections.}
\label{fig:narcres}
\end{figure}
%%%%%%%%%%%%%%%%%%%%%%%%%%%%%%%%%%%%%%%%%%%%%%%%%%%%%%%%%%%%%%%%%%%
%%%%%%%%%%%%%%%%%%%%%%%%%%%%%%%%%%%%%%%%%%%%%%%%%%%%%%%%%%%%%%%%%%%
\begin{figure}[!ht]
\resizebox{\hsize}{!}{\includegraphics{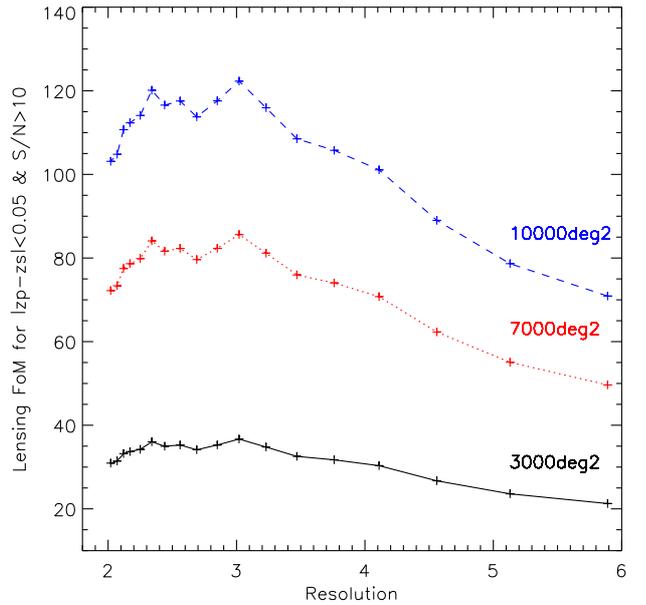}}
\caption{Lensing FoM as a function of the filter set resolution
  without Galactic absorption included, and a constant Zodiacal light.
  The dashed blue, dotted red, and solid black curves represent respectively 
  a survey area of 10000, 7000, and 3000 deg$^2$.}
\label{fig:fomres}
\end{figure}
%%%%%%%%%%%%%%%%%%%%%%%%%%%%%%%%%%%%%%%%%%%%%%%%%%%%%%%%%%%%%%%%%%%

{\bf Figure ~\ref{fig:narcres}} shows the number of galaxies and the
median$[zp-zs]$ for different photometric redshift quality
selections as a function of the filter resolution. On the one hand, if a strict photoz 
quality selection is used, the resolution giving the highest number 
of galaxies is in the range of $\mathcal{R}=3-4$.
On the other hand, the bias is smaller at higher filter resolution.
This shows the importance of an accurate spectroscopic
redshift calibration to estimate and correct for the bias of
the photometric redshift distribution.

To reach a definite conclusion about the filter resolution question,
{\bf Figure ~\ref{fig:fomres}} shows the FoM 
(defined in section \ref{sec:WLrequirements}) as a
function of the filter resolution for three different fractions of the sky
observed. A filter resolution of 3.2 provides the tightest
DE constraints. This FoM calculation does not take
into account the catastrophic redshift rate. However, this filter
resolution corresponds to the lowest rate of catastrophic redshifts (as shown
in Figure~\ref{fig:catares}), hence should be the optimal filter resolution 
in terms of the photoz accuracy of a WL analysis.

\section{CCD's blue sensitivity for catastrophic redshift} 
\label{sec:blue}
A possible way of reducing the number of catastrophic redshifts at low and high
redshift is to optimize the efficiency of visible detectors
in the near-UV. A higher sensitivity in the wavelength range $[3000-4000\AA]$ 
improves the photoz results at low redshift derived from either 
the Balmer or D4000 break.
This results in more accurate color gradient and photometric redshifts. 
In a similar way, it also helps to decrease the catastrophic
redshifts rate at high redshift related to the Lyman break feature. The
Lyman break is at $912\AA$ rest-frame and enters into the filter set for
galaxies at $z \sim 2.5$, which will help us to break the color degeneracy
between low and high redshift galaxies. \\

Thus, we explore the impact of the CCD quantum efficiency (QE)
and produce five QE curves differing in the near-UV wavelength
range. Each detector curve is then used with the eight-filter set of a
resolution $\mathcal(R)\approx3.2$ to derive noise properties that are
applied to generate a realistic mock catalog following the
prescription described in Section \ref{subsec:noise} using the CMC. Photometric
redshifts are thereafter calculated using the Le Phare photometric
redshift code.

\begin{figure}[!ht]
\vbox{
\resizebox{\hsize}{!}{\includegraphics{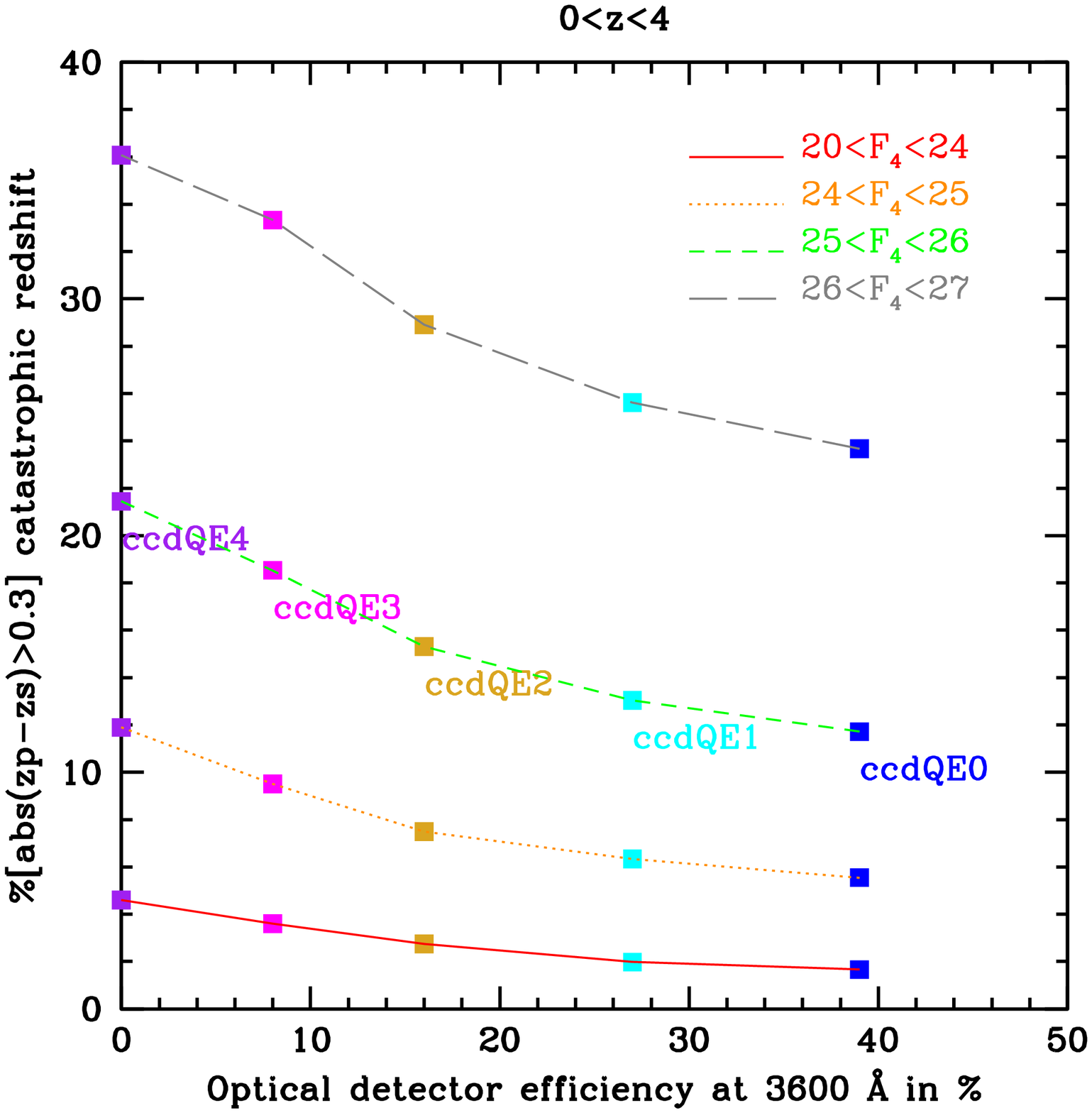}}
\resizebox{\hsize}{!}{\includegraphics{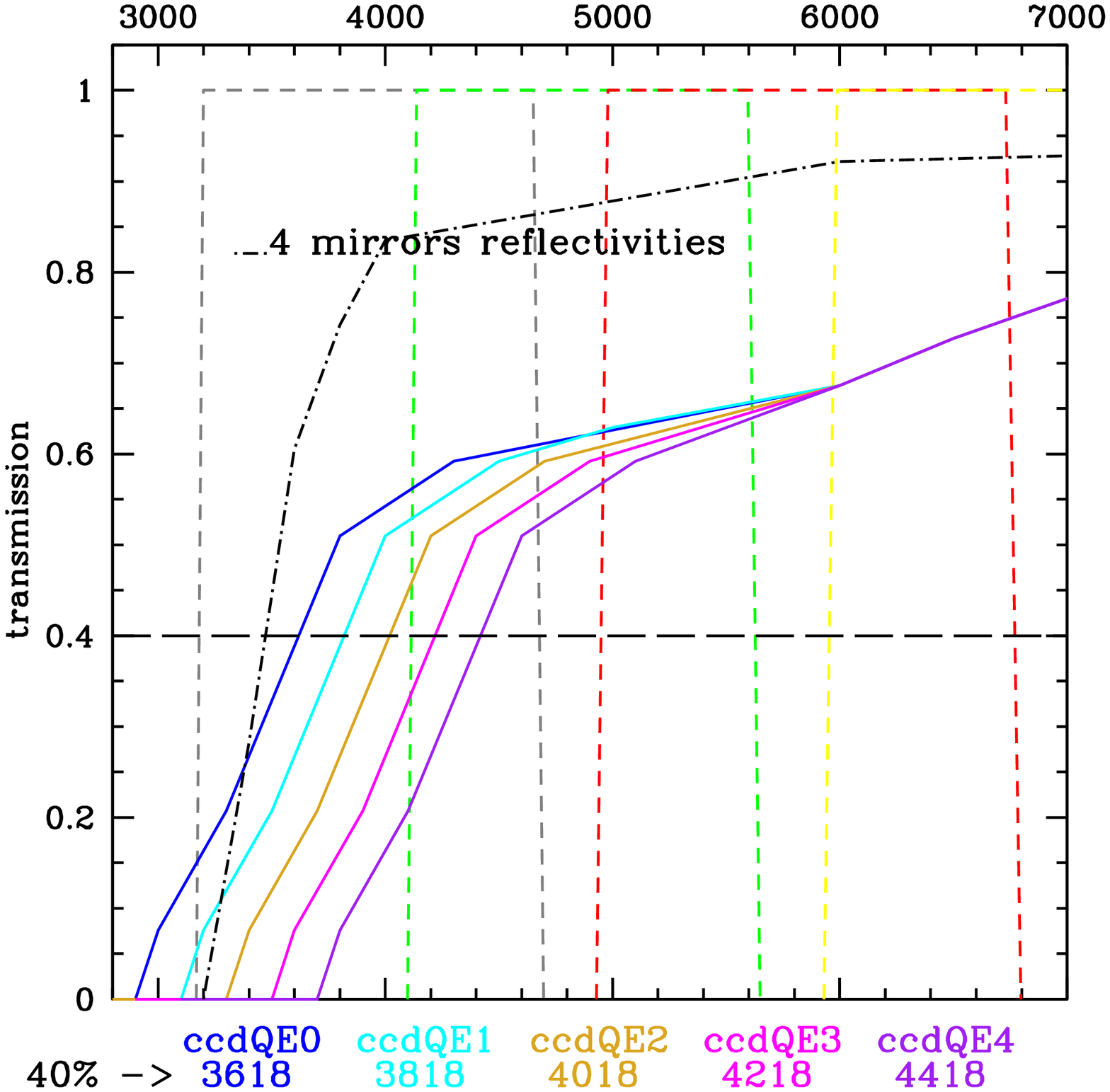}}
}
\caption{Bottom panel represents the transmission of optical dectectors (solid lines), 
filters (dahsed lines), and mirror reflectivities (dot-dashed lines) as a function of wavelength in \AA.  
The inscriptions below the bottom figure are the wavelength at 40\% efficiency for the different
optical detector curves named ccdQE$x$ where $x=[0,..,4]$. Top panel represents the percentage of catastrophic redshifts as a function of
the efficiency of the optical detector at 3600\AA. The colors of points are corresponding to the optical detector
of the same color that have been used in the photometric noise calculation.}  
\label{fig:eper}
\end{figure}

The bottom panel of {\bf Figure ~\ref{fig:eper}} shows the five CCD
QE curves used (solid multicolor lines) as a function
of wavelength. Each curve defines a mock catalog called ``ccdQE$x$'',
where $x=[0,..,4]$. The blue ''ccdQE0'' is the most efficient at blue
wavelengths, having a $40\%$ efficiency at $3620\AA$. The one with the
worst efficiency ``ccdQE4'' is shown in purple and has a $40\%$ efficiency at
$4420\AA$. The numbers below the figure are the wavelength at which
the CCD QE curves reach $40\%$ efficiency. The
dot-dashed line is the four mirror reflectivity. We note that the four
mirror reflectivity produces a cut-off of the bluest detector
curve. The dashed lines are the four first filters of the eight filter
set. The first two filters are the most affected by this gain in the CCD
QE in the near-UV/visible  wavelength range.

%%%%%%%%%%%%%%%%%%%%%%%%%%%%%%%%%%%%%%%%%%%%%%%%%%%%%%%%%%%%%%%%%%%%%%%%ù
%%%%%%%%%%%%%%%%%%%%%%%%%%%%%%%%%%%%%%%%%%%%%%%%%%%%%%%%%%%%%%%%%%%%%%%%%%
Figure \ref{fig:eper} top panel shows the percentage of
catastrophic redshifts as a function of the visible  detector efficiency
at $3600\AA$. Each square point is the same eight-filter set convolved
with a different efficiency of visible  detector and each curves
represent a different magnitude bin. We are mostly interested in the
magnitude range $25<F_4<26.5$ since this contains the very faint
galaxies that will be used in a WL analysis. For $25<F_4<26$, the green
curve shows that the mock catalog ccdQE1 contains $13\%$ of
catastrophic redshifts, while the catalog ccdQE4 contains around $21\%$.
A blue-optimised detector allows us to strengthen the
Balmer break signal in the bluest bands, which helps us to
decrease the percentage of catastrophic redshifts at low-z i.e. 
$0<z<1$ and high-z i.e. $z>2.5$.

The ccdQE4 catalog provides simular results to a seven-filter
configuration covering a wavelength range of $[4100-17000\AA]$. This illustrates how 
the removal of the first filter would affect the
percentage of contamination by the catastrophic redshifts in a
given magnitude range.  The fraction of catastrophic redshifts is about
two times higher in the magnitude range of interest for WL, $25<F_4<26$,
when not including the bluest filter.

We note that there is not much
improvement between using the catalogs ccdQE0 and ccdQE1, which is explained by
the cut-off in the four mirror reflectivity at $3200\AA$.
We note, however, that it is critical to improve the blue 
sensitivity of visible detectors (and possibly
the reflectivity of the mirrors at blue-UV wavelengths) to
help remove catastrophic redshifts. Ultimately, this can be done by using a
detector dedicated to the U-band, which would be blue-optimised.

%%%%%%%%%%%%%%%%%%%%%%%%%%%%%%%%%%%%%%%%%%%%%%%%%%%%%%%%%%%%%%%%%%%%%%%
\begin{figure}[!h]
  \resizebox{\hsize}{!}{\includegraphics{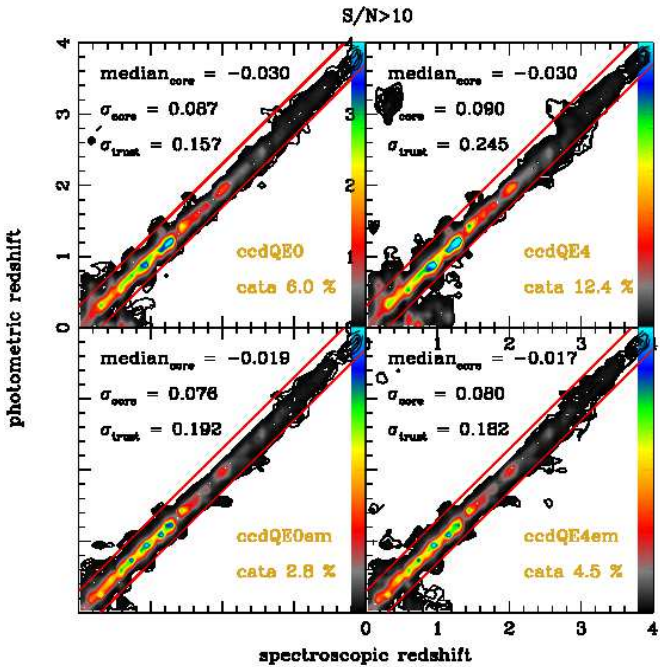}}
  \caption{zp-zs distribution in two cases of visible detector efficiency
    curves: ccdQE0 for panels on the left and ccdQE4 for panels on the right. 
    Bottom panels include the effect of emission lines in the library of templates used 
    for the photometric redshift estimate, while top panels do not. }
  \label{fig:zpzsblue}
\end{figure} 
%%%%%%%%%%%%%%%%%%%%%%%%%%%%%%%%%%%%%%%%%%%%%%%%%%%%%%%%%%%%%%%%%%%%%%%%%

{\bf Figure ~\ref{fig:zpzsblue}} shows the $z_p-z_s$ distribution of
ccdQE0 for left figures and ccdQE4 for right figures. We computed the 
photoz using a library of SED templates with emission lines for the 
bottom panels and without emission lines for the top panels. 
This last configuration is equivalent to a poor photometric redshift calibration. 
The mean and scatter are calculated for the galaxies whose redshift is 
located inside the two red lines as defined in section \ref{subsec:photozq}. 
The catastrophic redshift rate (gold writting) is the
percentage of galaxies outside the red lines. We also calculated the
scatter $\sigma_{trust}$ for galaxies meeting the 68\%
confidence interval criterion $\Delta^{68\%}z<0.5$ as defined in section~\ref{subsec:photozq}. 
A higher efficiency in the $3000-4000 \AA$  range helps to reduce the photoz scatter 
and minimize the number of catastrophic redshifts as shown in Figure ~\ref{fig:zpzsblue}. 
The catastrophic redshift rate is multiplied by a
factor of two when the U-band photometry is of poor quality. In the case of both a 
poor U-band photometry and a poor photometric redshift calibration, the
catastrophic redshift rate is multiplied by a factor of three as shown in Figure ~\ref{fig:zpzsblue}.

To summarize, improving the U-band photometry helps to similarly improve 
photometric redshift estimates and reduce the catastrophic redshift fraction 
by a factor of two in the magnitude range of interest for WL. 

\section{WL survey strategy: Number of filters and survey area}
\label{sec:surveystrategy}

\begin{figure}[!ht]
\resizebox{\hsize}{!}{\includegraphics{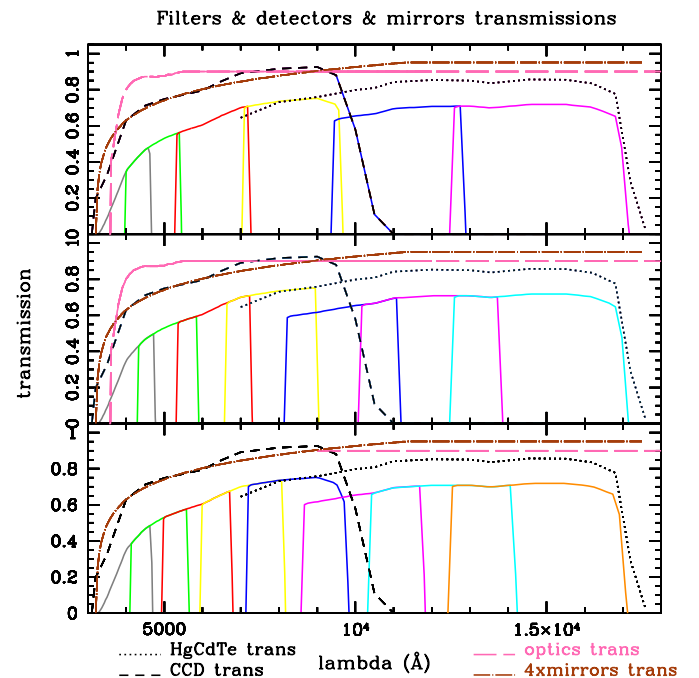}}
\caption{The 6-, 7- and 8-filter configurations using a filter
  resolution of $\mathcal{R}=3.2$. We include the
   transmission of optics (long-dashed pink), CCD (small-dashed black),
  NIR detector (dotted black), and the 4-mirror reflectivities (solid brown).}
\label{fig:filtersets}
\end{figure}

Future DE surveys plan to cover large areas to
detect a large number of galaxies. However, one should carefully
consider the survey strategy and the instrument design {\em together} to optimize the
areal coverage. Consequently, one has to choose the pixel scale and
the number of pixels that determine the 
FOV of the camera, the observation time, the number of exposures, and
the number of filters. Each of these choices affects the WL analysis
in terms of the number density of galaxy sources, the photometric redshift
accuracy, and the shape measurement quality, which defines the 
appropriate number of galaxies.
%%%%%%%%%%%%%%%%%%%%%%%%%%%%%%%%%%%%%%%%%%%%%%%%%%%%%%%%%%%%%%%%%%%%%%%
\begin{figure}[!h]
  \resizebox{\hsize}{!}{\includegraphics{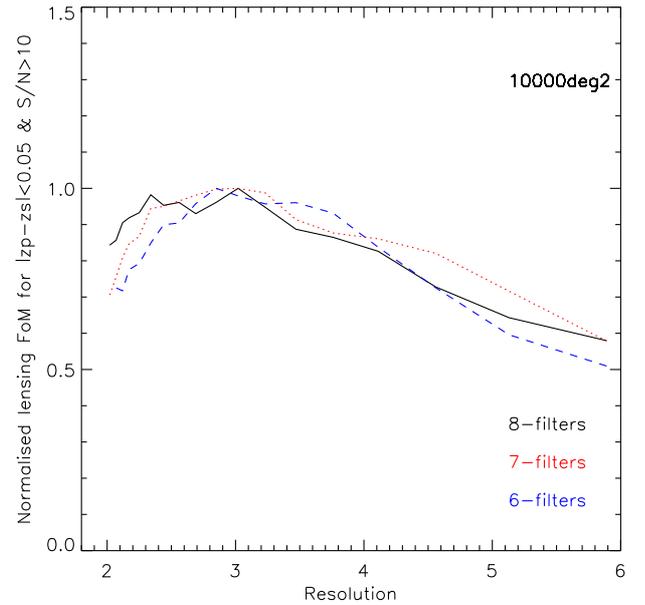}}
  \caption{Normalised lensing FoM as a function of filter resolution
    for a 6, 7, and 8 filter configurations in respectively dashed blue, dotted red,
    and black solid curves.}
  \label{fig:nfresol}
\end{figure}
%%%%%%%%%%%%%%%%%%%%%%%%%%%%%%%%%%%%%%%%%%%%%%%%%%%%%%%%%%%%%%%%%%%
In section \ref{subsec:tobs}, we have defined a minimal observation time
of 200s for each filter. In section \ref{subsec:res}, using this exposure 
time we found an optimal filter resolution of $\mathcal{R}\approx3.2$ for an eight filter
configuration. In {\bf Figure \ref{fig:nfresol}}, we show the WL dark
energy FoM normalised by their respective maximum for a six, seven, and eight
filter configuration as a function of filter resolution. The
resolution of 3.2 provides the highest values of FoM independently of the number
of filters. We thus use this optimal resolution to compare the efficiency of 
six, seven, and eight filter configurations, aiming to define a minimum number of
filters. With the CMC mock catalogs, we simulated three surveys
of six, seven, and eight filters using a resolution of
$\mathcal{R}\approx3.2$.  {\bf Figure~\ref{fig:filtersets}} shows
these filter configurations. To make a fair comparison
between survey configurations, we use the same total observation time
and divide this by the number of filters for a given survey
configuration. In all Figures of this section, we used a total
observation time of $2400s$, which is distributed into four exposures of 
respectively 150s, 150s, and
120s exposures per optical filters for the six, seven, and eight filter sets.
The 6-filter configuration has four optical filters and two NIR, while the 
7-filter and 8-filter configurations have three NIR filters each and respectively four 
and five optical filters, as described in Table~\ref{tab:filtersets}.
The I-band of the three configurations is F4 for the
8-filter catalog and F3 for the 7 and 6 filter catalogs. However, to
make fair selections and look at the same galaxy photometric redshift results
binned by magnitude, we simulated for each survey configuration a
magnitude in the F4 filter of the 8-filter survey configuration and used this
band as a reference to make different magnitude cuts for all 6, 7, 8-filter survey
configurations and allow simple comparisons.

%%%%%%%%%%%%%%%%%%%%%%%%%%%%%%%%%%%%%%%%%%%%%%%%%%%%%%%%%%%%%%%%%%%
%%%%%%%%%%%%%%%%%%%%%%%%%%%%%%%%%%%%%%%%%%%%%%%%%%%%%%%%%%%%%%%%%%%

The filter set properties are described in the Table~\ref{tab:filtersets}, 
where $\mathcal{R}$ and $\mathcal{R}_{eff}$ are respectively the filter resolution 
before and after convolution with the mirrors reflectivity and detectors QE. 
To have a close effective resolution $\mathcal{R}_{eff}$ for all filters, 
we constructed the first filter independently of the other filters. 
We note that the detectors QE and mirrors reflectivity have a large impact on 
$\mathcal{R}_{eff}$ for the first filter as shown in Table \ref{tab:filtersets}.

\begin{table}[!ht]
\caption{Filters characteristics.}
\begin{tabular}{cccccccc} \hline\hline
Camera & Filters   &  $\lambda_{mean}$(\AA) &  FWHM(\AA)  &  $\mathcal{R}$ & $\mathcal{R}_{eff}$ & $T_{obs}$(sec)\\
\hline\hline
Visible & 8-F0     &      3928.7  &    1490.0   & 2.64  &	4.78 & 120\\
& 8-F1     &      4869.8  &    1504.0   & 3.24  &	3.30 & 120\\
& 8-F2     &      5858.4  &    1809.9   & 3.24  &	3.28 & 120\\
& 8-F3     &      7047.6  &    2177.5   & 3.24  &	3.27 & 120\\
& 8-F4     &      8478.3  &    2618.3   & 3.24  &	3.27 & 120\\
NIR & 8-F5     &     10199.4  &    3150.3   & 3.24  &	3.27 & 200\\
& 8-F6     &     12269.9  &    3789.8   & 3.24  &	3.25 & 200\\
& 8-F7     &     14760.8  &    4559.6   & 3.24  &	3.27 & 200\\
\hline \hline
Visible & 7-F0   &    3963.8  &     1560.0  &  	2.54   & 4.53 & 150\\
& 7-F1   &    5091.6  &     1561.6  &  	3.26   & 3.31 & 150\\
& 7-F2   &    6302.9  &     1933.2  &  	3.26   & 3.30 & 150\\
& 7-F3   &    7802.3  &     2391.7  &  	3.26   & 3.28 & 150\\
NIR & 7-F4   &    9658.5  &     2961.1  &  	3.26   & 3.29 & 200\\
& 7-F5   &   11956.3  &     3666.1  &  	3.26   & 3.27 & 200\\
& 7-F6   &   14800.7  &     4538.0  &  	3.26   & 3.29 & 200\\
\hline \hline				       
Visible & 6-F0   &   3913.6  &    1460.0  &   2.68  & 4.89 & 150\\
& 6-F1   &   4703.5  &    1447.0  &   3.25  & 3.32 & 150\\
& 6-F2   &   6265.1  &    1927.1  &   3.25  & 3.29 & 150\\
& 6-F3   &   8345.1  &    2566.9  &   3.25  & 3.27 & 150\\
NIR & 6-F4   &  11115.8  &    3419.1  &   3.25  & 3.27 & 300\\
& 6-F5   &  14806.2  &    4554.9  &   3.25  & 3.29 & 300
\end{tabular}				  
\label{tab:filtersets}			  
\end{table}

\subsection{Photometric redshift quality vs number of filters}
\label{subsec:photoznbr}
%%%%%%%%%%%%%%%%%%%%%%%%%%%%%%%%%%%%%%%%%%%%%%%%%%%%%%%%%%%%%%%%%%%
\begin{figure}[!h]
  \vbox{ \resizebox{\hsize}{!}{\includegraphics{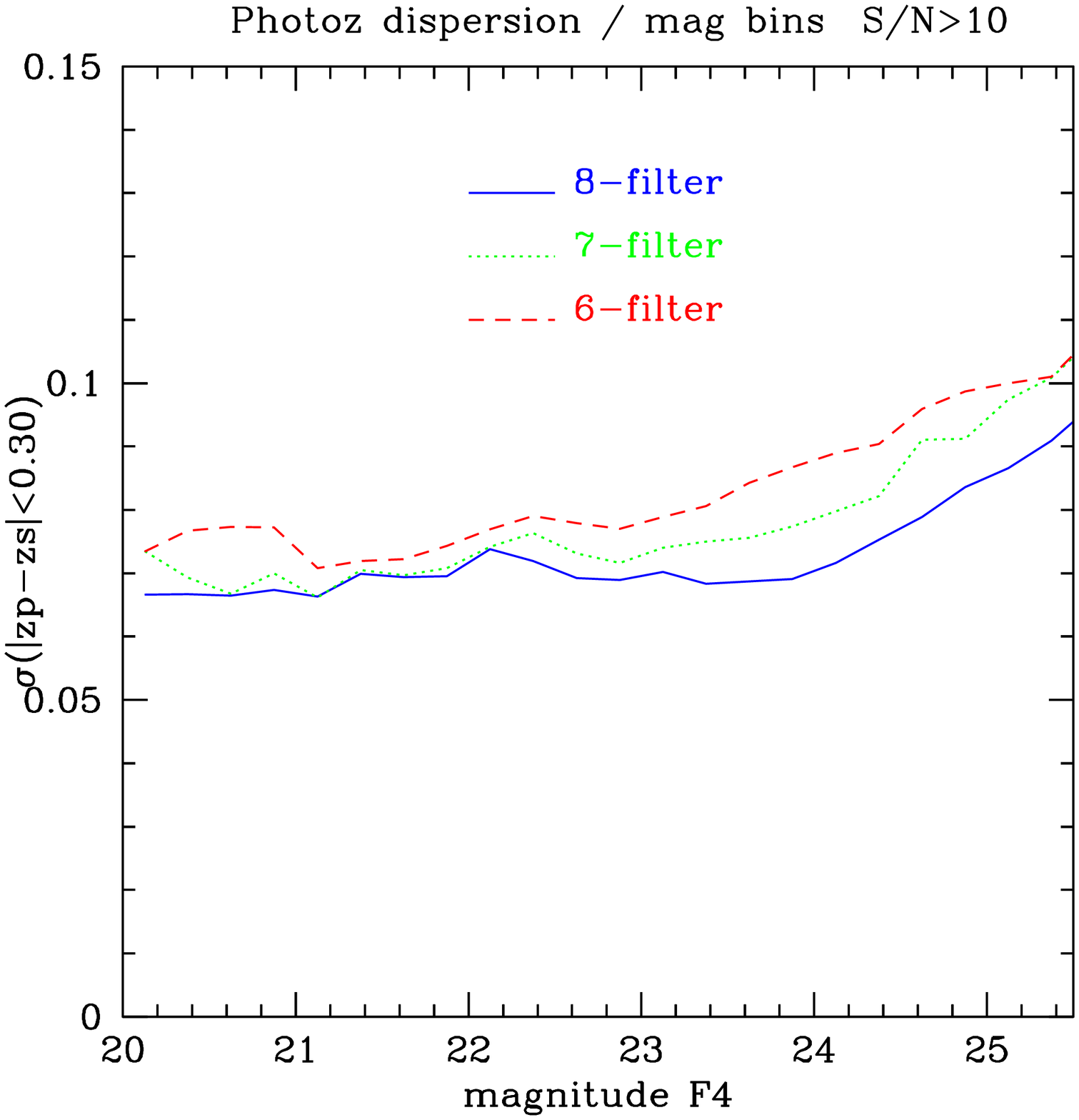}}
    \resizebox{\hsize}{!}{\includegraphics{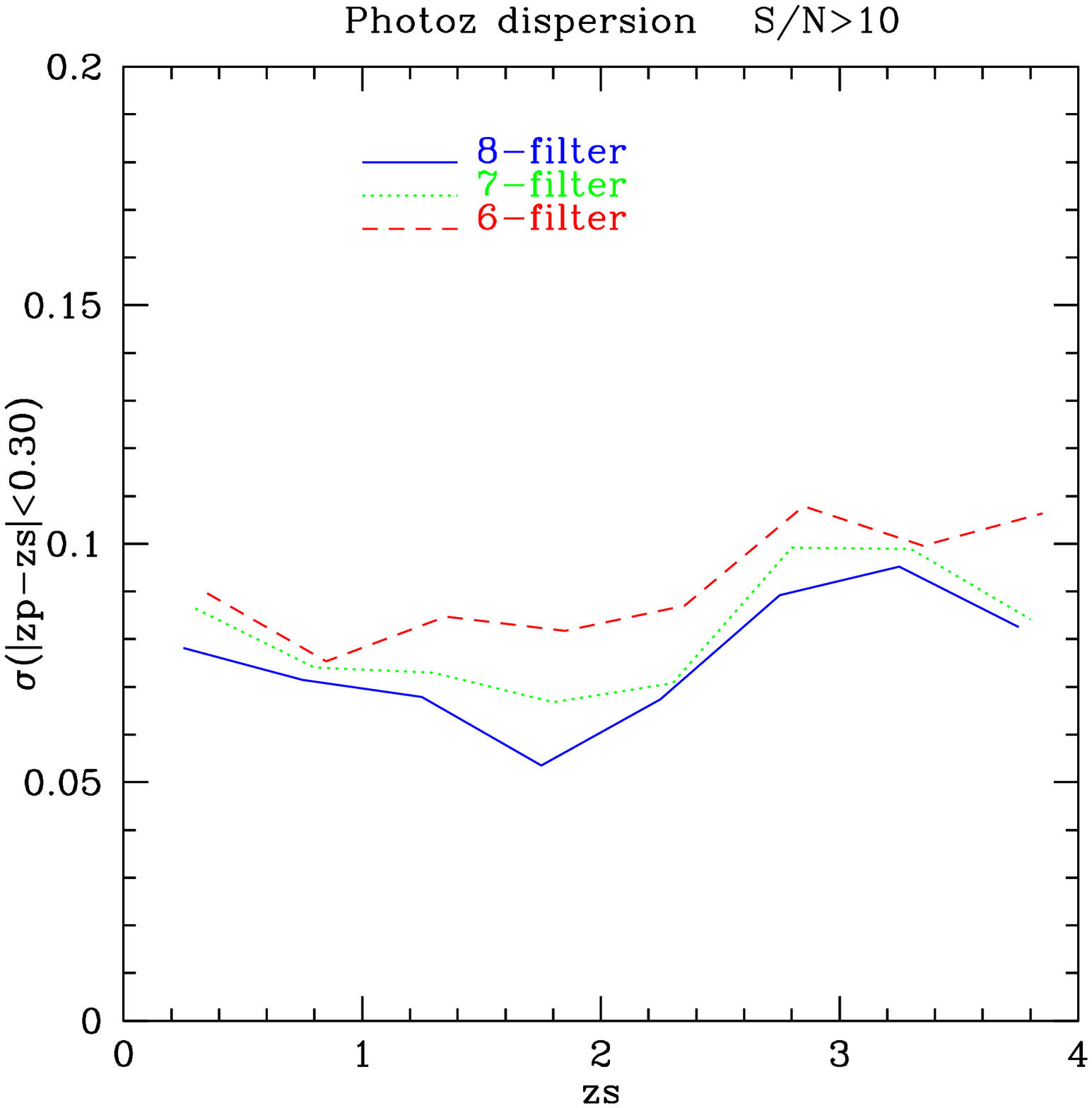}} }
  \caption{Photometric redshift scatter as a function of the spectroscopic redshift (top panel)
        and I-band magnitude (bottom panel) for the 6-, 7-, and 8- filter configuration.}
  \label{fig:sigset}
\end{figure}
%%%%%%%%%%%%%%%%%%%%%%%%%%%%%%%%%%%%%%%%%%%%%%%%%%%%%%%%%%%%%%%%%%%
%%%%%%%%%%%%%%%%%%%%%%%%%%%%%%%%%%%%%%%%%%%%%%%%%%%%%%%%%%%%%%%%%%%
\begin{figure}[!h]
\vbox{ \resizebox{\hsize}{!}{\includegraphics{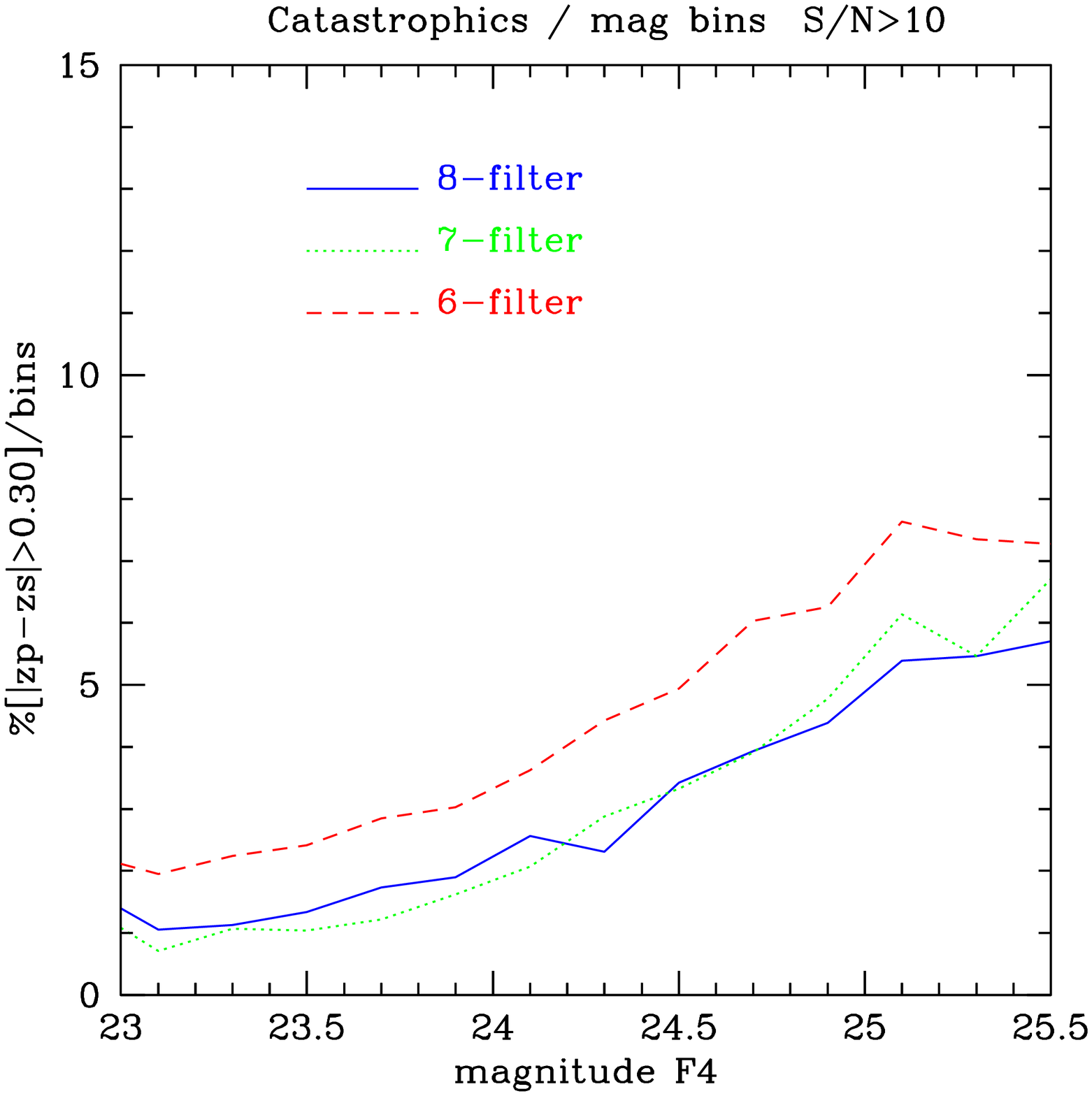}}
  \resizebox{\hsize}{!}{\includegraphics{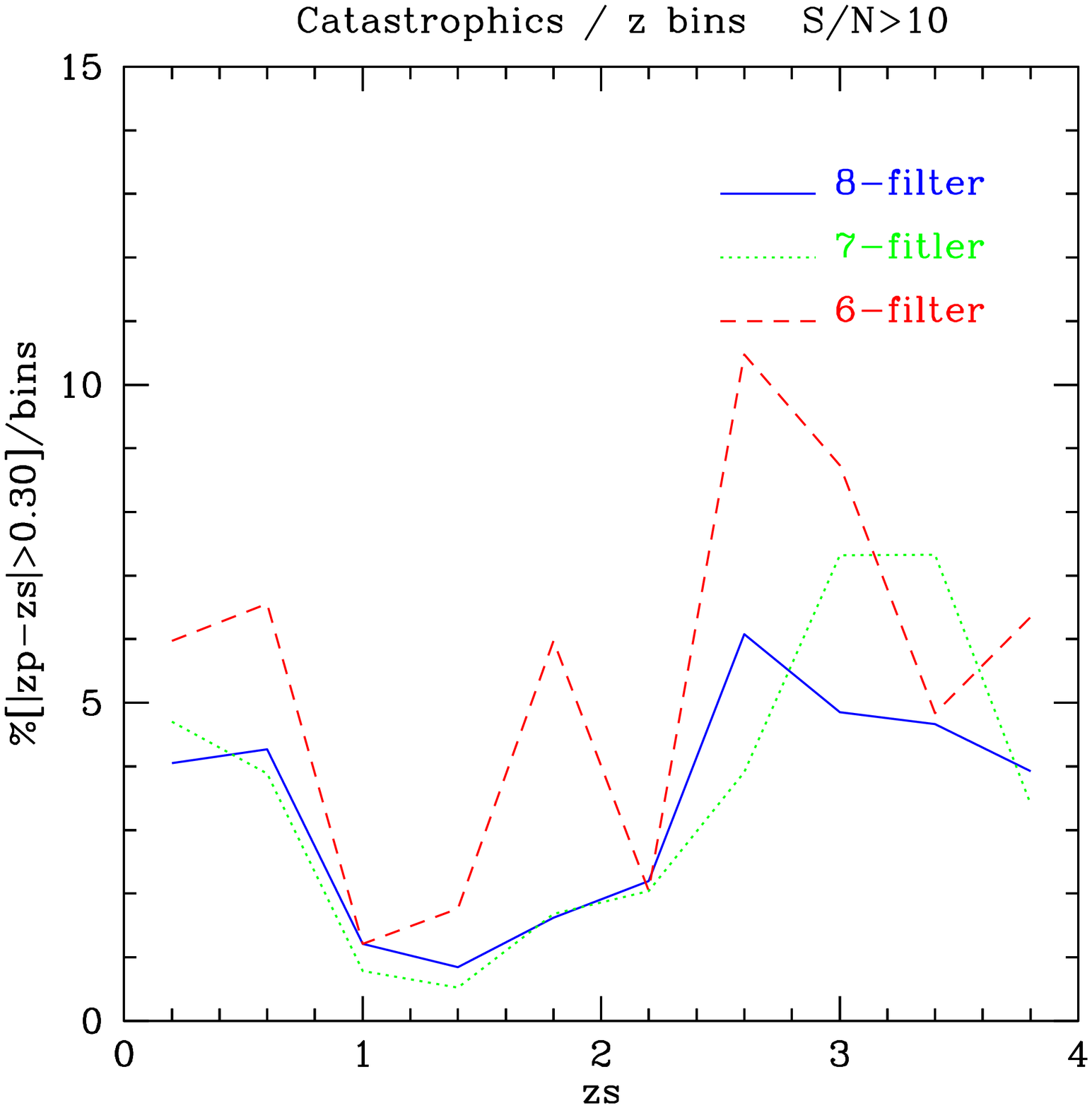}} }
\caption{Percentage of catastrophic redshifts as a function of the spectroscopic redshift (top panel)
        and I-band magnitude (bottom panel) for the 6-, 7-, and 8- filter configuration.}
\label{fig:cataset}
\end{figure}
%%%%%%%%%%%%%%%%%%%%%%%%%%%%%%%%%%%%%%%%%%%%%%%%%%%%%%%%%%%%%%%%%%%
%%%%%%%%%%%%%%%%%%%%%%%%%%%%%%%%%%%%%%%%%%%%%%%%%%%%%%%%%%%%%%%%%%%
\begin{figure}[!h]
\vbox{
\resizebox{\hsize}{!}{\includegraphics{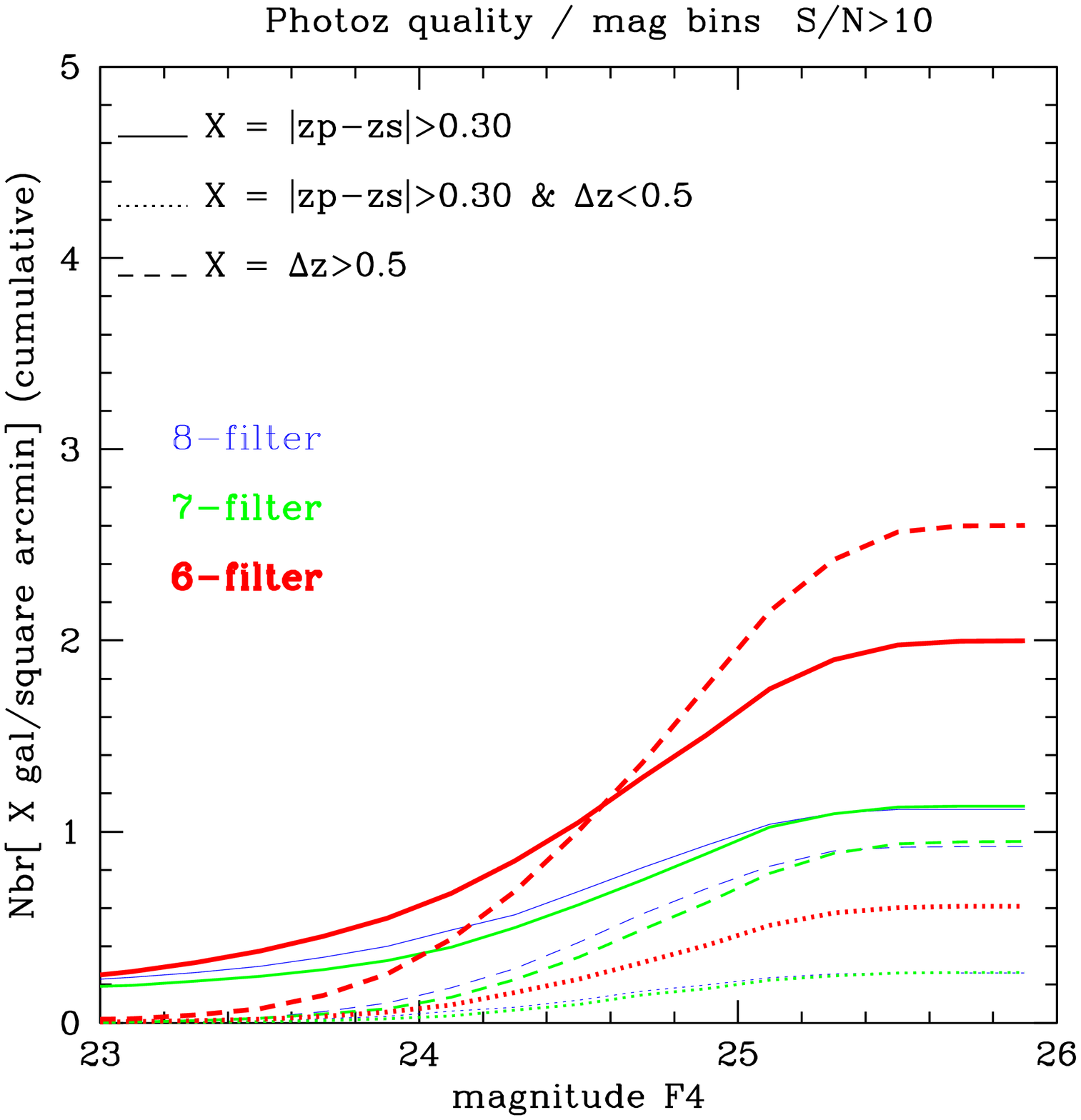}}
\resizebox{\hsize}{!}{\includegraphics{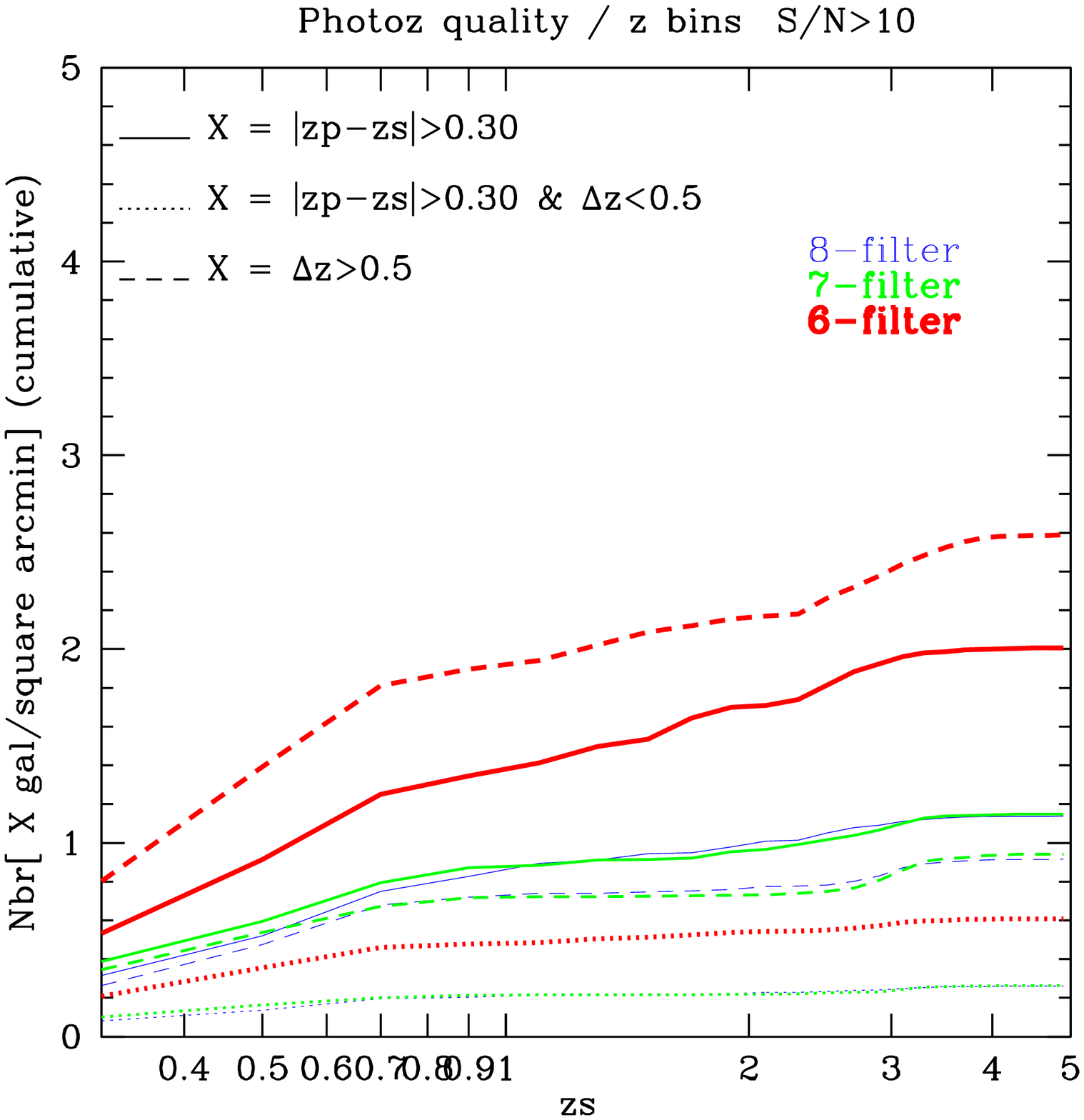}}
}
\caption{Density of catastrophic redshifts  as a function of the spectroscopic redshift (top panel)
        and I-band magnitude (bottom panel) for the 6-, 7-, and 8- filter configuration.}
\label{fig:densityset}
\end{figure}
%%%%%%%%%%%%%%%%%%%%%%%%%%%%%%%%%%%%%%%%%%%%%%%%%%%%%%%%%%%%%%%%%%% In
In {\bf Figures \ref{fig:sigset}}, {\bf \ref{fig:cataset}}, and
{\bf \ref{fig:densityset}}, we use the galaxies with $S/N>10$ in the I-band
in the photometric redshift analysis.  Figures~\ref{fig:sigset}
  and \ref{fig:cataset} show, respectively, the photometric redshift
dispersion and percentage of catastrophic redshifts as defined in
section~\ref{subsec:photozq}. Compared to the 8-filter configuration,
the 7-filter configuration has a photometric redshift scatter that is larger 
by 0.01, while the 6-filter configuration has a dispersion that is 0.02 larger
(top panel Figure \ref{fig:sigset}). However, the percentage of
catastrophic redshifts is lower in the 7-filter configuration than in
other configurations as shown in Figure \ref{fig:cataset}. This
last figure shows the catastrophic redshift rate as a function of
redshift (bottom panel) and magnitude (top panel). Both panels show
that the 7-filter configuration is very similar to the 8-filter
configuration. However, for faint galaxies at magnitude $I>24.5$, 
the 8 filter configuration has lower catastrophic redshift rates 
as shown in the top panel of figure  \ref{fig:cataset}.  
The 6-filter configuration provides results that are relatively 
worse than other configurations, 
with about 8\% of catastrophic redshift
contamination for galaxies at magnitude $I\approx 25.5$ at a
$S/N>10$, while the 7 and 8-filter set have about 5\% contamination in
this magnitude range. Even though the 6-filter configuration has 
deeper photometry than other configuration, this does not compensate for
the lack of color gradient information (from the amount of filters) needed
for a good photoz estimation. 
%for the lack of information on the galaxy colors that is needed to find
%the redshift of galaxies.

\citet{Ma08a}, \citet{Huterer06}, and \citet{Bernstein10} showed that the photoz scatter 
does not significantly degrade the estimated DE parameters. However,
the photoz scatter impacts the intrinsic alignment as
shown in \citet{Bridle07}. Thus, photoz scatter,
biases, and catastrophic redshifts may have a strong effect on the
estimated parameters. Consequently, the recommended minimum number of filters is seven
assuming our covering strategy in wavelength.
This configuration gives the highest accuracy and minimizes
the number of catastrophic redshifts binned in redshift and magnitude. We
also conclude that the 6-filter configuration does not seem a good option 
in terms of photometric redshift accuracy.

{\bf Figure~\ref{fig:densityset}} shows the cumulative number density
of catastrophic redshifts as a function of magnitude (top panel)
and redshift (bottom panel). The solid, dotted, and dashed lines
show respectively the galaxy number density for which
($|z_p-z_s|>0.3$),($|z_p-z_s|>0.3$ \& $\Delta^{68\%}z<0.5$), and
($\Delta^{68\%}z>0.5$). The 7 and 8-filter catalog have similar 
results with a number of catastrophic redshifts of 1.1
gal/arcmin$^2$, while the 6-filter configuration has 2
gal/arcmin$^2$. For the 7 and 8-filter configuration, most of the
catastrophic redshifts are flagged as a poor photoz estimates ($\Delta^{68\%}z>0.5$),
leaving only 0.3gal/arcmin$^2$ of contamination when excluding these.
For the 6-filter configuration, the number density is 0.6gal/arcmin$^2$
after excluding flagged objects. Thus, these dotted lines show the
number of catastrophic redshifts that the code selected as a reliable
estimate of galaxy redshifts that we define to be $\Delta^{68\%}z<0.5$
in Section~\ref{sec:WLrequirements}. These galaxies are the
ones that enter into a WL analysis and degrade the cosmological parameter estimation.

This work also demonstrates that for the same amount of telescope time the 7 and 
8-filter configurations are the best options in terms of photoz results. 
Both of these configurations would give an accurate photometric redshift 
estimate for most galaxies down to $I\approx26$. All these numbers are optimistic since
we are using the same library of galaxy spectra to generate mock
catalogs and calculate photometric redshifts. If the
photometric redshift calibration survey (PZCS) is perfectly
representative of the galaxy distribution in terms of redshift,
depth, color, and size, these results should be very close to what
we will obtain in future DE misssions. 

\subsection{Optimizing the WL strategies using the dark energy FoM}

We use the iCosmo package to compute
FoM using the telescope characteristics defined in section~\ref{sec:WLrequirements}.
To calculate the FoM from these configurations, we
assume a space-based mission with all filters onboard the satellite.

\subsubsection{Surveying the sky, limits from Galactic absorption and Zodiacal light}

%%%%%%%%%%%%%%%%%%%%%%%%%%%%%%%%%%%%%%%%%%%%%%%%%%%%%%%%%%%%%%%%%%%
\begin{figure}[!ht]
\resizebox{\hsize}{!}{\includegraphics{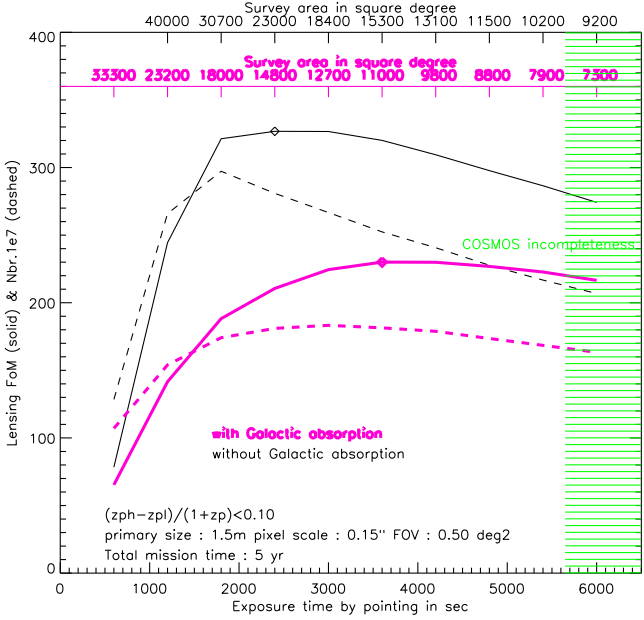}}
\caption{WL FoM as a function of exposure time by pointing with a primary mirror 
diameter of 1.5m, a pixel scale of 0.15'', and a FOV=0.5deg$^2$ for a 5-year mission. 
The magenta curves include the effect of the Galactic absorption and Zodiacal 
light variation, while the black curve do not. 
The dashed curves represent the number of galaxies$\times$1e7 for
both configurations. The green region shows the COSMOS incompleteness affecting 
the reliability of CMC galaxy densities, which may impact conclusions in these areas.} 
\label{fig:fom_galac}
\end{figure}
%%%%%%%%%%%%%%%%%%%%%%%%%%%%%%%%%%%%%%%%%%%%%%%%%%%%%%%%%%%%%%%%%%%

In {\bf Figure \ref{fig:fom_galac}}, we computed 
the WL FoM (solid lines) and total number of objects (dahsed lines) for a 
five-year mission as a function of the exposure time by pointing 
for an 8-filter configuration. 
We assume two cameras, with five filters for the visible 
camera and three filters for the IR camera. We assume the same total 
exposure time for the NIR and visible cameras, leading to different 
exposure times for each filter depending on the camera the filter 
belongs to (see Table~\ref{tab:filtersets}).
The magenta curves illustrate the effect of the Galactic absorption and Zodiacal 
light variations as discussed in section~\ref{subsec:area} eq. \ref{eq:tobs_pting}, 
while  black curves do not. 
The green dashed area shows the CMC incompleteness region, which
corresponds to the limiting depth of the COSMOS data. Because the incompleteness
region only occurs at the longest exposure time considered, it does not affect the
results presented here. This
green-dashed part of the graph will be affected by the COSMOS limiting depth.  
The axis on the top represents the area in square degrees for a five years WL mission,
including (magenta lines) or excluding (black lines) the variation in both 
Galactic absorption and Zodiacal light. The black
curves are what can be found if you neglect the Galactic absorption variation across 
the sky (as in \citet{Amara07}). The magenta curves represent a specific 
survey strategy consisting of compensating for the 
Galactic absorption and Zodiacal light variation by varying the exposure time in
a way that the photometry depth is equal for the whole area surveyed. 
This survey strategy has a strong impact on the survey speed of the mission.
When properly taken into account the Galactic absorption and the Zodiacal 
light variation, the often quoted 20,000 deg$^2$ is no longer optimal. 
The optimal WL survey we found covers $\sim$ 11\,000deg$^2$ for a five-year mission.

The survey strategy chosen here is to obtain a similar photometric 
depth across the survey area, thus increasing the exposure time at low 
Galactic latitudes. Another possible strategy is to use the same exposure 
time for every pointing, but this would result in an inhomogeneus 
photometric depth across the survey area.
The photometric redshift quality would then vary across the survey area. This would be difficult to 
compensate for and would likely degrade the DE constraints from the WL analysis as the survey beyond 
10000 deg$^2$ will not provide a much higher galaxies density than can be achieved from
the ground. Ultimately, it may be interesting to consider a trade-off between 
survey area and depth in more complexe suvey, especially if we include other probes 
in the FoM calculation, such as BAO or SN. 

We note that the impact of the Galaxy on the survey remains simplified, as we did 
not take into account the variation in the star densities. 
This will reduce the effective surface density of extragalactic sky observed
at low Galactic latitudes, and support the case for an even smaller 
survey area than the one suggested here.

\subsubsection{Photometric redshift quality $\epsilon$}

%%%%%%%%%%%%%%%%%%%%%%%%%%%%%%%%%%%%%%%%%%%%%%%%%%%%%%%%%%%%%%%%%%%
\begin{figure}[!ht]
\resizebox{\hsize}{!}{\includegraphics{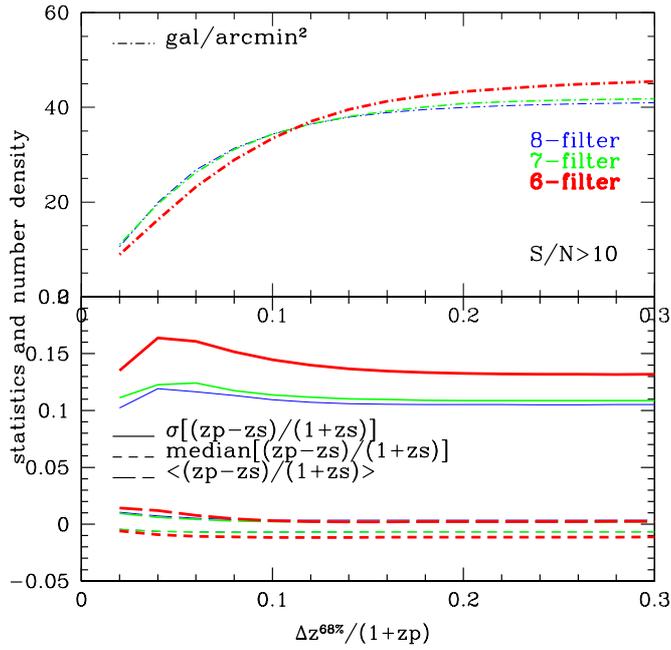}}
\caption{Statistics and number density for the 6-, 7-, and 8-filter survey 
configurations using the telescope properties described in Table~\ref{tab:jdem} 
with observation times described in Sect. ~\ref{tab:filtersets}. }
\label{fig:sigtrust}
\end{figure}
%%%%%%%%%%%%%%%%%%%%%%%%%%%%%%%%%%%%%%%%%%%%%%%%%%%%%%%%%%%%%%%%%%%
% Low number density because I used a pointing of 2400sec or 120s/filters in the optical... 
% I used 200sec/filters for the narcres figure...

%Figure~\ref{fig:fom_zerr} shows the WL  FoM and the number
%density of objects derived from the photometric redshift catalogs
%generated for a set of exposure time going from 50s to 500s using
%an arbitrary area of 10000deg$^2$. The different curves represent
%different selections in terms of photometric redshift error
%$\epsilon=[0.03..0.07]$ where $\epsilon$ is defined in section
%\ref{sec:WLrequirements}. We find an optimal photometric redshift
%error value of $\epsilon=0.06$, when the number density raising of
%galaxies is not enough to counterbalance the loss in photometric
%redshift accuracy. However, we have prefered to use a value of
%$\epsilon=0.05$ in the other figures which is also close to optimal.

{\bf Figure~\ref{fig:sigtrust}} shows the galaxy number density
as a function of the quality criterion we use for the photometric redshift selection:
$\epsilon = \Delta^{68\%}z/(1+zp)$ (top panel). 
%The 6-filter configuration benefit of a longer exposure time per filter. It thus shows more galaxies with a
%high signal-to-noise. 
In the bottom panel of Figure~\ref{fig:sigtrust}, we compute the dispersion $\sigma$, 
mean, and median of the $(zp-zs)/(1+zs)$ distribution as a function of the 
photometric redshift quality criterion. 
This shows that the 6-filter configuration has a dispersion that is
larger by 0.02 to 0.05 than the values for the 7 and 8-filter configurations and is also
more biased as shown by of mean and median of the photometric redshift versus 
the spectroscopic redshift distribution. We note that, again, the 7-filter 
configuration has similar results to the 8 filter-configuration.

%%%%%%%%%%%%%%%%%%%%%%%%%%%%%%%%%%%%%%%%%%%%%%%%%%%%%%%%%%%%%%%%%%%
\begin{figure}[!ht]
%\resizebox{\hsize}{!}{\includegraphics{lensing_fom_5tomobins_1.5m_expotime_zerr0.05_sets.ps}}
%\resizebox{\hsize}{!}{\includegraphics{lensing_fom_5tomobins_1.5m_expotime_zerr0.10_3yr_sets.ps}}
%\resizebox{\hsize}{!}{\includegraphics{lensing_fom_5tomobins_1.5m_expotime_zerr0.05and0.10_3yr_678sets.ps}}
\resizebox{\hsize}{!}{\includegraphics{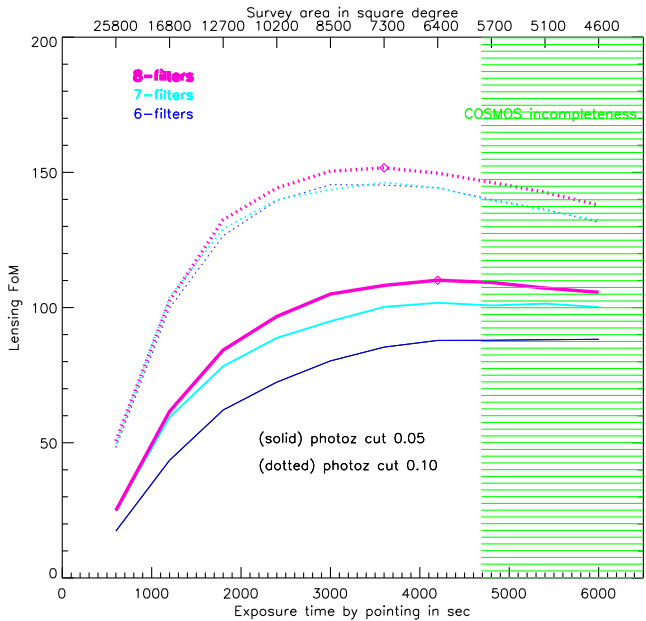}}
\caption{WL FoM as a function of exposure time by pointing for the 6-, 7-, and 
8-filter configurations using a primary mirror diameter of 1.5m and pixel scale of 
0.15'', and a FOV=0.5deg$^2$ for a 3-year mission.
The solid and dotted curves represent a photometric redshift selection 
of respectively (zph-zpl)/(1+zp)$<$0.05 and 0.01.}
\label{fig:fom_expo}
\end{figure}
%%%%%%%%%%%%%%%%%%%%%%%%%%%%%%%%%%%%%%%%%%%%%%%%%%%%%%%%%%%%%%%%%%%

We computed in {\bf Figure~\ref{fig:fom_expo}} the WL FoM for a three-year mission:
1) as a function of the photometric redshift quality $\epsilon$, and 
2) as a function of the exposure time by pointing as defined in section~\ref{subsec:area} 
(eq. \ref{eq:tobs_pting}), and 3) for different numbers of filters.

The axis at the top represents the area in square degrees for a three year WL mission,
including the effects of Galactic absorption. The magenta, cyan, and blue curves
represent the WL FoM for respectively the 8-, 7-, and 6-filter configurations as
defined in section~\ref{subsec:photoznbr}. The solid and dotted lines
present the WL FOM for respectively a photometric redshift quality selection of $\epsilon=0.05$ and
$\epsilon=0.1$. The green dashed area shows the CMC incompleteness region.
The incompleteness region varies for the different survey configurations. Hence, 
we present the region for the most restrictive case, which corresponds 
to the 6-filter configuration in this figure.

\begin{figure}[!ht]
\resizebox{\hsize}{!}{\includegraphics{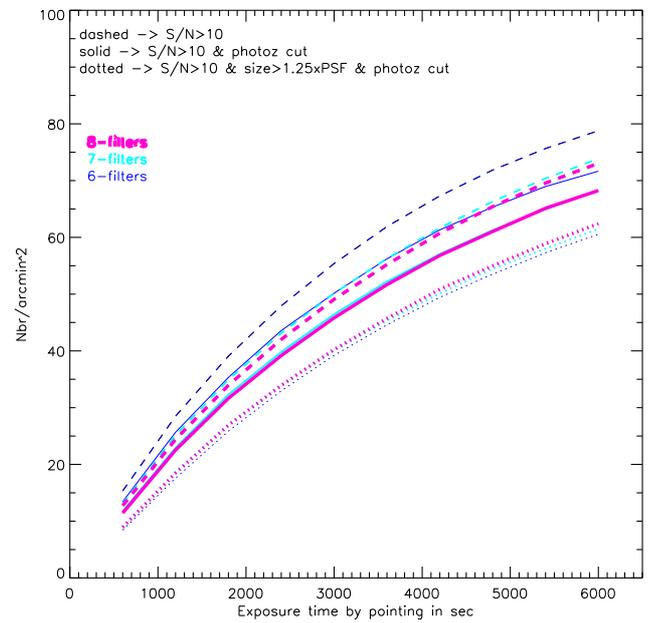}}
\caption{Density of galaxies in an I-band filter as a function of
  exposure time by pointing for S/N$>$10 (dashed), S/N$>$10 \& (zph-zpl)/(1+zp)$<$0.1 (solid),
   S/N$>$10 \& (zph-zpl)/(1+zp)$<$0.1 (solid) \& galaxy size$>$1.25$\times$ePSF (dotted).}
\label{fig:galdensity}
\end{figure}

The solid and dotted curves show that the eight-filter configuration provides the 
tighest DE constraints for an exposure time by pointing of 3600s
and 4000s, respectively. This corresponds to an exposure time by filters of 
180s-200s per filter for the visible camera
over an area close to 7300deg$^2$ for a three year mission. 
The number density of useful galaxies reaches 45 gal/arcmin$^2$ at 10$\sigma$ in I-band 
for the eight and six-filter configurations as shown 
in {\bf Figure \ref{fig:galdensity}}. This last figure represents the galaxy number 
density as a function of exposure time by pointing for the three configurations. 
The six-filter configuration has results close to the
eight-filter configuration, especially if we relax the photo-z quality
selection. However, this calculation does not include catastrophic
redshifts, the discrepancies between both configurations will be a larger, especially 
if we include the impact of a PZCS and the catastrophic redshift rate. 

The eight-filter configuration yields the best FoM (although 
not by a large factor). The accuracy needed to estimate the DE parameters requires
an ambitious survey. A 6-filter configuration may render the final measurement 
unreliable because of the systematic errors introduced by the larger 
number of catastrophic redshifts and a larger photometric redshift scatter and biases.

%%%%%%%%%%%%%%%%%%%%%%%%%%%%%%%%%%%%%%%%%%%%%%%%%%%%%%%%%%%%%%%%%%%
%%%%%%%%%%%%%%%%%%%%%%%%%%%%%%%%%%%%%%%%%%%%%%%%%%%%%%%%%%%%%%%%%%%

Figure \ref{fig:galdensity} also shows the impact of the photometric and
size selections on the three configurations. The most stringent
selection is the quality cut based on galaxy sizes with a loss of respectively 8 
gal/arcmin$^2$ for the 7- and 8-filter configurations and 13 gal/arcmin$^2$
for the 6-filter one. The 6-filter configuration has
deeper photometry, which raises the number of faint, small galaxies
thus causing this configuration to be more affected by the galaxy size
criterion. For the same reason, the 6-filter configuration is more
affected by the photometric redshift selection, $S/N>10$, and loses around
8 gal/arcmin$^2$, while other configurations lose around 5
gal/arcmin$^2$. 

In this comparison, we have used by default a 1.5m primary mirror
diameter. If were to consider a 1.2m mirror diameter the loss produced would
be a lot more significant because of the S/N decrease and the
larger galaxy-size selection. 

%%%%%%%%%%%%%%%%%%%%%%%%%%%%%%%%%%%%%%%%%%%%%%%%%%%%%%%%%%%%%%%%%%%
%\begin{figure}[!ht]
%\caption{Diagram photometric reshift vs true redshift util 26.5 AB
%magnitude in an I-band-like filter.}
%\resizebox{\hsize}{!}{\includegraphics{comp_set_zpzs_26.5_paper.ps}}
%\label{zpzset}
%\end{figure}

%%%%%%%%%%%%%%%%%%%%%%%%%%%%%%%%%%%%%%%%%%%%%%%%%%%%%%%%%%%%%%%%%%%
%\begin{figure}[!ht]
%\resizebox{\hsize}{!}{\includegraphics{lensing_fom_5tomobins_mir_expotime_zerr0.05_sets.ps}}
%\caption{WL  FoM as a function of exposure time and mirror size.}
%\label{fig:fom_mir}
%\end{figure}
%%%%%%%%%%%%%%%%%%%%%%%%%%%%%%%%%%%%%%%%%%%%%%%%%%%%%%%%%%%%%%%%%%%

\subsubsection{Effect of the mission duration $T_{mission}$}

%%%%%%%%%%%%%%%%%%%%%%%%%%%%%%%%%%%%%%%%%%%%%%%%%%%%%%%%%%%%%%%%%%%
\begin{figure}[!ht]
%\resizebox{\hsize}{!}{\includegraphics{lensing_8Bfom_5tomobins_1.5m_expotime_zerr0.10.ps}}
\resizebox{\hsize}{!}{\includegraphics{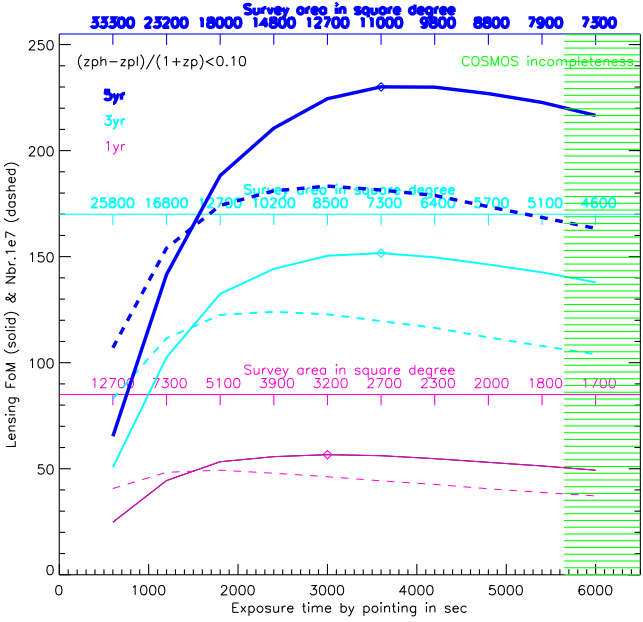}}
\caption{WL  FoM as a function of exposure time by pointing for a mission time of 
1 year (magenta), 3 years (cyan), and 5 years (blue). 
The dashed curves represent the number of
galaxies$\times$1e7 for the three mission durations.}
\label{fig:fomyr}
\end{figure}
%%%%%%%%%%%%%%%%%%%%%%%%%%%%%%%%%%%%%%%%%%%%%%%%%%%%%%%%%%%%%%%%%%%

We explore the influence of the mission duration for a 1.5m telescope
diameter and an eight-filter configuration with a photo-z quality selection of
$\epsilon=0.1$. In {\bf Figure \ref{fig:fomyr}}, we computed as a function of the 
exposure time per pointing the WL FoM (solid lines)
and the total number of galaxies (dashed lines) for a 1, 3, and 5-year mission.
The green dashed area shows the CMC incompleteness zone. 
This green-dashed part of the graph 
will be affected by the COSMOS limiting depth.
We find that the optimal configuration for a one year survey is a 3000s observing time by
pointing, which represents 150s per visible filter for a total survey area of
3200deg$^2$.
For a 3-year mission, the optimal observation time per pointing is 3600s, or 180s 
per visible filter, for a total survey area of 7300deg$^2$. 
For a 5-year mission, the optimal observation time per pointing is 
then 3600s or 180s per visible filter for a total survey 
area of 11000deg$^2$. This optimal exposure time corresponds to a number density 
of useful galaxies of 45gal/arcmin$^2$ at 10$\sigma$ in I-band. The shape of the curves indicate that 
longer missions reach their maximum FoM for slightly deeper exposures rather than shallower exposures. 
A one-year mission is likely limited by the RON regime (as the Galactic 
absorption is less important on scale of 3000 square degrees). 
As the mission duration increases, the Galactic absorption becomes 
more dominant and a deeper photometry provides more galaxies 
than a larger survey area.

We note that one can trade mission time with the size of the field-of-view 
(number of detectors assuming a fixed pixel scale) as discussed in 
section \ref{subsec:area}. For example, a 5-year mission with a 0.3 
square degrees FOV is in principle similar to a 3-year mission with 
a 0.5 square degrees FOV.

\subsubsection{Impact of the pixel scale $p^{vis}$}

\begin{figure}[!ht]
\resizebox{\hsize}{!}{\includegraphics{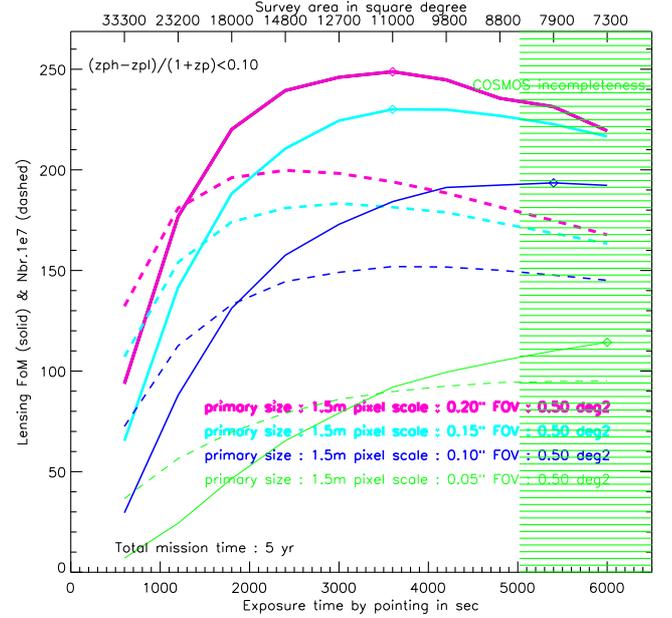}}
\caption{WL  FoM as a function of exposure time by pointing
  for a pixel scale of 0.2''(magenta), 0.15''(cyan), 0.1''(blue), and 0.05''(green) 
  using an 8-filter set with a resolution of $\mathcal{R}=3.2$. The dashed curves 
  represent the number of galaxies$\times$1e7 for all configurations.}
\label{fig:fompix}
\end{figure}

Next we compute the impact of the different pixel scales on the WL FoM as 
a function of the exposure time per pointing, assuming: a 1.5m
primary mirror diameter, an 8-filter configuration, a 5-year mission, 
and a FOV of 0.5deg$^2$. 
{\bf Figure \ref{fig:fompix}} shows that a larger pixel scale results in a larger FoM. 
We note that the maximum of the FoM is obtained at very different exposure 
times per pointing: smaller pixel scales 
require longer exposure times to reach the photon noise regime. 
The gain going from small to large pixels is significant in terms of FoM, 
although the gain from 0.15'' to 0.2'' is only 10\% in FoM. 
As there are concerns about how well the shape can be measured if the sub-dither 
is not optimal, we assume a pixel scale of 0.15'' to be an optimal choice.

%%%%%%%%%%%%%%%%%%%%%%%%%%%%%%%%%%%%%%%%%%%%%%%%%%%%%%%%%%%%%%%%%%%
%%%%%%%%%%%%%%%%%%%%%%%%%%%%%%%%%%%%%%%%%%%%%%%%%%%%%%%%%%%%%%%%%%%

\subsubsection{Impact of the primary mirror diameter $D_{1}$}

\begin{figure}[!ht]
\resizebox{\hsize}{!}{\includegraphics{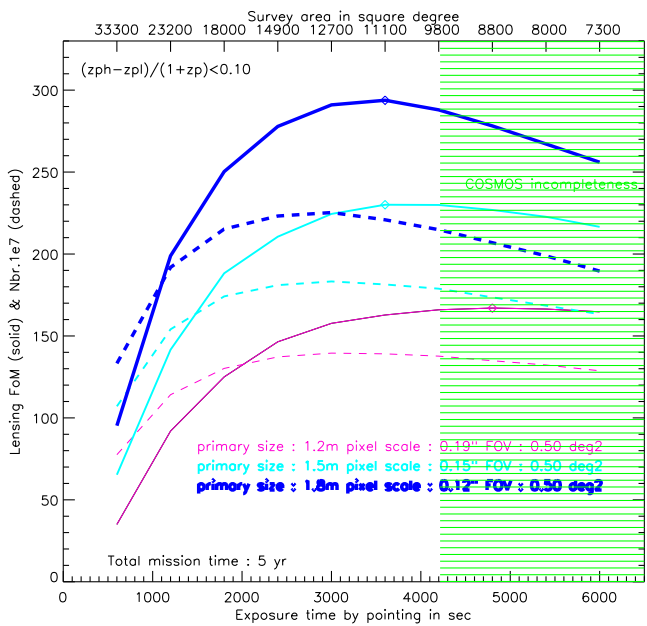}}
\caption{WL  FoM as a function of exposure time by pointing
  for a primary mirror diameter of 1.2m, 1.5m, and 1.8m with a pixel scale
  varying in order to have the same optimal effective PSF sampling using a
  8-filter set with a resolution of $\mathcal{R}=3.2$. The FOV is
  fixed at 0.5deg$^2$, which implies that the number of pixels decrease
  when the pixel scale is larger. The dashed curves represent the number of galaxies$\times$1e7 for all configurations.}
\label{fig:fom_fixedFOV}
\end{figure}

We then consider the impact of the primary mirror diameter on the WL FoM 
results in {\bf Figure \ref{fig:fom_fixedFOV}}. We use the optimal pixel 
scale of 0.15'' for a primary mirror diameter of 1.5m as argued in section 
\ref{subsec:ePSF}, and adapt the pixel scale 
to have the same sampling of the ePSF in the 1.2m and 1.8m case.

An interesting and not intuitive result is
that when using a primary mirror diameter of 1.5m, the mission time is
better employed in going deeper on smaller areas. In  5-year survey, we should then
cover about $\sim 10000$deg$^2$ with 4000sec by pointing, which represents 800s 
by filters in four exposures of 200s (at the Galactic pole) for the visible camera. 
With a primary mirror diameter of 1.8m, one can cover a wider area ($\sim 12000$deg$^2$) 
since a larger mirror
allows us to reach a higher galaxy density in a smaller exposure time. 
With a smaller primary mirror diameter of 1.2m, the optimal survey
covers a smaller area $\sim 8000$deg$^2$ with longer exposure time per pointing.

We note that we keep the FoV at 0.5deg$^2$ meaning that the number of pixels is
not fixed for different telescope diameters. 

%%%%%%%%%%%%%%%%%%%%%%%%%%%%%%%%%%%%%%%%%%%%%%%%%%%%%%%%%%%%%%%%%%%
%%%%%%%%%%%%%%%%%%%%%%%%%%%%%%%%%%%%%%%%%%%%%%%%%%%%%%%%%%%%%%%%%%%
%\begin{figure}[!ht]
%\resizebox{\hsize}{!}{\includegraphics{lensing_8Bfom_5tomobins_pix_equal_sampling_fixedNpix_expotime_zerr0.10_5yr_mir.ps}}
%\caption{WL  FoM as a function of exposure time by pointing
%  for a primary mirror diameter of 1.2m, 1.5m, 1.8m with a pixel scale
%  varying in order to have the same optimal PSF sampling using a
%  8-filter set with a resolution of $\mathcal{R}=3.2$. The number of
%  pixels is fixed which makes the FOV varying.}
%\label{fig:fom_equalconfig}
%\end{figure}
%%%%%%%%%%%%%%%%%%%%%%%%%%%%%%%%%%%%%%%%%%%%%%%%%%%%%%%%%%%%%%%%%%%

\section{Conclusion} 
\label{sec:conclusion}

On the basis of our realistic mock catalogs representative of faint galaxies 
in the Universe (Paper I), we have optimized the observational strategy and 
instrument parameters of possible future DE WL space missions, 
focussing in particular on the impact of the
photometric redshift quality on the DE constraints.

In section 2, we have described the photometric redshift technique and photometric 
uncertainties of a space survey. 
In particular, we have defined the parameters used to quantify the 
quality of the photometric redshifts used in the WL analysis.

To address the observational strategy of a WL space mission,
we then defined, in section 3, the key properties of the instrument, which consists 
of a visible camera and an infrared camera. We computed the effective PSF, the
overall throughput, and the sensitivities, which depend on the detector characteristics, 
the pixel size, the exposure time, the dithering strategy, and the value of 
the sky background.
In particular, we investigate the impact of the pixel scale on the sampling of the
effective PSF and place upper limits on the pixel scale values. We also suggest an 
optimal pixel scale based on simple arguments, which is further investigated 
in section 6 of the paper.

In section 4, we conducted a detailed analysis of the optimal filter set to minimize the
photometric redshift scatter and biases along with the number of catastrophic redshifts.
We found that square filters of resolution $\mathcal{R}\sim 3.2$ maximize 
the photometric redshift quality for the faint galaxy population used for 
WL measurements in the cases of 6, 7 or 8-filter sets. 
For a resolution of $\mathcal{R}\sim 3.2$, the filter width 
maximizes the S/N along with the galaxy colors resulting in an optimal 
photometric redshift quality. 
%without being to large to wash out the redshift information contained in the photometric data. 
We showed that this $\mathcal{R}\sim 3.2$ resolution maximizes the FoM of the 
WL tomography irrespective of the survey area of the mission.

In section 5, we addressed the issue of the blue sensitivity of the visible camera 
by looking at the quality of the photometric redshift and in particular 
the catastrophic redshift rate. We found that improving the blue sensitivity can 
reduce the catastrophic redshift rate by a factor of two in the magnitude range we 
are interested in for WL analysis. 
The detector blue sensitivity is then a critical parameter since it will help 
reduce biases in the WL
cosmological constraints coming from the photometric redshift quality. 
We propose a minimum throughput of 30\% at 3600 $\AA$  for the visible camera. 

Finally, in section 6, we investigated the impact of some key instrument parameters 
and the observational strategy to maximize the FoM of a WL survey. 
In particular, we showed  
that the instrument parameters and survey strategy are both important in optimizing a WL 
space mission and that they cannot be addressed separately.
We thus defined the survey strategy computing the survey area as a function 
of the exposure time per pointing, the field-of-view of the instrument, 
the mission duration, and the sky model, which includes both the 
Galactic absorption and Zodiacal light variation.

Quantitatively, assuming a 1.5m primary mirror telescope and investigating the ePSF 
sampling, we argued that a pixel scale of 0.15'' for the visible camera is the 
optimal choice to conduct an efficient WL survey. 
This choice is the best trade-off value that is small enough to ensure a 
sufficiently well-sampled ePSF after dithering and maximize the FoM unlike 
smaller pixel size. However, this result is based on simple arguments and 
assumes that the PSF of the instrument is sufficiently stable and can be 
well calibrated - this should be a strong requirement of the mission and
needs an in-depth optimisation.

Furthermore, assuming a 0.5 square degree field-of-view for both visible and 
infrared cameras, and an eight-filter configuration with a $R=3.2$ filter-set, 
we demonstrated that a $\sim$11\,000 sq.deg survey reaching a homogeneous depth 
of I$_{AB}$=25.6 (at 10$\sigma$) (which can be achieved with four exposures 
of $\sim$200s per filter, at the Galactic poles) 
is the optimal and safest combination to maximize the weak lensing FoM. 
The survey strategy consists of varying the exposure time to keep an equal 
galaxy number density across the survey area.
At this depth, the effective number density of galaxies that can then be used for WL 
is ~45gal/arcmin$^{2}$, a factor of two better than a WL ground-based survey. 

In particular, we show that:
\begin{itemize}
\item An 8-filter configuration is better than a either 6 or 7-filter configuration, 
which both provides poorer photometric redshift quality and lead to a larger 
number of catastrophic redshifts.
\item The proper calculation of the survey strategy, including the Galactic absorption and the Zodiacal light variations as a function of the position of the sky, strongly limits the optimal survey size of the WL probe to a maximum of $\sim$ 11\,000 sq.deg. 
The exact optimal survey size will depend on the total mission time, the pixel scale of 
the visible detector, the primary mirror diameter of the telescope, and the exposure 
time per pointing, which should all be adapted accordingly. 
\end{itemize}

These conclusions are drawn from a survey strategy that ensures an homogeneus 
depth across the survey area. Different survey strategies
may be interesting to consider. A trade-off between survey depth and sky area covered 
might lead to improved dark energy constraints.  
We stress that our analysis is still relatively simplified, and that further
work is needed to investigate some important issues. In particular:
1) our figure of merit remain simple, as they do not take into account errors
produced by catastrophic redshift; 2) the pixel size optimization is done here 
using simple arguments, and would benefit from a proper analysis conducted 
in the global optimization scheme described in this paper; 3) the high number 
density of stars at low Galactic latitude will impact the useful survey area and 
thus the number of useful galaxies to be used in the WL survey, which may shift the 
optimal area to even smaller size.

To conclude, our analysis addresses the complex optimization of future WL DE space 
survey by including both the instrument parameters and the observational strategy. 
Since some of our results were unexpected, we believe 
that a full optimization that includes both the instrument parameters and the 
observation strategy \emph{is required} to maximize the cosmological constraints 
of the future DE space mission.

\begin{acknowledgements}
  We thank Pierre-Yves Chabaud for his time in helping us with technical details. 
  We acknowledge useful discussions with
  members of the COSMOS and SNAP collaborations. St\'ephanie Jouvel
  thanks CNES and CNRS for her PhD studentship and thanks STFC 
  for her postdoctoral support. Jean-Paul Kneib thanks
  CNRS and CNES for support. Gary Bernstein is supported by grant AST-0607667
  from the National Science Foundation, Department of Energy grant
  DOE-DE-FG02-95ER40893, and NASA grant BEFS-04-0014-0018.
\end{acknowledgements}

\bibliographystyle{aa}
\bibliography{biblio}
%\newpage

\appendix

\section{Typical noise properties for a space based survey}

\subsection{The galaxy flux $e_{signal}$}
\begin{figure}[!ht]
  \resizebox{\hsize}{!}{\includegraphics{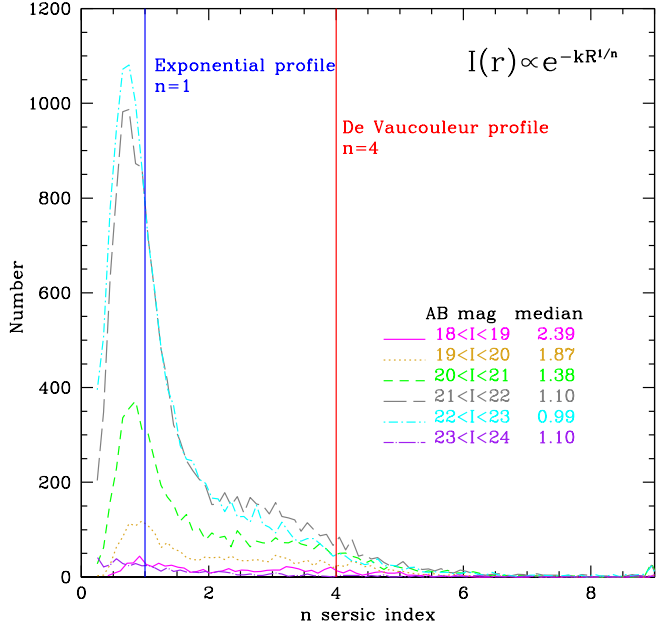}}
  \caption{Histogram of the S\'{e}rsic indices for the zCOSMOS
    survey.}
  \label{fig:sersic}
\end{figure}

To compute the galaxy flux $e_{signal}$, we derive the total
flux received by the telescope and the distribution of the light over
the detectors.  The former is given for each band by the magnitude in
the mock catalog, while the latter depends on the galaxy profile and
the observed effective radius $r_{eff}$, as well as the resolving
power of the telescope.  We assume for all galaxies an intrinsic
exponential profile following:
\begin{equation}\label{eq:galprofile}
I(r) = exp\bigg(-1.6785.\Big(\frac{r}{r_{eff}}\Big)\bigg).
\end{equation}
In analyzing the S\'{e}rsic indices (S\'{e}rsic 1968) of the
galaxy profiles of the zCOSMOS survey \citep{Tasca09}, we found that
most galaxy profiles tend to follow an exponential profile (S\'{e}rsic
index close to n=1) as shown in {\bf Figure~\ref{fig:sersic}}. This is particularly true
for the faint galaxies that represent the bulk of the galaxies used for WL measurement.

The resolving power of the telescope is the size the Airy disk of the
telescope and its FWHM is defined by
\begin{equation}
FWHM[Airy\ Disk]=FWHM_{tel} = 1.02. \frac{\lambda}{D_1},
\end{equation}
where $\lambda$ is the mean wavelength in the filter considered and
$D_1$ the diameter of the primary mirror. We assume that the FWHM of
the galaxy can be roughly derived from the convolution of five Gaussians
representing: 1) the PSF of the telescope $FWHM_{tel}$, 2) the pixel scale
$p=p^{ir/ccd}$, 3) the pixel diffusion $\sigma_d$, 4) the jitter of the
telescope $\sigma_j$ (see Table \ref{tab:jdem}), and 5) the galaxy
theoretical FWHM $2xr_{eff}$. Thus the observed FWHM of a galaxy is
expressed as

%% JPK - I change the 0.5 into a 0.2 in section 3 because FWHM=2.36\sigma
\begin{eqnarray}
\lefteqn{FWHM^{obs}_{gal} = }
\nonumber\\
& & {} \sqrt{FWHM_{tel}^2+0.5\left(p\right)^2+(2.r_{eff})^2+2.36\sigma_j^2+\left(2.36\sigma_d.\frac{p^{vis/ir}}{0.1}\right)^2},
\nonumber\\
\end{eqnarray}
% FWHM=2.r_eff
where the term containing the pixel scale contains a 0.5 factor
representing the combination of the rms size of a box and the 2.36
factor needed to convert the rms in FWHM. We defined the fraction of
photons $frac_{phot}$ as the luminosity enclosed inside $1.4\
FWHM^{obs}_{gal}$ that we calculated using the galaxy intensity
profile defined by Eq. \ref{eq:galprofile}. The fraction of photons received is then
converted into the number of electrons from the source. Thus, the total
number of electrons is
\begin{equation}\label{eq:esignal}
e_{signal} = \gamma_{signal}.\Big[T_{obs}.S.\eta.N_{expo}\Big].frac_{phot},
\end{equation}
where $N_{expo}$ and $S$ are listed in the Table \ref{tab:jdem} and
$\gamma_{signal}$ is the total number of galaxy photons arriving at
the telescope, $T_{obs}$ the observation time, and $\eta$ the total
transmission including the telescope, the mirror reflectivity, the
detector QE, and the filters and optics transmissions,
whose values are given in Table \ref{tab:mag_noise}.

\subsection{The background flux $e_{sky}$}

To calculate the sky emission $e_{sky}$, we used the Zodiacal
light parametrisation described in \cite{Aldering02}, which gives a
useful approximation of the sky background fluxes using a broken
log-linear relation in $erg\textrm{
}cm^{-2}s^{-1}\AA^{-1}arcsec^{-2}$
\begin{eqnarray}
log_{10}(f(\lambda))&=&17.755 \quad \textrm{for} \quad 0.4<\lambda<0.6 \mu m, \\
log_{10}(f(\lambda))&=&17.755-0.73(\lambda-0.61\mu m) \textrm{ for }0.6<\lambda<2.2.
\nonumber \\
\end{eqnarray}
We then need to derive the number of pixels corresponding to 
$frac_{phot}$, which corresponds to the number of pixels
within the circular area of radius $1.4\ FWHM^{obs}_{gal}$
\begin{equation}
N_{pix} = \pi \  \left(\frac{1.4\ FWHM^{obs}_{gal}}{p}\right)^2.
\end{equation}
We can thus derive the number of electrons from the Zodiacal light in
a similar way as we did for the number of electrons produced by the
galaxy flux $e_{signal}$
\begin{equation}
e_{sky} = \gamma_{sky}\Big[T_{obs}.S.\eta.N_{expo}.N_{pix}\Big]\Omega,
\end{equation}
where ($\Omega$,$N_{expo}$,$T_{obs}$,$S$) are defined in Table
\ref{tab:jdem} and $\eta$ is the total transmission of the
telescope. We note that the difference between the number of electrons
produced by both a galaxy $e_{signal}$ and the Zodiacal light
$e_{sky}$, lies in the terms $frac_{phot}$ for galaxies and the term
$N_{pix}.\Omega$ for the Zodiacal light. We assume a total
galaxy flux after being affected by the instrument efficiency for a given pixel scale,
PSF convolution, jitter that decreases the flux by a fraction of it
$frac_{phot}$, while the total Zodiacal light is computed in the
circular area defined by $1.4\ FWHM^{obs}_{gal}$ corresponding to
$N_{pix}\Omega$ arcsec$^2$.

\subsection{The detector noises}
The detector noises correspond to the dark current from the thermal radiation of
detectors and the read-out noise due to the electron motion when
reading the detector. The dark current is thus expressed in
$e^-/pix/s$, while the read-out noise is in $e^-/pix$ since this is not
time dependent. The visible  and IR detectors are a compound of different
materials and thus have different characteristics. The visible 
detectors are CCDs or charge coupled device using semi-conductor
materials such as silicon and work mainly in the visible wavelength
range. The HgCdTe are IR detectors with mercury cadmium telluride. The
wavelength response can be varied by adjusting the alloy composition
of this ternary compound. Both detectors need to be cooled to
decrease both types of noise. The values of the dark current and read-out
noise are listed in Table \ref{tab:jdem}.

%%%%%%%%%%%%%%%%%%%%%%%%%%%%%%%%%%%%%%%%%%%%%%%%%%%%%%%%%%%%%%%%%%%
\begin{table}[!ht]
  \caption{Telescope characteristics studied.}
\begin{tabular}{cccccccc} \hline\hline Quantity & abbrev(if necessary) & Values \\ 
 Primary mirror & $D_1$ & 1.5m \\ 
 Secondary mirror & $D_2$ & 0.6m \\ 
 Observation time & $T_{obs}$ & 200s\\
 Field of view & FOV & 0.5deg$^2$\\ 
 Nbre of exposure & $N_{expo}$ & 4 \\ 
 Collecting area & $S$ & $\pi\Big(\big(\frac{D_1}{2}\big)^2-\big(\frac{D_2}{2}\big)^2\Big)$\\
 Pixel solid angle & $\Omega$ & $p^2$ \\
 Telescope jitter & $\sigma_j$ & 0.02" \\
 Pixel diffusion scale & $\sigma_d$ & 0.4 pixel\\  %%JPK I change this value
 Pixel diffusion size & &2.6$\mu$\\
 CCD pixel scale & $p^{vis}$ & 0.15" \\ %%JPK I change this value
 CCD pixel size & & $10.5 \mu$m \\
 CCD read-out noise & $e^{vis}_{RON} $ & 6 $e^-/pix$ \\
 CCD dark current noise & $e^{vis}_{dark}$ & 0.03 $e^-/pix/s$\\
 HgCdTe pixel scale & $p^{ir}$ & 0.26" \\ %%JPK I change this value
 HgCdTe pixel size & & $18 \mu$m \\ 
 HgCdTe read-out noise & $e^{ir}_{RON}$ & 5 $e^-/pix$ \\
 HgCdTe dark current noise & $e^{ir}_{dark}$ & 0.05 $e^-/pix/s$\\
 \hline \hline
\label{tab:jdem}
\end{tabular}
\end{table}
%%%%%%%%%%%%%%%%%%%%%%%%%%%%%%%%%%%%%%%%%%%%%%%%%%%%%%%%%%%%%

\end{document}